\documentclass[hyper,a4paper,notoc]{JHEP3}

\pdfoutput=1
\usepackage{graphicx}
\usepackage{graphics}
\usepackage[mathscr]{eucal}
\usepackage{amssymb}
\usepackage{amsmath}
\usepackage{xspace}
\usepackage{verbatim}
\usepackage{float} 
\usepackage{listings}
\usepackage{longtable}
\usepackage{courier}

\newcommand\FeynArts{{\tt FeynArts}\xspace}
\newcommand\FormCalc{{\tt FormCalc}\xspace}
\newcommand\CalcHep{{\tt CalcHep}\xspace}
\newcommand\CompHep{{\tt CompHep}\xspace}
\newcommand\SARAH{{\tt SARAH}\xspace}

\newcommand\SPheno{{\tt SPheno}\xspace}
\newcommand\MicrOmegas{{\tt micrOMEGAs}\xspace}
\newcommand\HB{{\tt HiggsBounds}\xspace}
\newcommand\WHIZARD{{\tt WHIZARD}\xspace}
\newcommand\OMEGA{{\tt O'Mega}\xspace}
\newcommand\SSP{{\tt SSP}\xspace}
\newcommand\Mathematica{{\tt Mathematica}\xspace}
\newcommand\VAMP{{\tt VAMP}\xspace}
\newcommand\FeynRules{{\tt FeynRules}\xspace}
\newcommand\SINDARIN{{\tt SINDARIN}\xspace}

\newcommand\Neu{\tilde{\chi}^0}
\newcommand\Cha{\tilde{\chi}^-}
\newcommand\Chap{\tilde{\chi}^+}
\newcommand{\InM}[1]{\textit{\scriptsize{#1}}}

\lstdefinelanguage{bash}{morekeywords={cd, cp, mv, make,Start,MakeTeX,MakeSPheno,MakeFeynArts,MakeCHep,MakeWHIZARD,CalcRGEs,Vertex,CalcLoopCorrections,MakeVertexList,TadpoleEquation,MassMatrix},
  morecomment=[s]{In[}{]},commentstyle=\InM}
\catcode`\@=11
\font\manfnt=manfnt
\def\Watchout{\@ifnextchar [{\W@tchout}{\W@tchout[1]}}
\def\W@tchout[#1]{{\manfnt\@tempcnta#1\relax%
  \@whilenum\@tempcnta>\z@\do{%
    \char"7F\hskip 0.3em\advance\@tempcnta\m@ne}}}
\let\foo\W@tchout
\def\dubious{\@ifnextchar[{\@dubious}{\@dubious[1]}}

\def\@dubious[#1]{%
  \setbox\@tempboxa\hbox{\@W@tchout#1}
  \@tempdima\wd\@tempboxa
  \list{}{\leftmargin\@tempdima}\item[\hbox to 0pt{\hss\@W@tchout#1}]}
\def\@W@tchout#1{\W@tchout[#1]}
\catcode`\@=12

\title{A tool box for implementing supersymmetric  models } 
\author{Florian Staub$^{1,a}$, Thorsten Ohl$^{1,b}$, Werner Porod$^{1,c}$,  Christian Speckner$^{2,d}$ \\
$^1$Institut f\"ur Theoretische Physik und Astrophysik, Universit\"at W\"urzburg,\\
D-97074  W\"urzburg, Germany\\
$^3$%
Albert-Ludwigs-Universit\"at Freiburg\mbox{,} Physikalisches Institut\\
Hermann-Herder-Stra\ss{}e 3, D-79104 Freiburg, Germany%
\\
$^a$Email: \email{florian.staub@physik.uni-wuerzburg.de} \\
$^b$Email: \email{ohl@physik.uni-wuerzburg.de} \\
$^c$Email: \email{porod@physik.uni-wuerzburg.de} \\
$^d$Email: \email{Christian.Speckner@physik.uni-freiburg.de}
}

\preprint{FR-PHENO-2011-017}

\abstract{
We present a framework for performing a comprehensive analysis of a
large class of supersymmetric models, including spectrum calculation, dark
matter studies and collider phenomenology. To this end, the respective model
is defined in an easy and straightforward way using the \Mathematica package
\SARAH. \SARAH then generates model files for \CalcHep which can be used with
\MicrOmegas as well as model files for \WHIZARD and \OMEGA. In addition,
Fortran source code for \SPheno is created which facilitates the determination
of the particle spectrum using two-loop renormalization group equations and
one-loop corrections to the masses. As an additional feature, the generated
\SPheno code can write out input files suitable for use with \HB to apply
bounds coming from the Higgs searches to the model. Combining all program
provides a closed chain from model building to phenomenology.}

\begin{document}
\tableofcontents


\section{Introduction}
With the first collisions at the  Large Hadron Collider
(LHC), a new era of high
energy physics has started. The LHC is designed to get new and deeper insights
into the fundamental principles governing our physical world. It is not only supposed  to find
the last missing particle of the Standard Model (SM) of particle physics, the
Higgs boson, but also to discover possible physics beyond the SM.

Despite having proven itself as a
successful and precise description of all experiments in particle physics for
last 30 years, the SM suffers from several theoretical shortcomings, and
therefore a host of
extensions of the model which lead to such new effects have been devised over
the last years. Among the most prominent examples of such conceptual flaws which
plague the model are the hierarchy problem
\cite{Weinberg:1975gm,Weinberg:1979bn}
and the lack of a candidate for Dark Matter.

Supersymmetry (SUSY) is arguably among the best studied extensions of
the SM \cite{Ramond:1971gb,Wess:1974tw,Volkov:1973ix} which have been developed to
cure these problems. Already the
minimal, supersymmetric extension of the SM, the
MSSM (Minimal Supersymmetric Standard Model)
\cite{Inoue:1982pi,Haber:1984rc,Djouadi:1998di}, solves the hierarchy problem,
is capable of providing a
dark matter candidate \cite{Goldberg:1983nd, Ellis:1983ew}, leads to
gauge coupling unification
\cite{Langacker:1991an,Ellis:1990wk,Giunti:1991ta,Hall:1980kf} and relates
electroweak symmetry breaking to the
large top mass \cite{Martin:2001vx,Ibanez:1982fr}. Because of these appealing
features, the MSSM was extensively
studied in the last decades and many software tools were developed for that
purpose.

However, the MSSM might also not be the final answer.
One reason is that the MSSM like the SM can't explain neutrino masses
\cite{Fukuda:1998mi,Eguchi:2002dm,Ahmad:2002jz,Schwetz:2011qt}. 
In priniple one can find for every SM extension designed to explain
the observed neutrino data a supersymmetric version. The most popular
among them are the various seesaw models 
\cite{Minkowski:1977sc,seesaw,MohSen,Schechter:1980gr,Cheng:1980qt,Foot:1988aq} 
 leading to an effective
 dim-5 operator generating Majorana masses for the neutrinos
\cite{Weinberg:1979sa,Weinberg:1980bf,Ma:1998dn}. Moreover,
there is an intrinsic supersymmetric mechanism to explain
neutrino data, namely the breaking of R-parity via
Lepton number violating interactions 
\cite{Hall:1983id,Romao:1999up,Hirsch:2000ef,Dedes:2006ni}. 
Another reason is the strong
CP problem which is also still present in the MSSM. One way to solve this is to
postulate  an additional \(U(1)\) group as proposed by Peccei and Quinn
\cite{Peccei:1977hh}. 
In SUSY,
this would lead to the presence of a pseudo scalar Axion, the scalar Saxion
and the fermionic Axino \cite{Covi:2009pq}. Another intrinsic problem of the MSSM is
the so called
\(\mu\) problem: above the breaking scale, the \(\mu\) parameter in the
superpotential of the MSSM is the only dimensionful parameter in the model.
The natural scale
of that parameter would therefore be either of the GUT scale or exactly 0 if it
is forbidden by symmetry. However, we know that it must be of order of
the electroweak scale to explain precision data \cite{Kim:1983dt}
and to fullfill the existing bounds from collider searches
\cite{Nakamura:2010zzi}.
One idea to
solve the \(\mu\)
problem is to create an effective \(\mu\) term just after SUSY breaking like in
the NMSSM 
\cite{Ellwanger:2009dp}. 

Of course, this list can be still extended, and over the years, people have come up
with many different ways of modifying and extending the MSSM.
However, the way from the first idea about a SUSY model to numerical results is
normally long and exhaustive: all analytical expressions for masses, vertices
and renormalization group equations have to derived. Code has to be generated
to calculate the numerical values for the masses and, if the results are to be
reasonable, to add loop corrections. Before the existing software tools to
calculate widths or cross sections as well as perform Monte Carlo studied can
be used, the new model has to be implemented. That demands not only a good
knowledge about the different programs but is also a very time consuming task.

In this paper, we present a framework which covers all of the aforementioned steps in
an automatized way and which goes also further. Based on \SARAH
\cite{Staub:2008uz,Staub:2009bi,Staub:2010jh}, a tool chain can be
created for a large variety of SUSY models which covers the spectrum calculation
with \SPheno \cite{Porod:2003um,Porod:2011nf}, the calculation of cross sections
with \CalcHep \cite{Boos:1994xb,Pukhov:2004ca} and the determination of the
dark matter relic density  using \MicrOmegas \cite{Belanger:2006is}, the
analysis of constraints coming from the Higgs searches via \HB
\cite{Bechtle:2008jh,Bechtle:2011sb} and Monte Carlo simulations of collider
observables with \WHIZARD
\cite{Moretti:2001zz,Kilian:2007gr}. The basic
idea is that the user can implement
the model in \SARAH in an intuitive and fast way. Afterwards, \SARAH generates
Fortran source code for \SPheno as well as model files for \CalcHep, \OMEGA and
\WHIZARD. Since the implementation of the model in all of the programs is based
on the one implementation in \SARAH, the conventions are the same for all
programs, greatly simplifying the transfer of information between the
different tools. For example, the spectrum file of \SPheno can be directly used by
\MicrOmegas to calculate the relic density. In addition, the new \SPheno modules include
the possibility to write input file for \HB.

We start with an introduction to \SARAH in sec.~\ref{sec:SARAH},
demonstrating how models can be implemented and what information can be derived
by \SARAH. In sec.~\ref{sec:SPheno}, we discuss the details of the Fortran
output of \SARAH for \SPheno, and in sec.~\ref{sec:HB}, we show how \HB can be applied
to the model. Afterwards, the link to \CalcHep and
\MicrOmegas is discussed in sec.~\ref{sec:CH_MO}, and the output for \OMEGA and
\WHIZARD is presented in sec.~\ref{sec:WO}. Finally, we
introduce the package \SSP which is designed for performing parameter scans
using the tools supported by \SARAH in sec.~\ref{sec:SSP} before explaining all
steps necessary to combine \SARAH and \SSP together with those programs into a
closed tool chain in sec.~\ref{sec:Combination}.

\section{Building new models with \SARAH}
\label{sec:SARAH}
\subsection{Overview}
\begin{figure}[t]
\vspace{-2cm}
\centering
\includegraphics[scale=0.7]{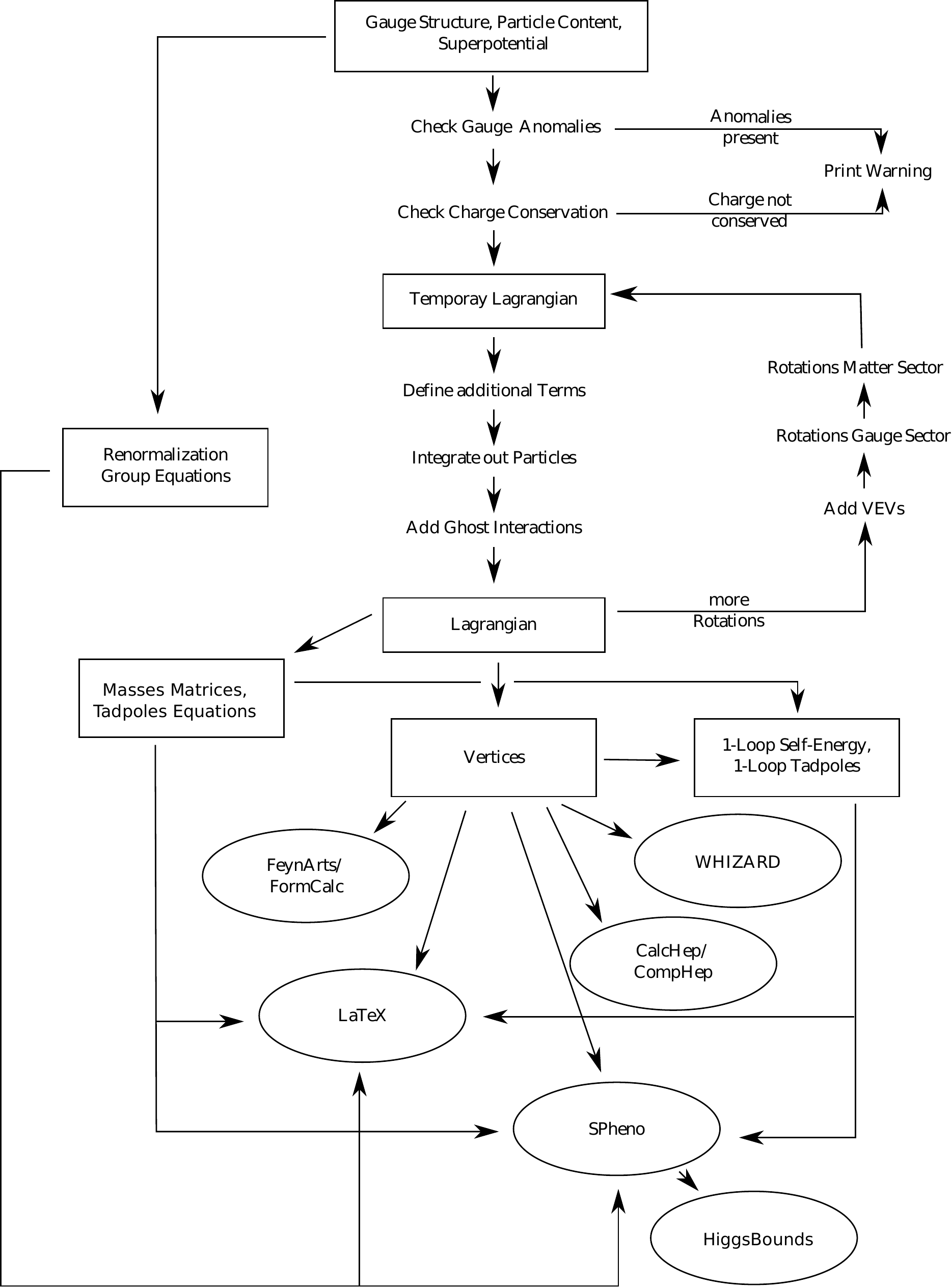}
\caption[Work-flow of \SARAH]{Schematic graph of the different steps performed
by \SARAH. The user has access to the calculated information shown in boxes. The
ellipses show the output which can be created by \SARAH.}
\label{fig:SARAH_workflow}
\end{figure}

\SARAH is a package for \Mathematica version 5.2 or higher and has been designed
to handle every \(N=1\) SUSY theory with an arbitrary direct product of \(SU(n)\) and/or
\(U(1)\) factors as gauge group. The chiral superfields can transform under arbitrary,
irreducible representations with regard to this gauge group, and all possible
renormalizable superpotential terms are supported. There are no restrictions
on either the number of gauge group factors, the number of chiral superfields or the
number of superpotential terms. Furthermore, any number of symmetry breakings or
field rotations is allowed. A schematic picture of the different steps performed
by \SARAH is shown in Fig.~\ref{fig:SARAH_workflow}.

\subsection{Download, installation and first evaluation}
\label{sec:down}
\SARAH can be downloaded from
\begin{lstlisting}
http://projects.hepforge.org/sarah/
\end{lstlisting}
The package should be extracted to the application directory of \Mathematica,
\begin{lstlisting}
$HOME/.Mathematica/Applications/
\end{lstlisting}
on a Unix system or
\begin{lstlisting}
[Mathematica-Directory]\AddOns\Applications\
\end{lstlisting}
in a Windows environment (\mbox{\texttt{\$HOME}} and
\mbox{\texttt{[Mathematica-Directory]}} should be substituted with the home
and \Mathematica installation directories respectively). 

Initially, the package itself consists  of three directories: the directory
\verb"Package" contains all \Mathematica package files, while in the directory
\verb"Models" the definitions of the different models are located. The third 
directory \verb"LaTeX" contains \LaTeX{} packages which are needed for the
appropriate output.  During execution, a fourth directory called \verb"Output"
is generated by \SARAH where the results of the different calculations as well
as the model files for the diagram calculators are stored. 

A comprehensive manual ({\tt 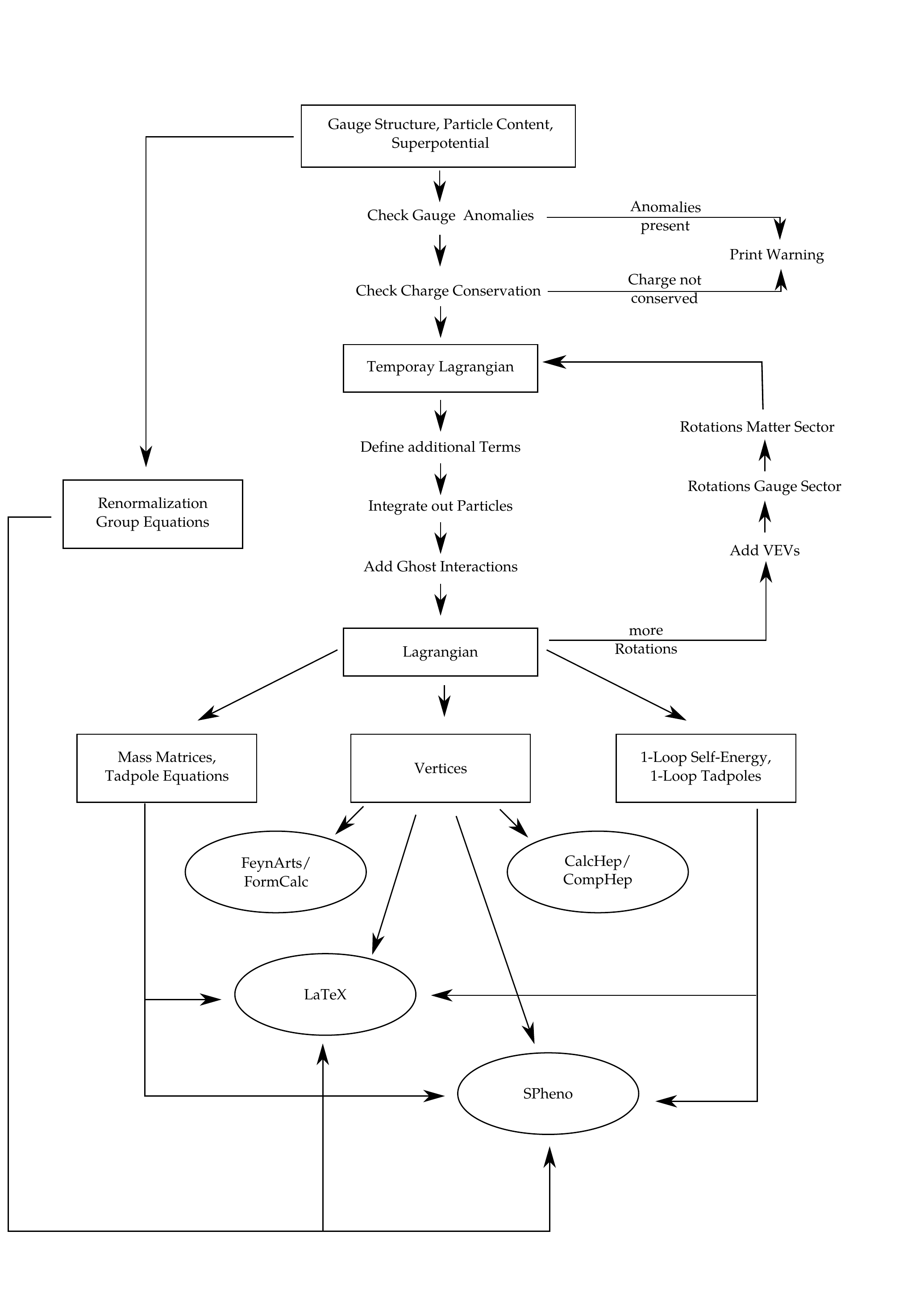}) is included in the package archive and
can also be found on the web page and on the arXiv \cite{Staub:2008uz}. In addition,
a file ({\tt models.pdf}) with information about all models delivered with the
package is part of the archive. Furthermore, a file with a short introduction to
the main commands is included ({\tt Readme.txt}) as well as an example
illustrating their use ({\tt Example.nb}).

After the installation, the package is loaded in \Mathematica via 
\begin{lstlisting}
In[1]: <<"sarah-3.0/SARAH.m" 
\end{lstlisting}
Subsequently, the model is initialized by  
\begin{lstlisting} 
In[2]: Start["Modelname"]; 
\end{lstlisting}
where \verb"Modelname" is the name of the corresponding model file.
As an example, the command would read
\begin{lstlisting} 
In[2]: Start["MSSM"]; 
\end{lstlisting}
for the minimal supersymmetric standard model and
\begin{lstlisting}
In[2]: Start["NMSSM","CKM"]; 
\end{lstlisting}
for the next-to-minimal supersymmetric model in CKM basis.

\subsection{Building new models with \SARAH}

As an example of how new models can be defined in \SARAH, we briefly go through
the existing MSSM model file and then show how this model can be augmented to
obtain an implementation of the
NMSSM (see \cite{Ellwanger:2009dp} and references therein).

\subsubsection{Model file for the MSSM}
\label{sec:MSSM_modelfile}
\lstset{frame=shadowbox}
In the following, we showcase the different components of the MSSM implementation in
\SARAH (see \cite{Staub:2010jh} for a summary of our conventions).
\begin{enumerate}
\item First, give a string as internal name for the model. Make sure to use only
numbers and letters and no spaces, because this name is later on used to
identify the model in \SPheno or \CalcHep. 
\begin{lstlisting}
ModelName = "MSSM";
ModelNameLaTeX ="MSSM";
\end{lstlisting}
{\tt ModelNameLaTeX} provides much more freedom: any symbol and spaces can be
used and even \LaTeX syntax is supported.
\item The gauge sector is \(U(1)\times SU(2)\times SU(3)\) and is defined
by declaring the corresponding vector superfields.
\begin{lstlisting}
Gauge[[1]]={B,   U[1], hypercharge, g1, False};
Gauge[[2]]={WB, SU[2], left,        g2, True};
Gauge[[3]]={G,  SU[3], color,       g3, False};
\end{lstlisting}
First, the name of the vector superfield is given. The second entry defines the
dimension of the group, the third is the name of the gauge group and the
forth gives the name of the corresponding gauge coupling. If the last entry is
set to {\tt True}, the sum over the group indices is expanded and the component
fields are distinguished; otherwise, the sum is left implicit.
In the above example, the color charges are written as indices, while the sum
over
the isospin multiplets is expanded.

Note that \SARAH automatically adds a soft-breaking gaugino mass for every
vector superfield.
\item The next step is to define the matter sector. That's done by the array
{\tt Fields}. The conventions are the following. First, the root of the names
for the component fields is given (e.g. {\tt X}): the derived names of the
fermionic components start with {\tt F} in front (i.e. {\tt FX}), while for
scalars a {\tt S} is used (i.e. {\tt SX}). At second position the number of
generations is defined and the third entry is the name of the entire
superfield. The remaining entries are the transformation properties with
respect  to the different gauge groups.

Using these conventions, the doublet superfields \(\hat{q},\hat{l}, \hat{H}_d,
\hat{H}_u\) are added by
\begin{lstlisting}
Fields[[1]] = {{uL,  dL},  3, q,   1/6, 2, 3};  
Fields[[2]] = {{vL,  eL},  3, l,  -1/2, 2, 1};
Fields[[3]] = {{Hd0, Hdm}, 1, Hd, -1/2, 2, 1};
Fields[[4]] = {{Hup, Hu0}, 1, Hu,  1/2, 2, 1};
\end{lstlisting}
While for the singlet superfields \(\hat{d}^c, \hat{u}^c, \hat{e}^c\) 
\begin{lstlisting}
Fields[[5]] = {conj[dR], 3, d,  1/3, 1, -3};
Fields[[6]] = {conj[uR], 3, u, -2/3, 1, -3};
Fields[[7]] = {conj[eR], 3, e,    1, 1,  1};
\end{lstlisting}
is used. 

Note that for scalars \SARAH also adds the soft masses automatically. 
\item The  superpotential of the MSSM is
\begin{equation}
\label{superpotential_MSSM}
W =  \hat{u}^c Y_u \hat{q} \hat{H}_u -  \hat{d}^c Y_d \hat{q} \hat{H}_d
  - \hat{e}^c Y_e \hat{l} \hat{H}_d  +\mu \hat{H}_u \hat{H}_d
\end{equation}
and represented in \SARAH by
\begin{lstlisting}
SuperPotential = { {{1, Yu},{u,q,Hu}}, {{-1,Yd},{d,q,Hd}},
                   {{-1,Ye},{e,l,Hd}}, {{1,\[Mu]},{Hu,Hd}}  };
\end{lstlisting}
\item There are two different sets of eigenstates: the gauge eigenstates before
EWSB and the mass eigenstates after EWSB. The internal names are
\begin{lstlisting}
NameOfStates={GaugeES, EWSB};
\end{lstlisting}
\item The gauge fixing terms for the unbroken gauge groups are
\begin{lstlisting}
DEFINITION[GaugeES][GaugeFixing]=
  { {Der[VWB],  -1/(2 RXi[W])},
    {Der[VG],   -1/(2 RXi[G]) }};	
\end{lstlisting}
This corresponds to
\begin{equation}
\mathscr{L}_{GF} = -\frac{1}{2 \xi_W}|\partial_\mu W^{\mu,i}|^2
   -\frac{1}{2 \xi_g}|\partial_\mu g^{\mu,i}|^2
\end{equation}
The gauge fixing terms are used for the calculation of the Ghost interactions
in general \(R_\xi\) gauge. These are used to write the vertices for \FeynArts
depending on gauge fixing constants, while the \CalcHep output is restricted to
unitary and 't Hooft gauge. Internal computations like those of the one-loop
self-energies are performed in 't Hooft gauge by \SARAH. 
\item The vector bosons and gauginos rotate after EWSB as follows
\begin{lstlisting}
DEFINITION[EWSB][GaugeSector] =
{ {{VB,VWB[3]},{VP,VZ},ZZ},
  {{VWB[1],VWB[2]},{VWm,conj[VWm]},ZW},
  {{fWB[1],fWB[2],fWB[3]},{fWm,fWp,fW0},ZfW}
};   
\end{lstlisting}
This encodes the common mixing of vector bosons and gauginos after EWSB
\begin{equation} 
\left(\begin{array}{c} 
B\\ 
W^3\end{array} \right) 
 =  \,Z^{\gamma Z}
\left(\begin{array}{c} 
\gamma\\ 
Z\end{array} \right),\thickspace
\left(\begin{array}{c} 
W^1\\ 
W^2\end{array} \right) 
 =  \,Z^{W}
\left(\begin{array}{c} 
W^-\\ 
(W^-)^*\end{array} \right),\thickspace 
\left(\begin{array}{c} 
\lambda_{{\tilde{W}},{1}}\\ 
\lambda_{{\tilde{W}},{2}}\\ 
\lambda_{{\tilde{W}},{3}}\end{array} \right) 
 =  \,Z^{\tilde{W}}
\left(\begin{array}{c} 
\tilde{W}^-\\ 
\tilde{W}^+\\ 
\tilde{W}^0\end{array} \right) \\ 
\end{equation} 
The mixing matrices can easily be parameterized in \SARAH to have the standard form
\begin{equation} 
Z^{\gamma Z}= \, \left( 
\begin{array}{cc} 
\cos\Theta_W  & - \sin\Theta_W   \\ 
 \sin\Theta_W  & \cos\Theta_W \end{array} 
\right),\thickspace  
Z^{W}= \, \left( 
\begin{array}{cc} 
\frac{1}{\sqrt{2}} & \frac{1}{\sqrt{2}} \\ 
 -i \frac{1}{\sqrt{2}}  & i \frac{1}{\sqrt{2}} \end{array} 
\right),\thickspace  
Z^{\tilde{W}}= \, \left( 
\begin{array}{ccc} 
\frac{1}{\sqrt{2}} & \frac{1}{\sqrt{2}} & 0 \\ 
 -i \frac{1}{\sqrt{2}}  & i \frac{1}{\sqrt{2}}  & 0 \\ 
 0 & 0 & 1\end{array} 
\right) \\ 
\end{equation}

\item The neutral components of the scalar Higgs receive vacuum expectation
values (VEVs) \(v_d\)/\(v_u\) and split into scalar and pseudo scalar components
\begin{align} 
H_d^0 =  \, \frac{1}{\sqrt{2}} \left( v_{d}  + i \sigma_{d}  +  \phi_d  \right)
\, , \hspace{1cm} 
H_u^0 =  \, \frac{1}{\sqrt{2}} \left( v_{u}    + i  \sigma_{u}  +  \phi_u
\right) 
\end{align} 
This is encoded in \SARAH by
\begin{lstlisting}
DEFINITION[EWSB][VEVs]= 
{{SHd0,{vd,1/Sqrt[2]},{sigmad,I/Sqrt[2]},{phid,1/Sqrt[2]}},
 {SHu0,{vu,1/Sqrt[2]},{sigmau,I/Sqrt[2]},{phiu,1/Sqrt[2]}}};
\end{lstlisting}
\item After EWSB the particles to new mass eigenstates
\begin{lstlisting}
DEFINITION[EWSB][MatterSector]= 
{{{SdL, SdR           }, {Sd, ZD}},
 {{SuL, SuR           }, {Su, ZU}},
 {{SeL, SeR           }, {Se, ZE}},
 {{SvL                }, {Sv, ZV}},
 {{phid, phiu         }, {hh, ZH}},
 {{sigmad, sigmau     }, {Ah, ZA}},
 {{SHdm, conj[SHup]   }, {Hpm,ZP}},
 {{fB, fW0, FHd0, FHu0}, {L0, ZN}}, 
 {{{fWm, FHdm}, {fWp, FHup}}, {{Lm,U},  {Lp,V}}},
 {{{FeL},       {conj[FeR]}}, {{FEL,ZEL},{FER,ZER}}},
 {{{FdL},       {conj[FdR]}}, {{FDL,ZDL},{FDR,ZDR}}},
 {{{FuL},       {conj[FuR]}}, {{FUL,ZUL},{FUR,ZUR}}} }; 
\end{lstlisting}
This defines the mixings to the mass eigenstates: first, a list with gauge
eigenstates is given, followed by the name of the new mass eigenstates and the
mixing matrix. Hence, the first line is interpreted as
\begin{align} 
\tilde{d}_{L,{i \alpha}} = \sum_{j=1}^3 Z^{D,*}_{j i}\tilde{d}_{{j \alpha}}\,,
\hspace{1cm} 
\tilde{d}_{R,{i \alpha}} = \sum_{j=1}^3 Z^{D,*}_{j+3 i}\tilde{d}_{{j \alpha}}
\end{align} 
while the 8th line defines the mixing in the chargino sector
\begin{align} 
\tilde{W}^- = \sum_j U^*_{j 1}\chi^-_{{j}}\,, \hspace{1cm} 
\tilde{H}_d^- = \sum_j U^*_{j 2}\chi^-_{{j}} \,, \hspace{1cm}  
\tilde{W}^+ = \sum_j V^*_{1 j}\chi^+_{{j}}\,, \hspace{1cm} 
\tilde{H}_u^+ = \sum_j V^*_{2 j}\chi^+_{{j}}
\end{align}

\item The new gauge fixing terms after EWSB are
\begin{eqnarray}
\nonumber
\mathscr{L}_{GF} &=&  - \frac{1}{2 \xi_\gamma} \left( \partial^\mu
\gamma_\mu\right)^2 - \frac{1}{2 \xi_Z} \left( \partial^\mu Z_\mu - \xi_Z M_Z
G^0 \right)^2  \\
\label{GFewsb}
&& - \frac{1}{\xi_{W^-}} \left| \partial^\mu W^-_\mu + i \xi_{W^-}
M_W G^-\right|^2   -\frac{1}{2 \xi_g}|\partial_\mu g^{\mu,i}|^2 \thickspace . 
\end{eqnarray}
That reads in \SARAH
\begin{lstlisting}
DEFINITION[EWSB][GaugeFixing]= 
{{Der[VP],                               - 1/(2 RXi[P])},	
 {Der[VWm]+ I Mass[VWm] RXi[W] Hpm[{1}], - 1/(RXi[W])},
 {Der[VZ] - Mass[VZ] RXi[Z] Ah[{1}],     - 1/(2 RXi[Z])},
 {Der[VG],                               - 1/(2 RXi[G])}};
\end{lstlisting}
Based on this definition, \(A^0_1\) and \(H^\pm_1\) are interpreted in all
calculations as Goldstone bosons.
\item No particles should be integrated out or deleted
\begin{lstlisting}
IntegrateOut={};
DeleteParticles={};
\end{lstlisting}
\item The Dirac spinors for the mass eigenstates are 
\begin{lstlisting}
DEFINITION[EWSB][DiracSpinors]={
 Fd - > {FDL, conj[FDR]},
 Fe  -> {FEL, conj[FER]},
 Fu  -> {FUL, conj[FUR]},
 Fv  -> {FvL, 0},
 Chi -> {L0, conj[L0]},
 Cha -> {Lm, conj[Lp]},
 Glu -> {fG, conj[fG]}
};
\end{lstlisting}
That leads to the replacements
\begin{equation}
d \rightarrow \left(\begin{array}{c} d_L \\ d_R \end{array} \right)\,, \dots \,,
\tilde{\chi}^- \rightarrow \left(\begin{array}{c} \lambda^- \\
(\lambda^+)^* \end{array} \right)\,, 
\tilde{g} \rightarrow \left(\begin{array}{c} \lambda_g \\ \lambda_g^*
\end{array} \right)\
\end{equation}
when going from four- to two-component formalism.
\end{enumerate}

\subsubsection{Creating a model file for the  NMSSM}
\label{sec:ToNMSSM}
Only a few changes are necessary to turn the above MSSM model definition into a
full-fledged implementation of the NMSSM:

\begin{enumerate}
\item Add a gauge singlet superfield
\begin{lstlisting}
Fields[[8]] = {sR, 1, s,    0, 1,  1};
\end{lstlisting}
\item Change the superpotential
\begin{lstlisting}
SuperPotential = { {{1, Yu},{q,Hu,u}}, {{-1,Yd},{q,Hd,d}},
                   {{-1,Ye},{l,Hd,e}}, 
                   {{1,\[Lambda]},{Hu,Hd,s}},
                   {{1/3,\[Kappa]},{s,s,s}}};
\end{lstlisting}
\item Give a VEV to the scalar component of the gauge singlet
\begin{lstlisting}
DEFINITION[EWSB][VEVs]= 
{...,
{SsR,{vS,1/Sqrt[2]},{sigmaS,I/Sqrt[2]},{phiS,1/Sqrt[2]}}};
\end{lstlisting}
\item Mix the scalar part of the gauge singlet with the Higgs and the fermionic part with the neutralinos
\begin{lstlisting}
DEFINITION[EWSB][MatterSector]= 
{...,
  {{phid, phiu, phiS}, {hh, ZH}},
  {{sigmad, sigmau,sigmaS}, {Ah, ZA}},
  {{fB, fW0, FHd0, FHu0,FsR}, {L0, ZN}},... }; 
\end{lstlisting}
\end{enumerate}

\lstset{frame=none}

\subsection{Calculations performed by \SARAH}

When a model is initialized using the {\tt Start} command, it is first
checked for gauge anomalies and charge conservation. If any of those checks
fail, a warning is printed. Afterwards, the calculation of the complete
Lagrangian at tree-level begins, and several tree-level results can be
obtained after it has finished.

\paragraph*{Masses and tadpole equations}
The masses and tadpole equations are derived automatically
during the evaluation of a model. The user has access to
both pieces of information through the command {\tt MassMatrix[Particle]}
for the mass matrix of  \verb"Particle" and {\tt  TadpoleEquation[VEV]}
for the tadpole equation of the corresponding field.

\paragraph*{Vertices}
\SARAH can be instructed to either calculate all vertices present in the model
or to extract only those for specific combinations of external particles.
The latter task is performed by
\begin{lstlisting}
 Vertex[{Particles},Options];
\end{lstlisting}
(the argument of this function being a list of external particles), while
all vertices for a set of eigenstates can be calculated via
\begin{lstlisting}
MakeVertexList[Eigenstates, Options];
\end{lstlisting}
This searches for all possible interactions present in the Lagrangian and
creates lists for the generic subclasses of interactions.
For more details about the calculation of a supersymmetric Lagrangian and the
conventions for extracting the vertices we refer to the appendix of the \SARAH
manual \cite{Staub:2008uz}. 

\paragraph*{Renormalization group equations} \SARAH calculates the RGEs for the
parameters of the superpotential, the soft-breaking terms, the gauge couplings
at one- and two-loop level and the VEVs. This is done by using the generic
formulas of \cite{Martin:1993zk}. In addition, to handle the case of several,
abelian gauge groups, the rules given in \cite{Fonseca:2011vn} are
implemented.
The calculation of the RGEs can be started after the initialization of a model
via
\begin{lstlisting}
CalcRGEs[Options];
\end{lstlisting} 
\paragraph*{Loop Corrections} \SARAH calculates the analytical expressions for
the one-loop corrections to the one- and two-point functions (tadpoles and self
energies of all particles).
These calculations are performed in \(\overline{\mbox{DR}}\)-scheme 
using the
't Hooft gauge. This is a generalization of the calculations for
the MSSM presented in \cite{Pierce:1996zz}. The command to start the calculation
is
\begin{lstlisting}
CalcLoopCorrections[Eigenstates];
\end{lstlisting}

\subsection{Export to external programs}

\SARAH can export the information derived from the model definition in a form
suitable for use with a number of external programs, among them
\FeynArts/\FormCalc, \CalcHep/\CompHep, \WHIZARD/\OMEGA, \SPheno and also plain
\LaTeX. In the next sections, we proceed with a detailed discussion of this
functionality and demonstrate how it can be used in order to facilitate a
comprehensive analysis of new models.

\section{Spectrum calculation with \SPheno}
\label{sec:SPheno}
\subsection{Introduction to \SPheno}
\SPheno \cite{Porod:2003um,Porod:2011nf} is a F95 program designed
for the precise calculation of the masses of supersymmetric particles.
For this 
the formulas of \cite{Pierce:1996zz} for the 1-loop masses have
been extended to account
for the flavour structures, e.g.\ to calculate the 1-loop corrected
$6\times 6$ mass matrices for squarks and charged leptons and the
1-loop corrected $3\times 3$ mass matrix for sneutrinos
\cite{Bruhnke:2010rh,Staub:2010ty}.
Here the complete one-loop contributions including general flavour
mixing and general CP-phases are taken into account. In addition,
the dominant two-loop contributions for the Higgs boson masses
are included based on \cite{Degrassi:2001yf,ADSZ1,ADSZ2,DedesA,AllanachA}
Moreover,
the decay rates for 2- and 3-body decays of supersymmetric particles
and Higgs bosons are calculated. 
The current version of \SPheno can be downloaded from 
\begin{lstlisting}
http://projects.hepforge.org/spheno/
\end{lstlisting}
\subsection{Combining \SPheno and \SARAH}
\SARAH is based on \Mathematica and therefore it is not sensible to do
exhaustive
numerical calculations in {\tt SARAH}'s native environment. As opposed to that,
\SPheno provides fast numerically routines for the evaluation of the RGEs,
calculating
the phase space of 2- and 3-body decays as well as Passarino Veltman integrals
and much more. Since these routines are model independent, they can be used 
 for all SUSY models implemented in \SARAH.

Our approach for combining all advantages of \SPheno and \SARAH in order to
create a very efficient and easy way from model building to numerical
results is depicted in Fig.~\ref{fig:SPheno_SARAH}: the model is defined in
\SARAH in the usual way. \SARAH calculates all analytical expressions needed for
a complete analysis of the model. This information is exported to Fortran code
in a way suitable for inclusion in \SPheno. This generates a fully functional
version of \SPheno for the new model without any need to change the source code
by hand.

The user has control over the properties of the generated \SPheno version
by means of a special input file for \SARAH. First, it is possible to
define the free parameters of the model. Those build later on the Block {\tt
MINPAR} in the LesHouches input file. Second, the boundary conditions at the
GUT-, SUSY- and electroweak scales as well as at possible threshold scales can be
set. Third, the parameters which are to be fixed by the solutions of the
tadpoles equations can be defined. An approximate solution to the tadpoles
can also be given,
if there isn't an analytic one.

\SARAH produces replacements for all model dependent files of \SPheno. These
files have to be copied to a new subdirectory of the \SPheno
directory. A {\tt Makefile} for
compiling the new model afterwards as well as a template for a LesHouches input
file are written by \SARAH.

The command to automatically calculate all necessary information like vertices
and RGEs and generate the source code is 
\begin{lstlisting} 
In[3]: MakeSPheno[Options];
\end{lstlisting} 
The name of the \SPheno specific input file of \SARAH can be given as an option.
This offers the possibility to easily create \SPheno versions for the same model
with changed boundary conditions or another set of free parameters.

\subsection{Features of the generated \SPheno version}

\paragraph{Calculation of the mass spectrum}
The \SPheno version generated by \SARAH calculates the complete mass spectrum
using 2-loop RGEs and 1-loop corrections to the masses, including the
full momentum dependence of all loop integrals.
In addition, for MSSM-like Higgs sectors, the
known two loop corrections to the Higgs masses and tadpoles can be included.
All calculations are performed with the most general flavor structure
and allow for the inclusion of CP phases.

The calculation of the mass spectrum happens in an iterative way: the gauge and
Yukawa couplings are fitted at \(M_Z\). Afterwards, a run to the GUT scale is
performed and the boundary conditions are enforced. The parameters are than evolved down
again to the SUSY scale and the mass spectrum is calculated. These steps are iterated
until the all masses have converged to a given relative
precision, by default
\(10^{-4}\).

The routines of \SARAH for calculating the gauge and Yukawa couplings closely
follow the procedures described in \cite{Porod:2003um}. The values
for the Yukawa couplings giving mass to the SM fermions and the gauge couplings
are determined at the scale \(M_Z\) based on the measured values of the quark and
lepton masses, the mass of the $Z$-boson $M_Z$, the Fermi constant
$G_F$ and the electromagnetic coupling in the Thompson limit 
$\alpha_{em}(0)$.  The 1-loop corrections to the mass of W- and Z-boson as well as
the SUSY contributions to muon-decay are taken into account into the calculation.
In addition, we include the complete 1-loop corrections to the self-energies
of SM fermions \cite{Pierce:1996zz} and re-sum the $\tan\beta$
enhanced terms in the calculation of the Yukawa couplings of the $b$-quark and
the $\tau$-lepton as described in \cite{Porod:2003um}.   The vacuum expectation
values \(v_d\) and  \(v_u\) are calculated with respect to the given value of
\(\tan\beta\) at \(M_Z\).
\paragraph{SUSY scale input} It is also possible to define the full set of free
parameters (i.e. gauge couplings, VEVs, superpotential and soft-breaking
parameters) of the model at a specific scale without RGE running. These
parameters  are afterward used to calculate the loop corrected mass spectrum and
the decays. To use this option with \SPheno, the LesHouches input file
\footnote{We extend and partly depart from the SUSY LesHouches
convention \cite{Skands:2003cj,Allanach:2008qq}
 to obtain a greater flexibility for the implementation
of new models.}  must
contain
\lstset{frame=shadowbox}
\begin{lstlisting}
Block MODSEL #
 1  0     #  Low scale input
12  2000. #  Renormalization scale
\end{lstlisting}
In this example all parameters are declared to be renormalized at 2~TeV. If the
no explicit renormalization scale is defined by the flag {\tt 12}, 1~TeV is
used. In addition, the numerical values of all parameters have to be given in
the LesHouches input file, e.g.
\begin{lstlisting}
Block GAUGEIN 	  #  
  1  0.384499E+00  # g1
  2  0.647209E+00  # g2
  3  1.121000E+00  # g3
Block YUIN     # 
  1  1     8.57631113E-06   # Y_u(Q)^DRbar
  2  2     3.63064589E-03   # Y_c(Q)^DRbar
  3  3     8.52980570E-01   # Y_t(Q)^DRbar 
...
\end{lstlisting}

\paragraph{Calculation of decay widths}
\label{sec:SPheno_decays}
The generated version of \SPheno is capable of calculating the widths and
branching
ratios of all non-SM scalars and fermions. While only two body decays are taken
in account for decaying scalars and vector bosons, the fermion widths also include the three body
decays to all-fermion final states. In the case of Higgs fields, the decays into
two-photon and two-gluon final states are included at leading
order \cite{Spira:1995rr}, as are the
gluonic QCD corrections for the Higgs decays into quarks and squarks.
\cite{Drees:1990dq,Djouadi:1995gt}.

\subsection{Defining the properties of the generated \SPheno version}
\label{sec:sphenoinputfile}
An additional file which defines the properties of the generated \SPheno function
is required by \SARAH for generating \SPheno output. This file must be
located in the same directory as the other models files for \SARAH. By default,
it is assumed that this file is named {\tt SPheno.m}, but it is also
possible to use other file names (see sec.~\ref{sec:GenerateSPheno}). This
way it is easily possible to generate different \SPheno implementations of the same model.

The content of the \SPheno specific input file for \SARAH is the following: 
  \begin{enumerate}
\item {\tt MINPAR}: A list of parameters which should be read by \SPheno from
the block {\tt MINPAR} in a LesHouches
 file. First, the number in the block is defined, afterwards the
variable. For example:
\begin{lstlisting}
 MINPAR = {{1,m0},
           {2,m12},
           {3,TanBeta},
           {4,SignMu},
           {5,Azero}};
\end{lstlisting}
Later, the values of those parameters can be communicated to \SPheno by
using an input file which contains
\begin{lstlisting}
 Block MINPAR #
    1  7.000000E+01 #  m_0
    2  2.500000E+02 #  M_1/2
    3  1.000000E+01 #  Tan(beta)
    4. 1.000000E+00 #  Sign(mu)
    5. 0.000000E+00 #  A_0 
\end{lstlisting}
\item {\tt EXTPAR}: It is also possible to define additional parameters for the
block {\tt EXTPAR} of the LesHouches input file by
\begin{lstlisting}
EXTPAR = {{Nr1,  Var1}, 
          {Nr2,  Var2},
          ...}; 
\end{lstlisting}
For instance, in order to give three additional VEVs as input, we can use
\begin{lstlisting}
EXTPAR = {{100, v1},
          {101, v2},
          {102, v3}};
\end{lstlisting}
and set the values later on in the input file by
\begin{lstlisting}
 Block EXTPAR #
    100  1.000000E-04 #  v_1
    101  1.500000E-04 #  v_2
    103  2.000000E-04 #  v_3
\end{lstlisting}
Note that there are no hard coded entries for {\tt MINPAR} or {\tt EXTPAR}. This
makes it necessary to define these blocks also for models with already existing
SLHA conventions. However, this also provides more freedom in varying the
model and the free parameters.
\item {\tt RealParameters}: By default, all parameters defined in {\tt MINPAR}
or {\tt EXTPAR} assumed to be complex, i.e. it is possible to use also the block
{\tt IMMINPAR} to define the imaginary part. However, some
Fortran functions like {\tt sin} can't be used with complex numbers, therefore
is is necessary to define parameters like \(\tan\beta\) explicitly as real, e.g.
\begin{lstlisting}
RealParameters = {TanBeta};
\end{lstlisting} 
\item {\tt ParametersToSolveTadpoles}: For each field which
can obtain a VEV, \SARAH derives the
corresponding minimum condition. These equations 
constrain as many parameters as there are VEVs in the model.  {\tt
ParametersToSolveTadpoles} defines which parameters the tadpole equations
will be solved for.

For example, to use the standard choice in the MSSM \(\mu, B_\mu\), the entry
reads:
\begin{lstlisting}
 ParametersToSolveTadpoles = {\[mu], B[\mu]};
\end{lstlisting}
\SARAH uses the {\tt Solve} command of \Mathematica to solve the tadpole
equations for the given set of parameters. If the solution is not unique because
a parameter \(X\) appears squared, \SARAH solves the equations for the absolute
squared. The phase is then defined by the automatically generated variable  {\tt
SignumX}, which is expected to be given as input. That's for instance the case
of the \(\mu\) parameter in the MSSM. 

The solutions for the tadpole equations are applied by \SPheno during the
numerical analysis at the SUSY as well at the electroweak scale. For that
purpose the running values of all parameters including the VEVs are taken as
input at the considered scale.
\item {\tt UseGivenTapdoleSolution}: In cases, in which \Mathematica won't find
an analytical solution for the tadpole equations for the given set of
parameters, this variable has to be set to {\tt True} and an approximated
solution can be given. These solutions are defined by
\begin{itemize}
\item {\tt SubSolutionsTadpolesTree}: For the solution at tree level
\begin{lstlisting}
 SubSolutionsTadpolesTree = {x1 -> sol1, x2 -> sol2,...};
\end{lstlisting}
Here, {\tt x1}, {\tt x2} are the names of the parameters which are fixed by the
tadpole equations and {\tt sol1}, {\tt sol2} are the approximated expressions
for them. 
\item {\tt SubSolutionsTadpolesLoop}: The solutions of the one loop corrected
tadpole equations. The one loop corrections to the different VEVs have to be
named {\tt Tad1Loop[i]}.
\end{itemize}
\item  {\tt RenormalizationScaleFirstGuess}: For the first run of the RGEs,
before any mass has been calculated by \SPheno, the squared renormalization
scale can be
defined by this entry. For example, for a mSugra scenario the common choice is
\begin{lstlisting}
  RenormalizationScaleFirstGuess = m0^2 + 4 m12^2;
\end{lstlisting}
This affects the running only if the SUSY scale is not fixed and SPA conventions
are disabled in the LesHouches input file.
\item {\tt RenormalizationScale}: For all further runs, another renormalization
scale can be given which is a function of the calculated masses, e.g.
\begin{lstlisting}
RenormalizationScale = MSu[1]*MSu[6]; 
\end{lstlisting}
\item Two loop contributions to the Higgs masses: if the Higgs sector of the
model is the same as for the MSSM, the original \SPheno routines for calculating
the two- loop tadpole equations and two-loop self energies to the the scalar and
pseudo scalar Higgs can be activated by setting
\begin{lstlisting}
 UseHiggs2LoopMSSM = True;
\end{lstlisting}
\item Condition for the GUT scale: to set a condition for a dynamically adjusted
GUT scale, use
\begin{lstlisting}
ConditionGUTscale = l.h.s == r.h.s; 
\end{lstlisting}
A common choice would be the unification point of \(g_1\) and \(g_2\). In that
case the condition reads
\begin{lstlisting}
ConditionGUTscale = g1 == g2; 
\end{lstlisting}
Note, that the value of the left-hand side must be smaller at scales below the
GUT scale as the right hand side. 
\item Boundary Condition: It is possible to define boundary conditions at three
different scales:
\begin{itemize}
\item Electroweak scale: {\tt BoundaryEWSBScale}
\item SUSY scale: {\tt BoundarySUSYScale}
\item GUT scale: {\tt BoundaryHighScale} 
\end{itemize}
In addition, if thresholds are involved, boundary conditions can be set at the
threshold scale, see sec.~\ref{sec:SPhenoThresholds}. It is also possible to
use a low scale input without any RGE running. In that case special boundary
conditions can be defined by  the array {\tt BoundaryLowScaleInput}.

All boundaries are defined by a two dimensional array. The first entry is the
name of the parameter, the second entry is the used condition at the considered
scale. The condition can be \dots
\begin{itemize}
\item \dots an input parameter from {\tt MINPAR} or {\tt EXTPAR}, e.g.
\begin{lstlisting}
 {MassB, m12};
\end{lstlisting}
\item \dots a block in the SLHA input file, e.g.
\begin{lstlisting}
 {Yv, LHInput[Yv]};
\end{lstlisting}
\item \dots a function of different parameters, e.g. 
\begin{lstlisting}
 {TYd, Azero*Yd};
\end{lstlisting}
\item \dots a diagonal matrix, e.g.
\begin{lstlisting}
 {md2, DIAGONAL m0^2};
\end{lstlisting}
\item \dots matrix  multiplications or the inverse of a matrix, e.g.
 \begin{lstlisting}
 {X, MatMul2[A,InverseMatrix[B], FortranFalse]};  
 \end{lstlisting}
For the matrix multiplication \verb"MatMul2" has to be used. The third argument
controls whether if only diagonal elements (\verb"FortranTrue") should be
considered or
not ( \verb"FortranFalse").

\item \dots a self defined function
\begin{lstlisting}
 {X, Func[A,B,C]};  
 \end{lstlisting}
It is also possible to use some self defined function. The Fortran code of that
function has to included in the array {\tt SelfDefinedFunctions} in {\tt
SPheno.m}. Later on it will be written to {\tt Model\_Data.f90}. Note, that the
standard functions needed for GMSB are already included \cite{Giudice:1998bp}:
\begin{itemize}
\item {\tt fGMSB[X]}:
\begin{eqnarray}
\nonumber f(x) &=& \frac{1+x}{x^2} \left(\ln(1+x) - 2 \text{Li}_2(\frac{x}{1+x})
+ \frac{1}{2} \text{Li}_2(2 \frac{x}{1+x}) \right) + \\
 && \frac{1-x}{x^2} \left(\ln(1-x) - 2 \text{Li}_2(\frac{x}{x-1}) + \frac{1}{2}
\text{Li}_2 (2 \frac{x}{x-1})\right)
\end{eqnarray}
\item {\tt gGMSB[X]}:
\begin{equation}
 g(x) = \frac{1+x}{x^2} \ln(1+x) + \frac{1-x}{x^2} \ln(1-x)
\end{equation}
\end{itemize}
\end{itemize}

Boundary conditions can be overwritten by assigning a value to a parameter in
the
LesHouches input file. For example, the Higgs soft breaking masses at
the GUT scale can be forced to have specific values instead of \(m_0^2\) by
declaring
\begin{lstlisting}
 Block MSOFTIN    #
  21  10000.000   # mHd2
  22  20000.00    # mHu2
\end{lstlisting}
in the SLHA file.

\paragraph*{Several sets of boundary conditions} In order to implement different
versions of a single model which differ only by the used boundary conditions,
{\tt BoundaryEWSBScale}, {\tt BoundarySUSYScale}, {\tt BoundaryHighScale} can be
also a nested list, e.g.
\begin{lstlisting}
BoundarySUSYScale = Table[{},{2}];
BoundaryGUTScale = Table[{},{2}];

BoundarySUSYScale[[1]] = {{KappaNMSSM, KappaInput},
                          {LambdaNMSSM, LambdaInput}};
BoundaryGUTScale[[1]]  = {};

BoundarySUSYScale[[2]] = {};
BoundaryGUTScale[[2]]  = {{KappaNMSSM, KappaInput},
                          {LambdaNMSSM, LambdaInput}};
\end{lstlisting}
In the first case, the input values for \(\lambda\) and \(\kappa\) are
taken at the SUSY scale, in the second one at the GUT scale. To
communicate to \SPheno which set of boundary conditions should be used
for a run, flag 2 in {\tt MODSEL} is used:
\begin{lstlisting}
Block MODSEL #
 2  X  # This uses the X. set of boundary conditions. 
\end{lstlisting}
The default value is 1. 
\item Lists for calculating decay widths:
\begin{itemize}
\item {\tt ListDecayParticles}: List of particles for which two-body decays
will be calculated. This can be a list of particles using the names inside
\SARAH, e.g.
\begin{lstlisting}
ListDecayParticles = {Sd,Su,Se,hh,Ah,Hpm,Chi};
\end{lstlisting}
or just {\tt Automatic}. If {\tt Automatic} is used, the widths of all particles
not defined as standard model particles as well as the top width are calculated
by \SPheno.
\item {\tt ListDecayParticles3B}; Three body decays of fermions. This can be a
list with the names of the particles and the corresponding files names, e.g.
\begin{lstlisting}
ListDecayParticles3B =  {{Chi,"Neutralino.f90"},
                         {Cha,"Chargino.f90"},
                         {Glu,"Gluino.f90"}}; 
\end{lstlisting}
or just {\tt Automatic}. If {\tt Automatic} is used, the widths of all fermions
not defined as standard model particles are calculated. The auto generated
file names are {\tt ParticleName.f90}.
\end{itemize}
\item Ordering of mass eigenstates: normally, all particles of one kind are
ordered in \SPheno by their mass. However, it might be desirable to override
this behavior and instead define another ordering scheme. For example, consider
several massless CP odd particles at tree level exist which can be assigned
to a Goldstone boson. For this purpose, a condition can be defined by using
\begin{lstlisting}
 ConditionForMassOrdering = { {Particle, Condition}, ... };
\end{lstlisting}
The condition has to be Fortran source code and is added to the corresponding
routine. For instance, a condition for the NMSSM would read
\begin{lstlisting}
ConditionForMassOrdering={
{Ah,
"If (Abs(ZA(1,3)).gt.Abs(ZA(1,2))) Then \n
   MAh2temp = MAh2 \n
   ZAtemp = ZA \n
   ZA(1,:) = ZAtemp(2,:) \n
   ZA(2,:) = ZAtemp(1,:) \n
   MAh2(1) = MAh2temp(2) \n
   MAh2(2) = MAh2temp(1) \n
End If \n \n"}
};
\end{lstlisting}
This example checks whether two massless pseudo scalars are present in the spectrum and,
if this is the case, uses the not singlet-like particle as Goldstone boson.

\end{enumerate}

\subsection{Example: The input file for the MSSM}
\label{sec:Example_SPheno_MSSM}
To generate the \SPheno output for the MSSM, {\tt SPheno.m} should include the
following information:
\begin{enumerate}
\item We want to have mSugra like boundary conditions. Therefore, we chose the
minimal
set of parameters defining the model as \(m_0,M_{1/2},A_0,\text{sign}\mu\)
and \(\tan\beta\). These will later be read from the {\tt MINPAR} block of a
LesHouches input file.
\begin{lstlisting}
MINPAR={{1,m0},
        {2,m12},
        {3,TanBeta},
        {4,SignumMu},
        {5,Azero}};
\end{lstlisting}
\item  {\tt TanBeta} has to be declared as a real parameter
\begin{lstlisting}
RealParameters = {TanBeta};
\end{lstlisting}
\item As usual in the MSSM, the tadpole equations should be solved with respect
to
\(\mu\) and \(B_\mu\)
\begin{lstlisting}
ParametersToSolveTadpoles = {\[Mu],B[\[Mu]]};
\end{lstlisting}
\item To study models with a dynamically adjusted SUSY scale, the expressions for
the
definition of the SUSY scale can be given. The first expression is used only
before the mass spectrum has be calculated the first time. Note, that these
definitions can easily disabled in the LesHouches input file by flag {\tt
MODSEL 12} and a fixed scale can be used. Also, when SPS conventions are
switched on in the LesHouches input file by {\tt SPhenoInput 2}, a fixed scale
of 1 TeV is used. 
\begin{lstlisting}
RenormalizationScaleFirstGuess = m0^2 + 4 m12^2;
RenormalizationScale = MSu[1]*MSu[6];
\end{lstlisting}
\item The GUT scale is the unification point of \(g_1\) and \(g_2\)
\begin{lstlisting}
ConditionGUTscale = g1 == g2; 
\end{lstlisting}
\item As said, we want to use mSugra like boundary conditions. These are
straightforward define by
\begin{lstlisting}
BoundaryHighScale={
  {T[Ye],   Azero*Ye},
  {T[Yd],   Azero*Yd},
  {T[Yu],   Azero*Yu},
  {mq2,     DIAGONAL m0^2},
  {ml2,     DIAGONAL m0^2},
  {md2,     DIAGONAL m0^2},
  {mu2,     DIAGONAL m0^2},
  {me2,     DIAGONAL m0^2},
  {mHd2,    m0^2},
  {mHu2,    m0^2},
  {MassB,   m12},
  {MassWB,  m12},
  {MassG,   m12}
};
\end{lstlisting}
\item It is also possible to use the generated \SPheno version with a low scale
input. This is enabled by setting {\tt MODSEL 1} to {\tt 0}. In that case,
input values for all free parameters of the model are expected. However, also in
this case, we can define a set of boundary conditions, e.g. for dynamically calculating
the SUSY VEVs
\begin{lstlisting}
BoundaryLowScaleInput={
 {vd,Sqrt[2 mz2/(g1^2+g2^2)]*Sin[ArcTan[TanBeta]]},
 {vu,Sqrt[2 mz2/(g1^2+g2^2)]*Cos[ArcTan[TanBeta]]}
};
\end{lstlisting}
\item Finally, we define that the code for the calculation of the two and three
body
decays is generated for all SUSY particles. That's done by using the flag {\tt
Automatic}.
\begin{lstlisting}
ListDecayParticles = Automatic;
ListDecayParticles3B = Automatic;
\end{lstlisting}
\end{enumerate}

\subsection{Generating the output}
\label{sec:GenerateSPheno}
After an input file with all necessary information has been created, the
generation of the source code for \SPheno is triggered by
\begin{lstlisting}[frame=none]
In[5]: MakeSPheno[Options]
\end{lstlisting}

The different options are:
\begin{itemize}
\item \verb"Eigenstates->Name of Eigenstates." If not specified, the last set of
eigenstates is used which corresponds to the  last entry of {\tt NameOfStates}
(see sec.~\ref{sec:MSSM_modelfile}).
\item \verb"ReadLists->True" can be used if all vertices and RGEs have already
been calculated for the model and the former results should be used to save
time.
\item \verb"InputFile". The name of the SPheno input file. If not defined,
\verb"SPheno.m" is used.
\end{itemize}

The generated source code is located in
\begin{verbatim}
[SARAH Directory]/Output/[Model]/[Eigenstates]/SPheno/ 
\end{verbatim}
It is sufficient to copy all files from this directory to a sub-directory {\tt
NAME} of \SPheno version 3.1.4 or later and compile it using  in the \SPheno root
directory
\begin{lstlisting}
> make Model=NAME
\end{lstlisting}
{\tt NAME} has to be the same as the name of the model defined in \SARAH (see
sec.~\ref{sec:MSSM_modelfile}).

\subsection{Including Thresholds}
\label{sec:SPhenoThresholds}
Using \SARAH it is possible to include thresholds in the RGE running performed
by \SPheno. 
\subsubsection{Thresholds without gauge symmetry breaking}
If all scales have the same gauge structure, it is possible  for \SARAH to
derive the RGEs for all scales from the RGEs for the highest scale by performing
the following steps:
\begin{itemize}
\item For those fields which should be integrated out during the run,
variables \(n_{gen}(\Phi_i)\) are introduced internally which define the number of
generation of the heavy field \(\Phi_i\). All gauge group constants like the
Dynkin index summed over chiral superfields, \(S(R)\), are expressed as function
of \(n_{gen}(\Phi_i)\). These \(n_{gen}(\Phi_i)\) are dynamically adjusted, when
the energy scale crosses a threshold.
\item When crossing a threshold, the couplings involving heavy fields are set
to zero. For example, the Yukawa type coupling of the
form \(Y^{ij} \Phi_i \phi_j H\) involves three generations of the heavy field
\(\Phi\). At the threshold of \(\Phi_k\), the \(k\)-th row of
\(Y\) is set to zero. That happens similarly for all other superpotential and
soft-breaking parameters. 
\item The masses of scalar and fermionic components of the heavy superfields are
assumed to be identical, i.e.\ the soft SUSY breaking terms are assumed to
be negligible. These masses are given by a bilinear
superpotential term.
\end{itemize}

In order to include thresholds without gauge symmetry breaking, the following
steps have to be performed:
\begin{enumerate}
\item The heavy fields must be marked for deletion in the \SARAH model definition:
\begin{lstlisting}
 DeleteFields = {...};
\end{lstlisting}
This ensures, that the decays, loop corrections, etc. at the SUSY scale
calculated by \SPheno receive no contributions from  the heavy fields
\item The thresholds have to be defined in \verb"SPheno.m" :
\begin{lstlisting}
 Thresholds = {{Scale1, {HeavyFields1}},
               {Scale2, ... }};
\end{lstlisting}
For all scales an entry in the array \verb"Thresholds" has to be added. Each
entry defines the threshold scale and at second position a list of the
heavy superfields which can be restricted to specific generations.
\end{enumerate}

It is possible to define boundary conditions at each threshold scale for running
up and down separately:
\begin{lstlisting}
 BoundaryConditionsUp[[x]] = { ...};
 BoundaryConditionsDown[[x]] = { ...};
\end{lstlisting}

\paragraph*{Threshold corrections}
Using 2-loop RGEs requires 1-loop boundary condition. Therefore, at each
threshold scale the one loop threshold corrections to gauge couplings and
gaugino masses  are calculated. The general
expressions are \cite{Hall:1980kf}
\begin{eqnarray}
\label{eq:shift1}
 g_i & \rightarrow & g_i \left( 1\pm \frac{1}{16 \pi^2} g_i^2 I^i_2(r) 
\ln\left(\frac{M^2}{M_T^2}\right)\right)  \thickspace ,\\
\label{eq:shift2}
 M_i & \rightarrow & M_i \left( 1\pm \frac{1}{16 \pi^2} g_i^2 I^i_2(r) 
\ln\left(\frac{M^2}{M_T^2}\right)\right) \thickspace .
\end{eqnarray}
\(I^i_2(r)\) is the Dynkin index of a field transforming as
representation \(r\) with respect to the gauge group belonging to the
gauge coupling \(g_i\), \(M\) is the mass of this particle and \(M_T\)
is the threshold scale.

\paragraph*{Example}
As an example, a version of \SPheno implementing the seesaw type~II and
type~III models can be generated by adding the following entries to
\verb"Spheno.m"
\begin{enumerate}
 \item Seesaw II: 
\begin{lstlisting}
Thresholds={
  {Abs[MTMIN],{s,sb,t,tb,z,zb}}
};
\end{lstlisting}
 \item Seesaw III:
\begin{lstlisting}
Thresholds={
  {Abs[MWM3IN[1,1]],{Hx3[1],Hxb3[1],Hg3[1],Hb3[1],Hw3[1]}},
  {Abs[MWM3IN[2,2]],{Hx3[2],Hxb3[2],Hg3[2],Hb3[2],Hw3[2]}},
  {Abs[MWM3IN[3,3]],{Hx3[3],Hxb3[3],Hg3[3],Hb3[3],Hw3[3]}}
};
 \end{lstlisting}
\end{enumerate}
For a more comprehensive discussion of the model files for the Seesaw I, see
appendix~\ref{app:Seesaw}.

\subsubsection{With gauge symmetry breaking}

If the gauge structure at the different scales are different, each set of RGEs
is calculated separately and this information is then combined
into one consistent version of \SPheno which includes
routines for calculating finite shifts in the gauge couplings, gaugino and
scalar mass parameters. As an example, the implementation of a left-right
supersymmetric model with one symmetry breaking scales is shown in
sec.~\ref{app:Omega}.

In order to implement such a model, the following steps are necessary:
\begin{enumerate}
\item For each regime, a separate model file for \SARAH has to be created.
These model file have to be saved in the subdirectories \verb"Regime-1",
\verb"Regime-2", \dots of  
\begin{lstlisting}
[SARAH Directory]/Models/[Model]/
\end{lstlisting}
beginning with the highest scale.
\item The \SPheno input file for the higher scales must provide the following
information:
\begin{itemize}
\item {\tt IntermediateScale = True}
\item {\tt RegimeNr = X}
\item A list of the heavy fields, which should be integrated out, the gauge
sector below the threshold as well as the corresponding quantum numbers of the
fields which are to be integrated out. This is needed to calculate the finite
shifts at the threshold scale, for instance
\begin{lstlisting}
HeavyFields = {Field_1, Field_2,..};
NextGauge =   {U[1],   SU[2],   SU[2],    SU[3]};
NextQN = {
       {Field_1, 0,       2,       1,         1},
       {Field_2, 1/3,     1,       2,         4},
 ...
};
\end{lstlisting}
Care has to be taken to use the same ordering of the $SU(2)$ gauge
groups as in the orignal definition.
\end{itemize}
\item All necessary information for combining the regimes to one \SPheno is
given in \verb"SPheno.m" of the lowest scale.
\begin{itemize}
\item {\tt IntermediateScale = False}
\item {\tt RegimeNr = X}
\item The threshold scales: {\tt ThresholdScales = {...} }
\item The boundary conditions for running up and down at each threshold scale:
\begin{lstlisting}
 BoundaryConditionsUp[[x]] = {...};
 BoundaryConditionsDown[[x]] = {...};
\end{lstlisting}
In the boundary conditions \verb"index1", \verb"index2", \dots can be used for
defining sums over indices.
\item The usual information for \SPheno, defined in the sec.
\ref{sec:sphenoinputfile}.
\end{itemize}
\end{enumerate}

When starting the \SPheno output of the lowest scale, all other scaler are
evolved automatically. Note that, in order to calculate the RGEs of the different regimes,
\SARAH starts one additional \Mathematica kernel.
For passing the information between the different \Mathematica kernels
 a directory \verb"Meta" in the model directory is created by \SARAH. The
screen output of \Mathematica during the evaluation of the higher regimes
is written to that directory (\verb"Output-Regime-X.m"), allowing the user to
supervise the progress and see potential error messages. The necessary
information of each regime for writing the combined source code for \SPheno at
the end is saved by \SARAH in the files \verb"Regime-X.m".

\subsection{Low energy \SPheno version}
\label{sec:LowScaleSPheno}
It is also possible to create a \SPheno version with much less features which
only accepts low energy input. That means, the RGEs are not written out and
also the fit to the electroweak data is not performed in the numerical
evaluation of one point. It just solves the tadpole equations, calculates the
tree- and one-loop masses as well as the decay widths and branching ratios. 
The advantage of such a \SPheno version is that it
works with a larger set of models, e.g. also non-SUSY models or other models
not supported by a full evaluation as explained in
sec.~\ref{sec:SPheno_restrictions}. To get a \SPheno version without
RGE evolution, insert
\begin{lstlisting}
OnlyLowEnergySPheno = True;
\end{lstlisting}
in {\tt SPheno.m}. The remaining information needed by \SARAH is only a small
subset of the settings discussed above and consists of
\begin{itemize}
 \item {\tt MINPAR}
 \item {\tt ParametersToSolveTadpoles}
 \item {\tt BoundaryLowScaleInput}
 \item {\tt ListDecayParticles} and {\tt ListDecayParticles3B}. Note that the {\tt
Automatic} statement for automatically deriving the decays of all non-SM particles
does not work in
this case as \SARAH doesn't differ between SUSY or Non-SUSY
particle in order to make the output as generic as possible. Therefore, the lists
of the decaying particles have to be supplied manually. 
\end{itemize}

\subsection{Differences to \SPheno {\tt 3.1.4}}
A few things are handled in a slightly different way in a \SPheno MSSM module
created by \SARAH in comparison to \SPheno 3.1.4. Also, some things are not yet
implemented. We give an overview in Tab.~\ref{tab:SPheno_SARAH}.

\begin{table}[t]
\begin{tabular}{|l|c|c|}
\hline
  & \SPheno {\tt 3.1.4} & \SPheno by \SARAH \\
\hline
\parbox[0pt][8em][c]{4cm}{Loop corrections in Higgs sector} &
\parbox{5cm}{Complete one-loop corrections and dominant two-loop
corrections.} & 
\parbox{5cm}{One-loop corrections are calculated by \SARAH. Dominant
two-loop corrections implemented in \SPheno can be linked.} \\
\hline
\parbox[0pt][5em][c]{4cm}{Tadpole equations} &
\parbox{5cm}{Solved at SUSY scale. Solutions are evaluated at the EW due to RGE
running} & 
\parbox{5cm}{Solved at SUSY and EW scale.} \\
\hline
\parbox[0pt][3em][c]{4cm}{Three body decays} &
\parbox{5cm}{Three body decays of fermions and stop } & 
\parbox{5cm}{Three body decays of fermions} \\
\hline
\parbox[0pt][3em][c]{4cm}{Loop induced\\ neutralino decays} &
\parbox{5cm}{One-loop decay into photon and neutralino included} & 
\parbox{5cm}{No loop induced decays included} \\
\hline
\parbox[0pt][3em][c]{4cm}{Loop induced\\ gluino decays} &
\parbox{5cm}{One-loop decay into gluon and neutralino included} & 
\parbox{5cm}{No loop induced decays included} \\
\hline
\parbox[0pt][3em][c]{4cm}{$e^+ e^-$ collisions} &
\parbox{5cm}{Calculates cross sections} & 
\parbox{5cm}{Cross section calculation not included} \\
\hline
\end{tabular}
\caption{Different handling and implementation status in \SPheno 3.1.4 vs.
the \SPheno modules created by \SARAH.}
\label{tab:SPheno_SARAH}
\end{table}

\subsection{Supported models and known issues}
\label{sec:SPheno_restrictions}
While \SARAH can create valid \SPheno code for many different
models, there are some requirements on the model and some minor restrictions
on the functionality  of the resulting \SPheno module. At the moment, those are

\begin{itemize}
 \item {\bf Fit to low energy data}: in order to perform a fit to low energy data
(e.g.\ for fermion masses, $m_Z$, $G_F$ and $\alpha_{em}$) as
starting point of the RGE evaluation, the following parameters must be present
in the model: Yukawa couplings for lepton and quarks, two Higgs VEVs and, of
course, the three SM gauge couplings and the SM particle content. However, it
is still possible to use at least some features of the \SPheno output of \SARAH
by manually supplying model parameters for \SPheno. In that way, the RGE evaluation and
the fit the electroweak data is skipped, but the one-loop corrected masses as well as the decay widths
and branching ratios are calculated.
\item {\bf Flavor decomposition}: with \SARAH it is possible to assign a unique
name to each generation of a particular field and this way treat the individual generations as
independent fields. That is not yet supported in the \SPheno output. Furthermore, mixing
matrices generated with the option {\tt NoFlavorMixing} can not yet be handled
by the numerical code. 
\end{itemize}
\lstset{frame=none}

\section{Checking Higgs constraints with \HB}
\label{sec:HB}
\HB \cite{Bechtle:2008jh,Bechtle:2011sb} is a tool to test the neutral and
charged Higgs 
sectors against the current exclusion bounds from the Higgs
searches at the LEP, Tevatron and LHC experiments. The required input consists
of the masses, width and branching ratios of the Higgs fields. In addition, it
is either possible to provide full information about production cross sections
in \(e^+ e^-\) and \(p p\) collisions, or to work with a set of effective
couplings. \HB can be downloaded from
\begin{lstlisting}
http://projects.hepforge.org/higgsbounds
\end{lstlisting}

Although \HB supports the LesHouches interface, this functionality is restricted
so far to at most 5 neutral Higgs fields, and therefore, we don't use it.
Instead, \SPheno modules generated by \SARAH can create all necessary input
files needed for a run of \HB with effective couplings (option {\tt
whichinput=effC}). To write these files, the flag {\tt 75} in the block {\tt
SPhenoInput} in the LesHouches input file has to be set to {\tt 1}.
\begin{lstlisting}[frame=shadowbox]
 Block SPhenoInput #
 75 1 # Write files for HiggsBounds
\end{lstlisting}
Unfortunately, we can not provide all information which can be used by \HB to
check the constraints. In particular, the effective couplings \(H \rightarrow \gamma
Z\) and \(H \rightarrow g g Z\) are not yet calculated by \SPheno and therefore
set to zero in the output. In addition, as already mentioned, the \SPheno
version created by \SARAH does not support calculating the \(e^+ e^-\) cross
sections. For this reason, also the LEP production cross section of charged
Higgs fields is not available for \SPheno and it sets this value also to 0.
However, it is of course possible to calculate this cross section as well as all
other cross sections needed for the options ({\tt whichinput=hadr} or {\tt
whichinput=part}) of \HB using \CalcHep or \WHIZARD with the corresponding
model files created by \SARAH (cf. sec.~\ref{sec:CH_MO}
and \ref{sec:WO}).

The following files are written by \SPheno
\begin{itemize}
 \item {\tt MH\_GammaTot.dat}: \\
 Masses and widths of all neutral Higgs fields
 \item {\tt MHplus\_GammaTot.dat}: \\
Masses and widths of all charged Higgs fields
 \item {\tt BR\_H\_NP.dat}: \\
Branching ratios of neutral Higgs fields into invisible and other neutral
Higgs fields.
 \item {\tt BR\_Hplus.dat}: \\
 Branching ratios of charged Higgs fields into \(c \bar{s}\), \(c \bar{b}\) and
\(\tau \bar{\nu}\) final states
 \item {\tt BR\_t.dat}: \\
 Top quark branching rations into a bottom quark plus either a \(W\) boson or a charged
Higgs
 \item {\tt effC.dat}:
Effective couplings among neutral Higgs fields and of neutral Higgses to
\(s\bar{s}\), \(c\bar{c}\), \(b\bar{b}\), \(t\bar{t}\), \(\mu\bar{\mu}\), 
\(\tau\bar{\tau}\), \(\gamma \gamma\), \(g g\), \(\gamma Z\), \(g g Z\) 
 \item {\tt LEP\_HpHm\_CS\_ratios.dat}: \\
LEP cross sections for the production of charged Higgs bosons (set to zero, see above) 
\end{itemize}
\HB can be run on the generated files by invoking it as
\begin{lstlisting}[frame=none]
> ./HiggsBounds LandH effC [NN] [NC] [SPheno Directory]
\end{lstlisting}
where {\tt [NN]} has to be replaced by the number of neutral Higgses
and {\tt [NC]} by the number of the charged ones. Please consult the \HB manual
for further information. The results of the check are written to the file {\tt
HiggsBounds\_results.dat} which is located in the same directory as the input
file, i.e. in our case in the \SPheno root directory.

\section{Calculation of cross sections, widths and relic density using \CalcHep
and \MicrOmegas}
\label{sec:CH_MO}
\CalcHep \cite{Pukhov:2004ca,Boos:1994xb} is a package for the calculation of
Feynman amplitudes and their integration over multi-particle phase space. It is
designed to provide a direct transition from the Lagrangian to the cross sections
and distributions. The \CalcHep homepage is located at
\begin{lstlisting}
http://theory.sinp.msu.ru/~pukhov/calchep.html
\end{lstlisting}

\subsection{Using \CalcHep with \SARAH and \SPheno}
\label{sec:CalcHep}
The \CalcHep model files produced by \SARAH support both the Feynman and unitarity
gauges. Furthermore, \SARAH can split  interactions between four colored
particles as required by \CalcHep/\CompHep. Models
with CP violation are also supported. The model files for \CalcHep/\CompHep are
created by  
\begin{lstlisting}
In[5]: MakeCHep[Options]; 
\end{lstlisting}
Options exist for specifying the gauge ({\tt FeynmanGauge \(\rightarrow\) True/False})
and for activating the support for CP violation ({\tt CPViolation \(\rightarrow\) True}).
In addition, the splitting of
specific four-scalar interactions can be suppressed as long as they are not
colored ({\tt NoSplitting} \(\rightarrow\) list of fields) and the
running of the strong coupling constant can be included as it is usually  done
in the standard \CalcHep files ({\tt UseRunningCoupling \(\rightarrow\) True}).

Recently, it became also possible to use SLHA files with \CalcHep  to provide
the numerical values of the parameters \cite{Belanger:2010st}. \SARAH uses this
option if the flag {\tt SLHAinput \(\rightarrow\) True} is used. Thus, the
parameters  calculated by \SPheno can be directly passed to \CalcHep.

In order to calculate the same process for different points in parameter space, the so
called blind mode of \CalcHep can be used for performing calculations without the
need to start the graphical interface. The blind mode is most conveniently put
to work using the following steps:
\begin{enumerate}
 \item Start \CalcHep, insert a process and choose {\tt make n\_calchep}. This
creates an executable file called {\tt n\_calchep} in the {\tt results}
subdirectory of your \CalcHep model.
 \item In order to keep a copy of the files, copy the content of {\tt results}
to another directory  (make sure that the subdirectory {\tt aux} is included in
the copy). The spectrum file generated by \SPheno goes into the same directory. 
 \item Change to the directory containing {\tt n\_calchep} and start it via
 \begin{lstlisting}
> ./n_calchep +blind
\end{lstlisting}
 \item Make the desired adjustments and start the numerical integration using
{\tt Vegas}.
 \item When {\tt n\_calchep} exits, a line of the form \verb"[{]][{{[]]"
is returned which mimics the keyboard commands you did  
 \item Now, the same calculation can be repeated without the
\CalcHep frontend using the command
 \begin{lstlisting}
> ./n_calchep -blind "[{]][{{[]]"
\end{lstlisting}
\end{enumerate}
This way, a number {\tt n\_calchep} instances
for several processes (e.g. sparticle or Higgs production at LEP, Tevatron or
LHC) can be created and called non-interactively from a script, thus facilitating
automatized scans over parameter space.
 
\subsection{Relic density calculations with \MicrOmegas}
\label{sec:MicrOmegas}
\MicrOmegas \cite{Belanger:2006is} is a well known tool for the calculation of
the relic density of a dark matter candidate. The download is located at 
\begin{lstlisting}
http://lapth.in2p3.fr/micromegas/
\end{lstlisting}
As \MicrOmegas uses CalcHep for the calculation of (co-)annihilation cross
sections, a model file for \CalcHep must be generated first before
\MicrOmegas can be used.

\SARAH writes two files for \MicrOmegas which can serve as so-called main
files, i.e. they can be compiled with \MicrOmegas and executed to perform
calculations. While {\tt CalcOmega.cpp} calculates only \(\Omega h^2\) and
writes the result to the file {\tt omg.out}. \linebreak {\tt
CalcOmega\_with\_DDetection.cpp} computes also direct detection signals.
As the SLHA+ import functionality of \CalcHep can also be used with \MicrOmegas,
it is sufficient to simply copy the spectrum file written by \SPheno to the
directory of \MicrOmegas and start the calculation.

\section{Monte Carlo studies with \WHIZARD}
\label{sec:WO}

\subsection{Introduction}
\WHIZARD \cite{Kilian:2007gr} is a fast tree-level Monte Carlo generator for
parton level
events. A particular strength of
the code is the efficient generation of unweighted events for high multiplicity final
states (simulations with 8 final state particles have been performed
successfully) using exact matrix elements. This makes it
particularly useful for the study of supersymmetric models which generically
feature complicated multiparticle final states arising from long decay chains.

Behind the scenes, \WHIZARD builds on the optimizing matrix element generator
\OMEGA \cite{Moretti:2001zz}. In
order to deliver the fast tree level matrix elements required by \WHIZARD,
\OMEGA builds up matrix elements as directed acyclical graphs of
one-particle off shell wave functions (1POW, Green's functions with all but one
legs
amputated). These graphs are then transformed into highly optimized {\tt FORTRAN
90} code which is called by WHIZARD to calculate helicity matrix elements by
recursively fusing 1POWs. This algorithm is guaranteed to avoid any redundancies
arising from the repeated evaluation of subdiagrams from the start and can be
shown to grow only exponentially in complexity with the number of external legs
(as opposed of the factorial growth of any traditional Feynman diagram based
approach).

For the treatment of color, \WHIZARD and \OMEGA leverage the color flow
decomposition \cite{Maltoni:2002mq}. The amplitudes are decomposed
into all possible different color flows,
for which the amplitudes are then calculated using {\OMEGA}s recursive algorithm and
finally combined into the squared and color summed matrix elements. In addition
to the fast calculation of the color trace, this algorithm also provides the color
connection information later required by fragmentation and hadronization algorithms.

The integration and event generation in \WHIZARD is performed using the
adaptive multichannel Monte Carlo code \VAMP \cite{Ohl:1998jn}. For each process,
\WHIZARD dynamically determines a set of suitable phase space maps. \VAMP
assigns
a grid to each map, all which are then linearly combined to form the phase space
parameterization. During the integration process, \VAMP adapts both the grids and
the weights of the different channels. After integration, the optimized grids
are used to facilitate the efficient generation of unweighted events.

Recently, \WHIZARD has received a major upgrade, going from the (now legacy) 1.x 
version branch (currently 1.97) to the modern 2.x branch (currently 2.0.5).
While the interface between \SARAH and \WHIZARD supports both branches, the
new version features both many new physics features (including
factorized matrix elements incorporating full spin and color correlations) and
technical improvements, so using the newer version is highly advised. The
\WHIZARD package can be downloaded from
\begin{lstlisting}
http://www.hepforge.org/archive/whizard/
\end{lstlisting}

\subsection{The interface between \SARAH and \WHIZARD}

The interface between \SARAH and \WHIZARD shares significant parts of its code
with the interface between \FeynRules \cite{Christensen:2008py}, with a thin layer on top to
interface with \SARAH. As such, the feature set is identical with that of the
\FeynRules interface and, although we will now give a short overview, we point
the reader to above reference for more information.

At the moment, the interface can handle spin $0$, $\frac{1}{2}$ and $1$ fields, with
support for spin $\frac{3}{2}$ being being planned for a future revision. The supported
interactions are a subset of the operators currently available in \OMEGA.
Specifically, up to very few exceptions, nearly all dimension $3$ and $4$
interactions are available, together with a couple of higher dimension
operators. In particular, this list is sufficient to handle all interactions
which are generated in a typical application of \SARAH to a supersymmetric
model without higher-dimensional terms in the Lagrangian.

The list of supported color structures is more complicated, but in practice
all color tensors arising from QCD gauge interactions which lead to
supported Lorentz structures are supported. Again, this covers all interactions
which can be generated by \SARAH. A full list of the supported color structures
can be found in \cite{Christensen:2010wz}.

The 2.x branch of \WHIZARD supports running the strong coupling,
and the interface supports this functionality. In order to use it, the model must
define a parameter called {\tt aS} which is treated by the interface as $\alpha_S$
at the $Z$ pole. In addition, the QCD coupling constant $g_S$ must be derived from
$\alpha_S$ and have the description {\tt "Strong-Coupling"}. Once these criteria
are met, the generated model will automatically evolve all vertex factors which
depend on either $g_S$ or $\alpha_S$ if the running coupling is activated
in \WHIZARD. For example, all model files included in \SARAH support the
running by deriving all vertices
as function of \(g_3\) and including the following definitions into
the parameter list:
\begin{lstlisting}[frame=shadowbox] 
{g3,     {    ...,
             Description -> "Strong-Coupling",
             DependenceNum -> Sqrt[AlphaS 4 Pi]  }}, 
             
{AlphaS, {   ...,
             OutputName-> aS  }}, 
\end{lstlisting} 

\subsection{Generating model files and using them with \WHIZARD}

\begin{table}
\centerline{\begin{tabular}{|l|l|}
\hline \texttt{WOWhizardVersion} & {\sc Whizard} versions supported \\\hline\hline
\texttt{"1.92"} & 1.92 \\\hline
\texttt{"1.93"} & 1.93 -- 1.95 \\\hline
\texttt{"1.96"} & 1.96+ \\\hline
\texttt{"2.0"} & 2.0 -- 2.0.2 \\\hline
\texttt{"2.0.3"} (default) & 2.0.3+ \\\hline
\end{tabular}}
\caption{Currently available version choices when generating \WHIZARD model
files, together with the respective \WHIZARD versions supported by them.}
\label{tab-wowhizardversion}
\end{table}

\paragraph*{Calling the interface}\mbox{}\\
In order to generate model files for \WHIZARD, the \verb"ModelOutput" command is
used
\begin{lstlisting}[frame=shadowbox] 
In[3]: ModelOutput[Eigenstates, WriteWHIZARD->True];
\end{lstlisting} 
This writes the model file using the default options. An equivalent syntax which
allows for passing options to the interface is
\begin{lstlisting}[frame=shadowbox] 
In[3]: ModelOutput[Eigenstates];
In[4]: MakeWHIZARD[Options];
\end{lstlisting} 
The different options accepted by \verb"MakeWHIZARD" are:
\begin{enumerate}
\item \verb"Exclude", Values: list of generic vertex types, Default: \verb"{SSSS}" \\
Prevents vertices matching the generic types from being generated in order to
speed up the program and reduce the complexity of the generated model.
\item \verb"WOModelName", Values: string, Default: predefined model name \\
Gives the possibility to change the model name. If output for \WHIZARD 1.x is generated,
the name should start with \verb"fr_" in order for the model to be picked up
automatically by the \WHIZARD build system.
\item \verb"MaximalCouplingsPerFile", Values: Number, Default: 500\\
Defines the maximal number of couplings written to one file. Adjusting this
eases the workload on the compiler when compiling the model.
\item \verb"Version", Values: String, Default: latest version\\
Defines the version of \WHIZARD for which the model file is generated. A list of
all values for this setting currently valid can be found in
Tab.\ref{tab-wowhizardversion}. In addition, you can get a list via
{\tt?WO`WhizardVersion} in Mathematica after \SARAH has been loaded.
\item \verb"ReadLists", Values: \verb"True" or \verb"False", Default: \verb"False" \\
This setting controls whether the cached results from a previous calculation
should be used.
\end{enumerate}

While generating the model files, the interface will print status messages and
information about potential incompatibilities of the model with \WHIZARD to the
screen. It is highly advised to read through this information carefully.

\paragraph*{Using the generated model files with \WHIZARD}\mbox{}\\
After the interface has completed, the generated files can be found in the
\verb"WHIZARD_Omega" subdirectory of {\SARAH}'s output directory. In order to
use
the model with \WHIZARD 2.x, the generated code must be compiled and installed.
For most applications, this is done by simply issuing (inside the output
directory)
\begin{lstlisting}
> ./configure
> make
> make install
\end{lstlisting} 
By default, the third command installs the compiled model into \verb".whizard"
in current user's home directory where it is automatically picked up by
\WHIZARD. Alternative installation paths can be specified using the
\verb"--prefix" option to \WHIZARD.
\begin{lstlisting}
> ./configure --prefix=/path/to/installation/prefix
\end{lstlisting} 
If the files are installed into the \WHIZARD
installation prefix, the program will also pick them up automatically, while
{\WHIZARD}'s \verb"--localprefix" option must be used to communicate any other
choice to \WHIZARD. In case \WHIZARD is not available in the binary search
path, the \verb"WO_CONFIG" environment variable can be used to point
\verb"configure" to the binaries
\begin{lstlisting}
> ./configure WO_CONFIG=/path/to/whizard/binaries
\end{lstlisting} 
More information on the available options and their syntax can be obtained with the
\verb"--help" option.

In the case of \WHIZARD 1.x output, the generated files must be patched into the
\WHIZARD source tree. To this end, the interface creates a script called
\verb"inject". In most cases, it is sufficient to simply call the script as
\begin{lstlisting}
> ./inject /path/to/whizard
\end{lstlisting} 
(from within the output directory). Issuing \verb"./inject --help" will display
a list of options which can be used to adapt the script to more complicated
usage scenarios.

\paragraph*{Spectra and parameters}\mbox{}\\
In order to communicate the numerical values of the parameters calculated by
\SPheno to \WHIZARD, each \SPheno version generated by \SARAH is capable of writing
out a separate file providing this information as \SINDARIN code which can
be directly included into the \WHIZARD input script. The user just has to
make sure to use
\begin{lstlisting}[frame=shadowbox]
 Block SPhenoInput #
 76 1 # WHIZARD file
\end{lstlisting}
in the LesHouches input file of \SPheno. The created file {\tt
WHIZARD.par.[Model]} can be included in the input script for \WHIZARD by means of
the \verb"include" statement. For example, for the MSSM, the corresponding line
would read
\begin{lstlisting}[frame=shadowbox]
include ("WHIZARD.par.MSSM")
\end{lstlisting}

\section{Parameter scans with \SSP}
\label{sec:SSP}
\begin{figure}[hbt]
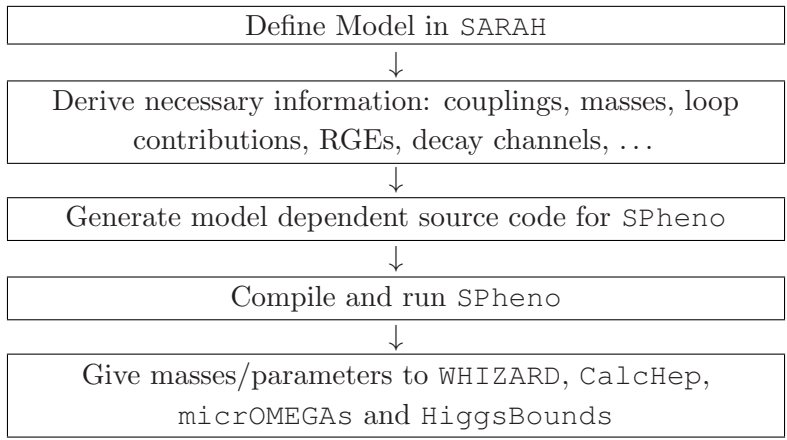

 \begin{center}
 \fbox{\parbox{10cm}{\centering
 Define Model in \SARAH
 }} \\
 \(\downarrow\) \\
 \fbox{\parbox{10cm}{\centering
 Derive necessary information: couplings, masses, loop contributions, RGEs,
decay channels, \dots
 }}\\ 
\(\downarrow\) \\
\fbox{\parbox{10cm}{\centering
Generate model dependent source code for \SPheno
}} \\
\(\downarrow\) \\
\fbox{\parbox{10cm}{\centering
Compile and run \SPheno
}}\\
 \(\downarrow\) \\
 \fbox{\parbox{10cm}{\centering
 Give masses/parameters to \WHIZARD, \CalcHep, \MicrOmegas and \HB
 }}
\end{center}
\caption[Automatized way from model building to phenomenology]{The model is
defined in \SARAH. Afterwards, \SARAH generates all necessary files to implement
this model in \SPheno, \MicrOmegas and \WHIZARD. In addition, the \SPheno
version writes a file which can be used as input for \HB. This provides an
completely automatized way from model building to phenomenology. }
\label{fig:SPheno_SARAH}
\end{figure}

In the previous sections, we have discussed the implementation of SUSY models into
different tools which can be combined to cover much of the analysis of new
models. The corresponding work flow is depicted in Fig.~\ref{fig:SPheno_SARAH}.
However, in order to use this tool chain it is still necessary to run all
programs and pass the information between them. To ease this task, we have
created the \SSP ({\tt SARAH Scan and Plot}) \Mathematica package. This tool
facilitates a quick overview of a new model by performing
the different analysis steps automatically for many different points in parameter space.
Furthermore, it is possible to define additional parameter space constraints
and to use \Mathematica's intrinsic functions for creating 2D parameter space
plots.

\subsection{Installation and running scans}
The package can be downloaded from
\begin{lstlisting}
http://projects.hepforge.org/sarah/SSP.html
\end{lstlisting}
After extracting the package archive to the \Mathematica application directory,
\SSP is loaded via
\begin{lstlisting}
In[1]: <<"SSP/SSP.m" 
\end{lstlisting}
A scan is started by
\begin{lstlisting}
In[2]: Start["Inputfile"]; 
\end{lstlisting}
where {\tt Inputfile} is a file containing all necessary information for a scan. 

\subsection{The input files}
\label{SSP_input}
Each scan performed by \SSP is based on the information given in two input
files. In a first file, which has to be located in the root directory of \SSP,
the location of the different programs is given and some basic information about
the different tools like the name of the in- and output file has to be
defined. For instance, the necessary entries for \SPheno read
\begin{lstlisting} [frame=shadowbox]
DEFAULT[SPheno] = "[SPheno Directory]/bin/SPheno[Model]";
DEFAULT[SPhenoInputFile] = "LesHouches.in.[Model]";
DEFAULT[SPhenoSpectrumFile] = "SPheno.spc.[Model]";
\end{lstlisting}

The second file contains all information defining a specific scan: the name of the
requested settings file, the programs which should be included into the scan,
input parameters and ranges, any constraints which should be applied and the
definitions of the actual plots. As several different scans can be defined in
one input
file, each scan has first to be assigned an unique identifier, e.g. 
\begin{lstlisting}[frame=shadowbox]
RunScans = {ScanM0,ScanM12};
\end{lstlisting}
In order to toggle the inclusion of the different tools, the flags {\tt
IncludeWHIZARD}, \linebreak {\tt IncludeHiggsBounds}, {\tt IncludeCalcHep} or
{\tt IncludeMicrOmegas} are used
\begin{lstlisting}[frame=shadowbox]
DEFINITION[ScanM0][IncludeHiggsBounds] = True;
DEFINITION[ScanM12][IncludeHiggsBounds] = False;  
\end{lstlisting}
The information on the parameters is defined by different lists for
each block of the LesHouches input file which might look like
\begin{lstlisting}[frame=shadowbox]
DEFINITION[ScanM0M12][MINPAR]={
  {{1}, {Min->0,Max->1000, Steps->50,Distribution->LINEAR}},
  {{2}, {Min->0,Max->1000, Steps->50,Distribution->LINEAR}},
  {{3}, {Value->10}},
  {{4}, {Value->1}},
  {{5}, {Value->0}}  };
\end{lstlisting}
Other possible distributions are {\tt LOG} and {\tt RANDOM}. 

\subsection{Features}
\paragraph*{Fit to constraints}\mbox{}\\
\SSP can be instructed to adjust one or several model parameters such that a
list of constraints is fulfilled.
To this end, the \Mathematica function {\tt NMinimize} is utilized to find a
combination of free parameters which leads to the smallest \(\chi^2\) for the
given constraints. For instance, to fix the Higgs mass between 114.9 and
115.1~GeV by a variation of \(A_0\) and \(\tan\beta\), the necessary input is 
\begin{lstlisting}[frame=shadowbox]
DEFINITION[FITHIGGS][FitValues]={
 {MASS[25],115,0.1}  };

DEFINITION[FITHIGGS][FreeParameters]={
 {TANBFIT,{5,15}},
 {A0FIT,{0,100}} };

DEFINITION[FITHIGGS][MINPAR]={
  {{1}, {Min->0.,Max->1000, Steps->10,Distribution->LINEAR}},
  {{2}, {Value->500.}},
  {{3}, {Value->TANBFIT}},
  {{4}, {Value->1.}},
  {{5}, {Value->A0FIT}}   };
\end{lstlisting}
In order to tune the behavior of {\tt NMinimize}, e.g.
manually choosing a fit algorithm, options can be passed to the function 
\begin{lstlisting}[frame=shadowbox]
 DEFINITION[FITHIGGS][FitOptions]={Method->"NelderMead"};
\end{lstlisting}
As most of the fit algorithms cannot deal with a finite parameter range
\([a,b]\), \SSP transforms to infinite boundaries via stretching
\begin{equation}
 P_{ext} = a + \frac{b-a}{2} +  \frac{\mbox{atan}(P_{int})}{\pi}(b-a)
\end{equation}
with \(P_{int} \in [-\infty,+\infty]\) and \(P_{ext} \in [a,b]\). For
more details on {\tt NMinimize} and its various options, we refer to the
\Mathematica manual.

\paragraph*{2D parameter sampling}\mbox{}\\
One common problem is the sampling of a two
dimensional parameter space, e.g. when checking the dark matter relic density in
the \((m_0,M_{1/2})\) plane. However, in many cases,
a fixed grid or random scan might not be the
best choice because certain areas should be sampled more precisely than others.
The {\tt ContourPlot} function of \Mathematica was developed exactly for such
purposes and therefore, \SSP can be instructed to use it.

\begin{figure}[htb]
\begin{minipage}{\textwidth}
\includegraphics[width=0.3\textwidth]{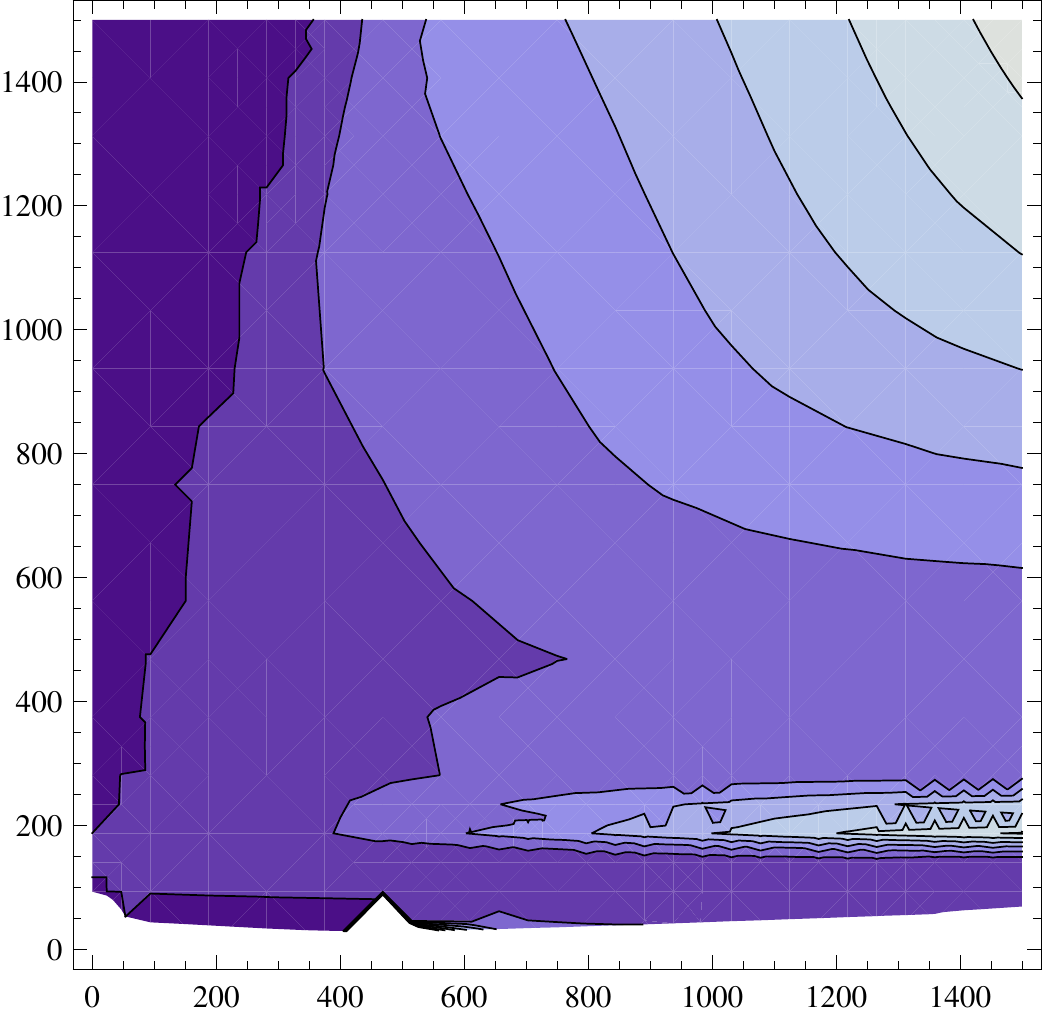} \hfill
\includegraphics[width=0.3\textwidth]{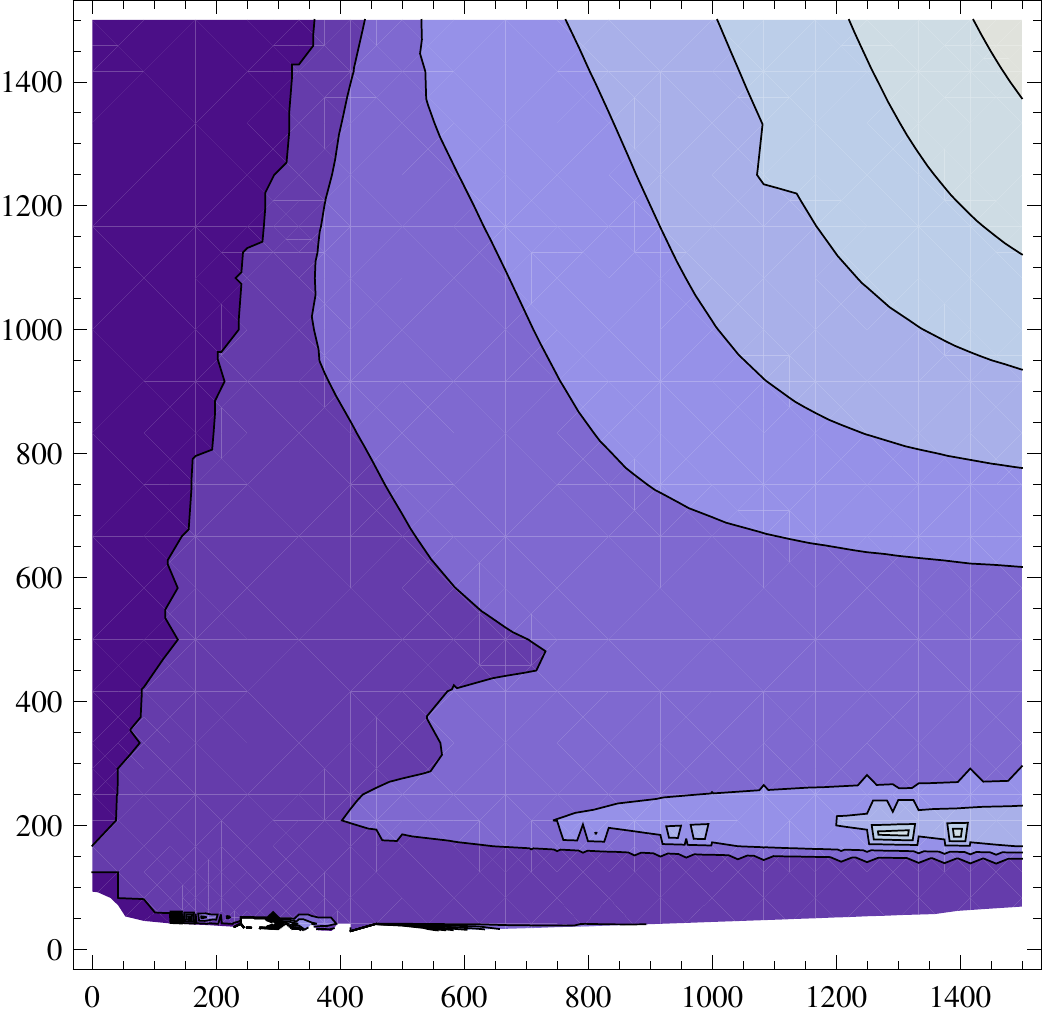} \hfill
\includegraphics[width=0.3\textwidth]{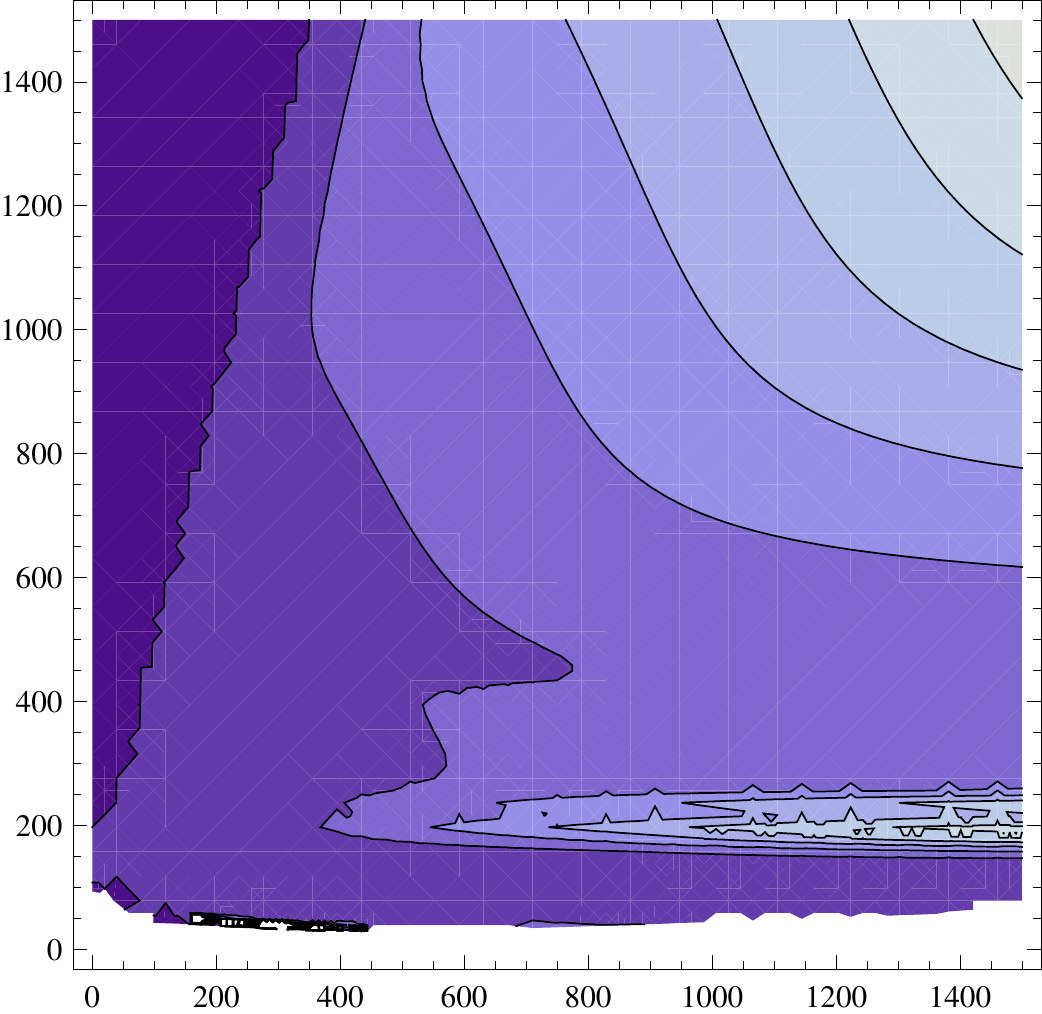} \\
\includegraphics[width=0.3\textwidth]{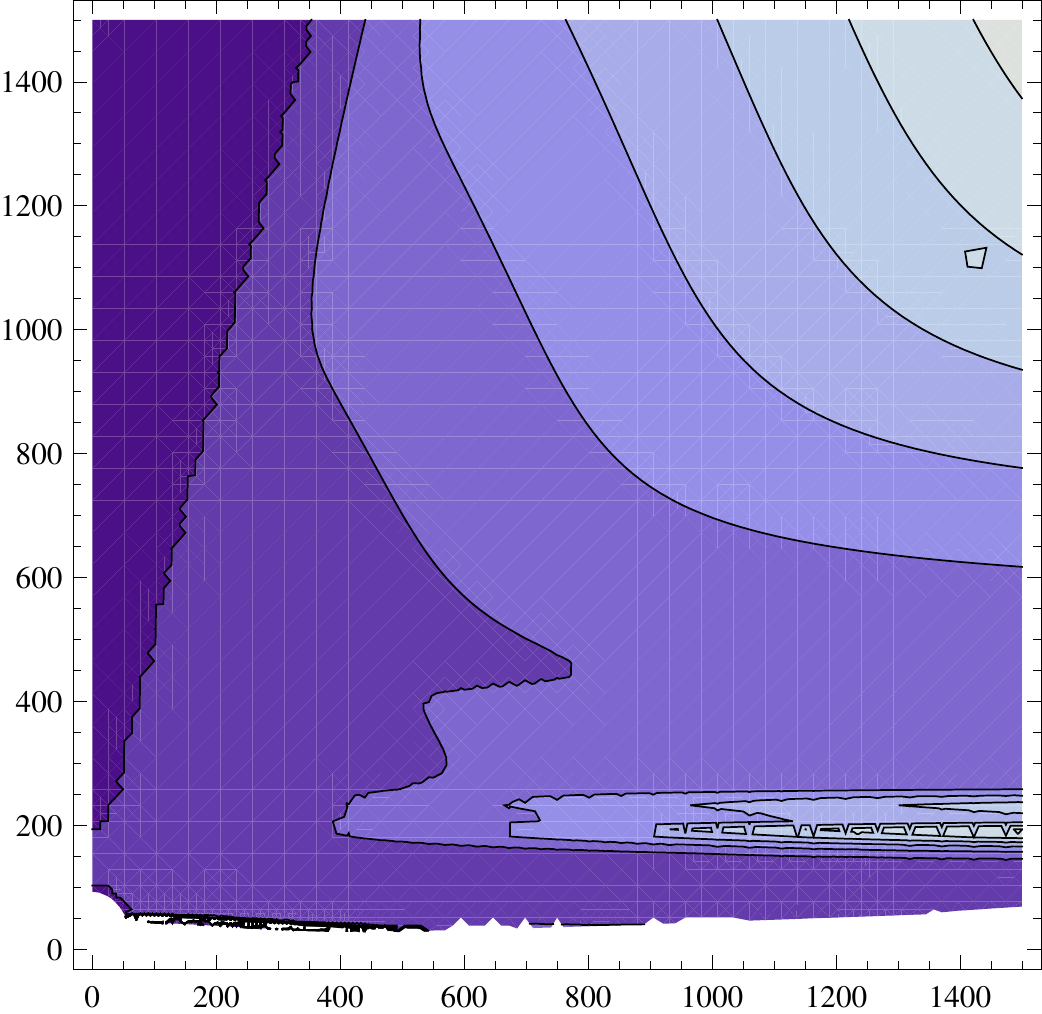} \hfill
\includegraphics[width=0.3\textwidth]{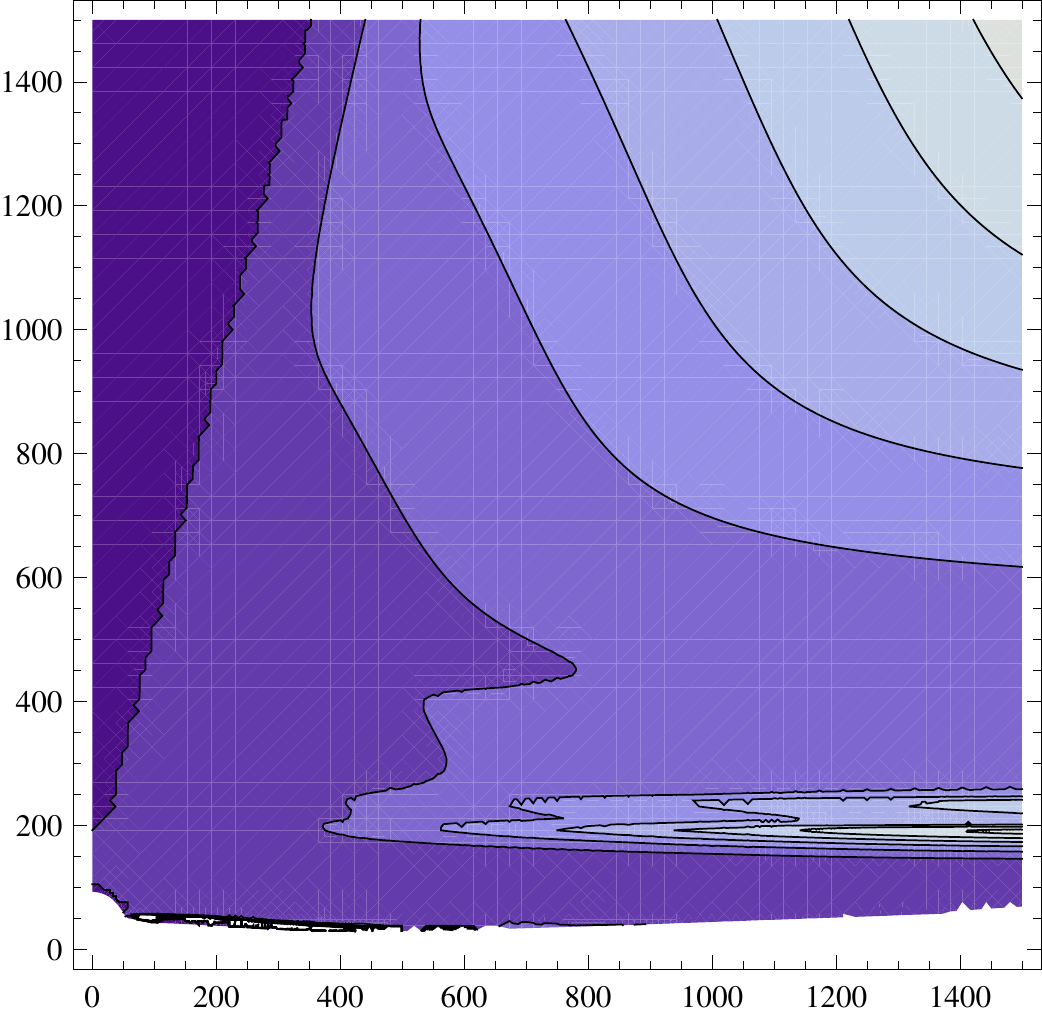} \hfill
\includegraphics[width=0.3\textwidth]{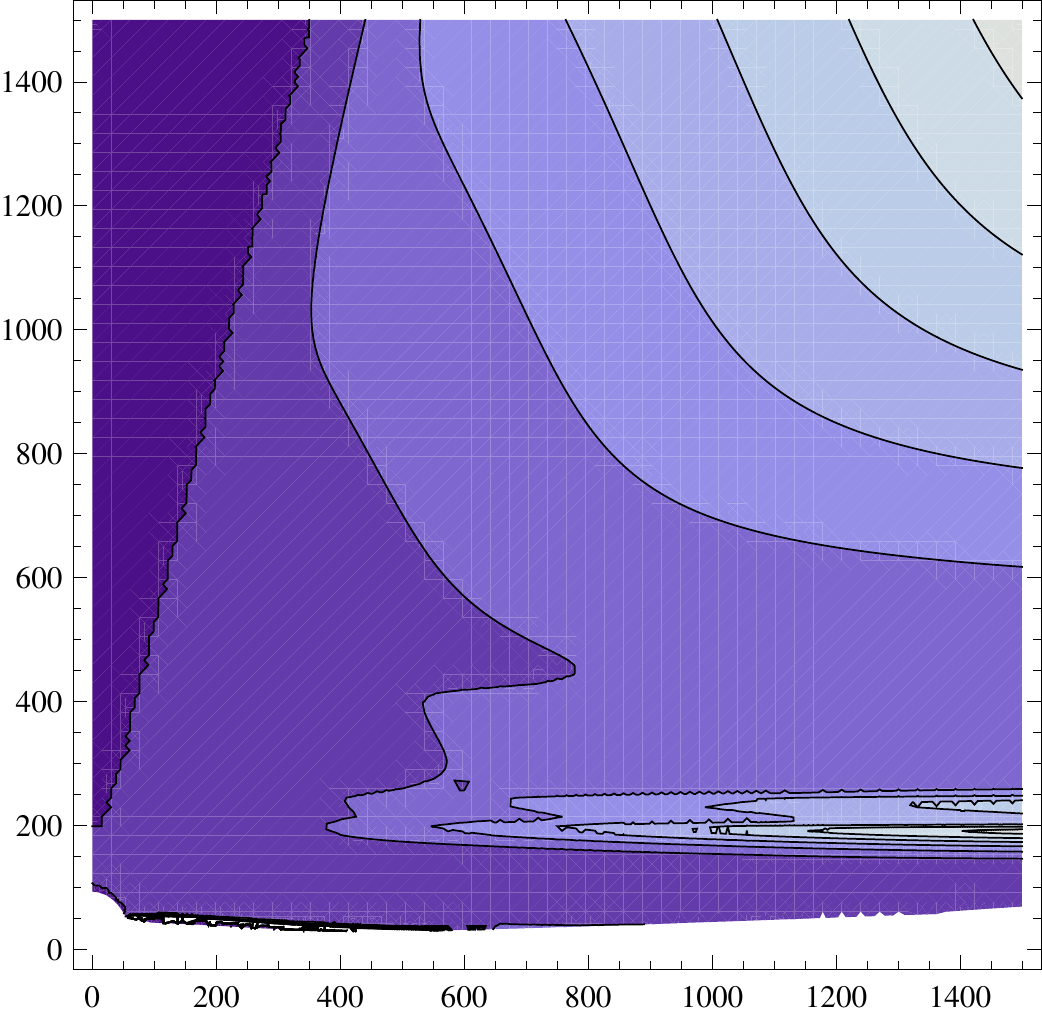} \\
\end{minipage}
 \caption{$(m_0,M_{1/2})$ plane for different values of {\tt PlotPoints}. First
row (from left to right): 5, 10, 20. Second row (from left to right): 30, 40,
50. The mesh gives an insight, where \Mathematica has decreased the distance
between the interpolation points.}
\label{fig:SSP_DM_PP}
\end{figure}

The input for using {\tt ContourPlot} with \SSP to do a \((m_0,M_{1/2})\) scan
is
\begin{lstlisting}[frame=shadowbox]
DEFINITION[M0M12][CountourScan]=
  {DARKMATTER[1],   
    {CONTOURSCANPARAMTER[1],0,1500},  
    {CONTOURSCANPARAMTER[2],0,1500},
    ContourPlotOptions,"DM_A0.eps"};

DEFINITION[M0M12][MINPAR]={
  {{1}, {Value->CONTOURSCANPARAMTER[1]}},
  {{2}, {Value->CONTOURSCANPARAMTER[2]}},
  {{3}, {Value->10}},
  {{4}, {Value->1}},
  {{5}, {Min->1,Max->1000,Steps->3,Distribution->LOG}}   };
\end{lstlisting}
With this setup, contour plots for the relic density in the \((m_0,M_{1/2})\) plane
are created for \(A_0=1,10,1000\)~GeV and \(\tan\beta=10, \mbox{sign}\mu>0\). As
options for the scan, the usual options of {\tt
ContourPlot} can be used. For instance, in Fig.~\ref{fig:SSP_DM_PP}, we used {\tt
PrecisionGoal->"Quality"} and varied the {\tt PlotPoints} between 5 and 50. The
number of valid parameter points which have been evaluated can be found in
Tab.~\ref{tab:DM_points}.

\begin{table}[htb]
\centering
 \begin{tabular}{|c|c|c|c|c|c|c|}
  \hline
   {\tt PlotPoints} & 5 & 10 & 20 & 30 & 40 & 50\\
\hline
  Calculated points & 990 & 2077 & 4589 & 9242 & 13324 & 17416 \\
 \hline
 \end{tabular}
\caption{Number of evaluated parameter space points in the
$(m_0,M_{1/2})$-plane for a given value of {\tt PlotPoints}.}
\label{tab:DM_points}
\end{table}

\section{Putting the programs to work}
\label{sec:Combination}

\subsection{Automated setup and model implementation}
\label{sec:UsingTheScripts}
In order to simplify the setup of the different programs and the generated model
files, we provide a script to setup and install the complete environment and
another script for the automated implementation of new models. These scripts
can be downloaded from 
\lstset{frame=none}
\begin{lstlisting} 03
http://projects.hepforge.org/sarah/Toolbox.html
\end{lstlisting}
After downloading the current version of the package, untar it using
\begin{lstlisting} 
> tar -xzf toolbox-$Version.tar.gz
\end{lstlisting}
To download and install the different packages, create and change to a working
directory (which will later contain the installed packages) and call the {\tt
configure} script from there, e.g.
\begin{lstlisting} 
> mkdir build
> cd build
> ../configure
\end{lstlisting}
The {\tt configure} script will now proceed to check for the requirements of the
different packages and download any missing files. All downloaded archives will
be placed in the {\tt tarballs} subdirectory of the directory containing the
{\tt configure} script from where they will be reused in all subsequent runs.
Command line options can be used to disable specific packages and to point the
script to custom locations of compilers and of the \Mathematica kernel; a full
list of those can be obtained by calling {\tt configure} with the {\tt --help}
option.

After {\tt configure} finishes successfully, {\tt make} can be called to build
all configured packages
\begin{lstlisting} 
> make
\end{lstlisting}
After {\tt make} has finished, the different packages are ready to use and can
be found in subdirectories of the build directory, together with a suitable
setup file for \SSP.

Also created by {\tt configure} is a script which automates the implementation
of a new models into the different packages
\begin{lstlisting} 
> ./butler [model]
\end{lstlisting}
where {\tt [model]} is the models' \SARAH name. For
instance,
\begin{lstlisting} 
> ./butler NMSSM
\end{lstlisting}
resp.
\begin{lstlisting} 
> ./ImplementModel.sh OmegaShort/Regime-2
\end{lstlisting}
can be used to install the NMSSM / resp. the left-right symmetric model
discussed in sec.~\ref{app:Omega}. A list of different command line options can
be obtained with
\begin{lstlisting} 
> ./butler --help
\end{lstlisting}
After {\tt butler} has completed, the new model is implemented into all selected
packages and ready for use.

\subsection{Preparing a model by hand}
If more flexibility is required than is offered by butler, the model files must be
generated and integrated into the various packages by hand.
In order to demonstrate how the corresponding workflow looks like, we show how the
NMSSM version included in \SARAH is
implemented into the other programs\footnote{
For the details on the installation of the different packages we refer to the
respective documentation of the programs.}.

In order to implement the MSSM and prepare for an automatic parameter scan using
SSP, the following steps are necessary
\begin{enumerate}
 \item Prepare \SARAH and generate all output
\begin{enumerate}
 \item Define the model in \SARAH and evaluate it
\begin{lstlisting}
In[1]: << SARAH.m; 
In[2]: << Start["NMSSM"];
\end{lstlisting}
 \item Create the source code for \SPheno 
\begin{lstlisting}
In[3]: MakeSPheno[]
\end{lstlisting}
 \item Create model files for \CalcHep and \MicrOmegas
\begin{lstlisting}
In[6]: MakeCHep[]; 
\end{lstlisting}
\item Create model files for \WHIZARD 
\begin{lstlisting} 
In[5]: MakeWHIZARD[];
\end{lstlisting}
The last three steps can also be done at once by using {\tt MakeAll[]}. This
produces the \SARAH output for \SPheno, \CalcHep, \WHIZARD, \FeynArts and
\LaTeX.
\end{enumerate}

\item Prepare \SPheno
\begin{enumerate}
 \item Create a subdirectory in your installation  of \SPheno {\tt 3.1}
or later and copy the gnerated code there
\begin{lstlisting}
> cd [SPheno Directory]/
> mkdir NMSSM
> cp [SARAH Directory]/Output/NMSSM/EWSB/SPheno/* \
                             [SPheno Directory]/NMSSM/
\end{lstlisting}
 \item Compile \SPheno
\begin{lstlisting}
> make Model=NMSSM
\end{lstlisting}
\end{enumerate}

\item Prepare \CalcHep
\begin{enumerate}
\item Create a new project in \CalcHep
\begin{lstlisting}
> cd [CalcHep Directory]/
> ./mkUsrDir NMSSM 
\end{lstlisting}
\item Copy the model files from \SARAH to \CalcHep
\begin{lstlisting}
> cp [SARAH Directory]/Output/NMSSM/EWSB/CHep/*  \
             [CalcHep Directory]/NMSSM/models/
\end{lstlisting}
\item Start \CalcHep and create an executable {\tt n\_calchep} program for
a specific process using {\tt make n\_calchep}
\begin{lstlisting}
> cd NMSSM/
> ./calchep
\end{lstlisting}
\item Copy {\tt n\_calchep} and the other created files including the
subdirectory {\tt aux} to a new directory
\begin{lstlisting}
> cp -R results/* [Process Directory]/
\end{lstlisting}
\item Repeat the last two steps for other processes if necessary
\end{enumerate}

\item Prepare \MicrOmegas
\begin{enumerate}
\item Create a new project in \MicrOmegas
\begin{lstlisting}
> cd [MicrOmegas Directory]
> ./newProject NMSSM 
\end{lstlisting}
\item Copy the \CalcHep model files to the project directory
\begin{lstlisting}
> cp [SARAH Directory]/Output/NMSSM/EWSB/CHep/*  \
             [MicrOmegas Directory]/NMSSM/work/models/
\end{lstlisting}
\item Move the \MicrOmegas main file to the main directory and compile it
\begin{lstlisting}
> cd [MicrOmegas Directory]/NMSSM/
> mv work/models/CalcOmega.cpp .
> make main=CalcOmega.cpp
\end{lstlisting}
\end{enumerate}

\item Prepare \WHIZARD
\begin{enumerate}
\item Configure the model source
\begin{lstlisting}
> cd [SARAH Directory]/Output/NMSSM/EWSB/WHIZARD_Omega/
> ./configure WO_CONFIG=[WHIZARD Directory]/bin/
\end{lstlisting}
\item Compile and install the new model (with this configuration, the files will
end up in {\tt\$HOME/.whizard})
\begin{lstlisting}
> make install
\end{lstlisting}
\end{enumerate}
\end{enumerate}

\subsection{Running the chain}
Now, everything is in its place, and the model can be used with the different
programs:
\begin{itemize}
 \item Go the the \SPheno directory, edit the LesHouches file and run \SPheno
\begin{lstlisting}
> cd [SPheno Directory]/NMSSM/
> edit LesHouches.in.NMSSM
> ./../bin/SPhenoNMSSM
\end{lstlisting}
Make sure that the flags
\begin{lstlisting}[frame=shadowbox]
 Block SPhenoInput #
 75 1 # HiggsBounds files
 76 1 # WHIZARD file
\end{lstlisting}
are set in {\tt LesHouches.in.nMSSM} in order to create the parameter file for
\WHIZARD as well as the input files for \HB.
\item Copy the spectrum file to the \MicrOmegas directory and the file for
\WHIZARD to your \WHIZARD working directory
\begin{lstlisting}
> cp SPheno.spc.NMSSM [MicrOmegas Directory]/NMSSM/
> cp WHIZARD.par.NMSSM [WHIZARD Working Directory]/  
\end{lstlisting}
\item Run \MicrOmegas
\begin{lstlisting}
> cd [MicrOmegas Directory]/NMSSM/
> ./CalcOmega
\end{lstlisting}
\item Create a \WHIZARD input file (e.g {\tt Input.sin}) with your process
setup and include the parameter file of \SPheno by using {\tt
Include("WHIZARD.par.NMSSM")}.
\item Run \WHIZARD
\begin{lstlisting}
> whizard Input.sin
\end{lstlisting}
\item Copy the \SPheno spectrum files to the directory of your \CalcHep
executables and run {\tt n\_calchep}
\begin{lstlisting}
> cp [SPheno Directory]/bin/SPheno.spc.NMSSM [CH Process Directory]/
> cd [CH Process Directory]/
> ./n_calchep  
\end{lstlisting}
To run {\tt n\_calchep} using a script you can also use the blind mode,
see sec.~\ref{sec:CalcHep}.
\end{itemize}

\section{Conclusions}
We have presented a toolchain for studying extensions of the MSSM. The model
under consideration is
implemented in \SARAH which creates then source code for \SPheno, thus
facilitating the implementation of new models into \SPheno in a modular way.
The new \SPheno
modules calculate the  mass spectrum using two-loop RGE running and one-loop
corrections to the masses. The precision of the calculation of the mass
spectrum for a large range
of SUSY models is comparable to the known precision of the MSSM provided by the
established spectrum generators (excluding two loop effects in the Higgs sector).
Therefore, \SARAH can be regarded as a 
'spectrum-generator-generator', extending the availability of spectrum
generators for different models dramatically.

Apart from creating output for \SPheno, \SARAH is capable of creating model
files suitable for use with \CalcHep / \MicrOmegas, \FeynArts, \HB and
\WHIZARD.
The masses and parameters calculated by \SPheno can be used directly as input
for those programs which then facilitate further studies of the models'
phenomenology. In particular, the dark matter relic density predicted by the model
can be simulated using \CalcHep / \MicrOmegas, while \WHIZARD is well suited for
examining collider observables involving the long decay chains typical for
supersymmetric models. The scan of observables over parameter space can be
automatized using the \SSP package.

As \SARAH works in a very generic way, a wide range of
possible models is covered. Models with new fields and/or an extended gauge
sector at the SUSY scale are supported as well as models with thresholds and/or gauge
symmetry breaking at higher energy scales. This flexibility, together with
the integration of different existing tools together with \SARAH and \SSP into
the largely automated toolchain presented in this paper should greatly simplify
the study of new supersymmetric models.

\section*{Acknowledgments}
We thank for beta testing and useful feedback Martin Hirsch, Avelino Vicente,
Bj\"orn Herrmann, Laslo Reichert, Benedikt Vormwald and Ben O`Leary. W.P.\ and
F.S.\ have been supported by the DFG project No. PO-1337/1-1. CS has been
supported by the Deutsche Forschungsgemeinschaft through
the Research Training Group GRK\,1102 \textit{Physics of Hadron Accelerators}.

\begin{appendix}
 \lstset{basicstyle=\scriptsize, frame=shadowbox}
\section{Input files for \SARAH to generate a \SPheno version with
thresholds}
\subsection{Seesaw I}
\label{app:Seesaw}
\subsubsection*{Seesaw1.m}
In the case of  seesaw~I, the particle content is extended by three generations
of a gauge singlet \(\hat{\nu}_R\). 
\begin{lstlisting}
Fields[[8]] = {vR, 3, v, 0, 1, 1};
\end{lstlisting}
The new field is assigned a Yukawa like interaction and a mass term. In addition,
there is an effective operator which is generated by integrating out the
right handed neutrino. 
\begin{lstlisting}
SuperPotential =
  {...,{{1,Yv},{v,l,Hu}},{{1/2,Mv},{v,v}},{{1/2,WOp},{l,Hu,l,Hu}}};
\end{lstlisting}
Since the field is heavy, it should not be included in the calculation of the
vertices and masses at the SUSY scale. Therefore, we ``delete'' it. Note that,
although deleted particles do not appear in the low energy spectrum and Feynman
rules, they do contribute to the RGEs above the threshold scale.
\begin{lstlisting}
DeleteParticles={v};
\end{lstlisting}

\subsubsection*{SPheno.m}
We choose a unification of the soft-breaking mass of the scalar singlet with the
other soft-breaking masses at the GUT scale. In addition, we want to define the
values of the superpotential parameters at the GUT scale as input values in the
LesHouches file. Furthermore, we also impose a mSugra like condition for the
trilinear soft-breaking coupling. As the bilinear soft-breaking term does
not influence the RGE running of the other parameters, we can safely set it to
zero. Altogether, the additional boundary conditions at the GUT scale are
\begin{lstlisting}
BoundaryHighScale={
  ...,
  {mv2,    DIAGONAL m0^2},
  {Mv,     LHInput[Mv]},
  {Yv,     LHInput[Yv]}
  {B[Mv],  0},
  {T[Yv],  Azero*LHInput[Yv]}
};
\end{lstlisting}
We want to include three threshold scales: each generation of the gauge singlet
should be integrated out at energies similar to their mass. Therefore, a good
choice is
\begin{lstlisting}
Thresholds={
  {Abs[MvIN[1,1]],{v[1]}},
  {Abs[MvIN[2,2]],{v[2]}},
  {Abs[MvIN[3,3]],{v[3]}}
};
\end{lstlisting}
When thresholds are included, the boundary conditions for running up and down
the RGEs can be defined separately for each threshold scale. To this end, it is
first necessary to initialize the corresponding arrays
\begin{lstlisting}
BoundaryConditionsUp=Table[{},{Length[Thresholds]}];
BoundaryConditionsDown=Table[{},{Length[Thresholds]}];
\end{lstlisting}
When a gauge singlet is integrated out, the Wilson coefficient of
the effective dim-5 operator receives a
contribution of the form
\begin{equation}
\label{eq:mnuSeesawI}
 \kappa = - Y_\nu^T \, M_N^{-1} \, Y_\nu \thickspace . 
\end{equation}
When heavy superfields are integrated out, the mass splitting between the fermionic
and scalar component is neglected. The masses are calculated individually at
each threshold scale and saved in arrays with the names
\begin{lstlisting}
 MassOf <> Name of Superfield
\end{lstlisting}
Therefore, the contributions to the effective operator at the different scales
are given by
\begin{lstlisting}
BoundaryConditionsDown[[1]]={
{WOp[index1,index2],WOp[index1,index2] - Yv[1,index1] Yv[1,index2]/MassOfv[1]}
};
BoundaryConditionsDown[[2]]={
{WOp[index1,index2],WOp[index1,index2] - Yv[2,index1] Yv[2,index2]/MassOfv[2]}
};
BoundaryConditionsDown[[3]]={
{WOp[index1,index2], - Yv[3,index1] Yv[3,index2]/MassOfv[3]}
};
\end{lstlisting}

\subsection{Implementation of a model with a gauge symmetry breaking scale in
\SARAH and \SPheno}
\label{app:Omega} 
This section is devoted to a detailed discussion of
the implementation of a left-right supersymmetric model
(based on that presented in \cite{Esteves:2010si})
into \SPheno and \SARAH. Since we are mainly interested in the
implementation in \SARAH, it is sufficient to simplify the model a bit by just
choosing one threshold scale and not two as discussed in \cite{Esteves:2010si}.
Adding the second threshold scale is straightforward, but would lead to some
redundancy in the following discussion. 

\subsubsection{Summary of the model} 
We here give only a short summary about the model and refer to
\cite{Esteves:2010si} and references therein for more details.
As adread discussed, we introduce
only one threshold scale at which \(\times SU(2)_R \times
U(1)_{B-L}\) gets broken
\begin{equation}
 SU(2)_L \times SU(2)_R \times U(1)_{B-L} \rightarrow  SU(2)_L \times U(1)_Y
\end{equation}

\paragraph*{From GUT scale to $SU(2)_R \times U(1)_{B-L}$ breaking scale}
The MSSM particle content above the threshold is extended by the presence of
four fields which are triplets under \(SU(2)_L\) or \(SU(2)_R\) and which carry
a \(B-L\) charge. In addition, there are two triplets which are uncharged under
\(B-L\). Furthermore, the right handed neutrino are part of the spectrum and
the Higgs fields are arranged in so-called bi-doublets \(\Phi\). To get a
non-trivial CKM matrix, we need at least two generations of \(\Phi\) fields.
The particle content can be summarized as follows \\
\begin{center}
\begin{tabular}{c c c c c c}
\hline
Superfield & generations & $SU(3)_c$ & $SU(2)_L$ & $SU(2)_R$ & $U(1)_{B-L}$ \\
\hline
$Q$ & 3 & 3 & 2 & 1 & $\frac{1}{3}$ \\
$Q^c$ & 3 & $\bar{3}$ & 1 & 2 & $-\frac{1}{3}$ \\
$L$ & 3 & 1 & 2 & 1 & -1 \\
$L^c$ & 3 & 1 & 1 & 2 & 1 \\
$\Phi$ & 2 & 1 & 2 & 2 & 0 \\
$\Delta$ & 1 & 1 & 3 & 1 & 2 \\
$\bar{\Delta}$ & 1 & 1 & 3 & 1 & -2 \\
$\Delta^c$ & 1 & 1 & 1 & 3 & -2 \\
$\bar{\Delta}^c$ & 1 & 1 & 1 & 3 & 2 \\
$\Omega$ & 1 & 1 & 3 & 1 & 0 \\
$\Omega^c$ & 1 & 1 & 1 & 3 & 0 \\
\hline
\end{tabular} 
\end{center}
The superpotential for the model reads
\begin{eqnarray} \label{eq:Wsuppot1}
{\cal W} &=& Y_Q Q \Phi Q^c 
          +  Y_L L \Phi L^c 
          - \frac{\mu}{2} \Phi \Phi
          +  f L \Delta L
          +  f^* L^c \Delta^c L^c \nonumber \\
         &+& a \Delta \Omega \bar{\Delta}
          +  a^* \Delta^c \Omega^c \bar{\Delta}^c
          + \alpha \Omega \Phi \Phi
          +  \alpha^* \Omega^c \Phi \Phi \nonumber \\
         &+& M_\Delta \Delta \bar{\Delta}
          +  M_\Delta^* \Delta^c \bar{\Delta}^c
          +  M_\Omega \Omega \Omega
          +  M_\Omega^* \Omega^c \Omega^c \thickspace.
\end{eqnarray}

\paragraph*{Below $SU(2)_R \times U(1)_{B-L}$ breaking scale} Here, we are
left with the MSSM plus the effective Weinberg operator which causes neutrino
masses after EWSB.  

\paragraph*{Boundary conditions} In order to link both scales, we need the following set
of boundary conditions:
\begin{eqnarray}
\label{eq:boundary_LR_1}
Y_d = Y_Q^1 \cos \theta_1 - Y_Q^2 \sin \theta_1 \thickspace,
&\qquad& Y_u = - Y_Q^1 \cos \theta_2 + Y_Q^2 \sin \theta_2  \thickspace, \\
Y_e = Y_L^1 \cos \theta_1 - Y_L^2 \sin \theta_1 \thickspace,
&\qquad& Y_\nu = - Y_L^1 \cos \theta_2 + Y_L^2 \sin \theta_2 \thickspace,
\end{eqnarray}
where $R = \sin (\theta_1 - \theta_2)$. For the soft-trilinear couplings,
\(Y\) is replaced with \(T\) in the expressions. 
For the sfermionic soft masses, we have 
\begin{eqnarray}
m_{q}^2 =m_{u^c}^2 = m_{d^c}^2 &=& m_{Q^c}^2 \thickspace,\\ 
m_{l}^2=m_{e^c}^2  &=& m_{L^c}^2 \thickspace, \\ 
M_L = M_R &=& M_2 \thickspace.
\end{eqnarray}
while in the Higgs sector we need the relations
\begin{eqnarray}
m_{H_d}^2 &=& \cos^2 \theta_1 (m_\Phi^2)_{11} + \sin^2 \theta_1
(m_\Phi^2)_{22} - \sin \theta_1 \cos \theta_1 \left[ (m_\Phi^2)_{12} +
(m_\Phi^2)_{21} \right] \thickspace,  \\ 
m_{H_u}^2 &=& \cos^2 \theta_2
(m_\Phi^2)_{11} + \sin^2 \theta_2 (m_\Phi^2)_{22} - \sin \theta_2 \cos
\theta_2 \left[ (m_\Phi^2)_{12} + (m_\Phi^2)_{21} \right] \thickspace, 
\end{eqnarray}
In the gauge sector, we have to express the hypercharge coupling and the
corresponding gaugino through
\begin{eqnarray}
g_1 & = & \frac{\sqrt{5} g_2 g_{BL}}{\sqrt{2 g_2^2 + 3 g_{BL}^2}} \thickspace ,
\\
\label{eq:boundary_LR_2}
M_1 & = & \frac{2 g_2^2 M_{BL} + 3 g_{BL}^2 M_R}{2 g_2^2 + 3 g_{BL}^2} \thickspace.
\end{eqnarray}

\subsubsection{Model files for \SARAH}
We now discuss the input files for \SARAH required for defining the model at
the different scales. For shortness, we concentrate on the parts necessary
for \SPheno output and skip the gauge fixing terms and definition of Dirac
spinors above the threshold scale. 
\subsubsection*{From GUT scale to $SU(2)_R \times U(1)_{B-L}$ breaking scale} 
The vector and chiral superfields at the highest scale defining the gauge sector
and particle content are defined by
\lstset{basicstyle=\scriptsize,
frame=shadowbox}
\begin{lstlisting}
Gauge[[1]]={B,   U[1], bminl,       gBL,False};
Gauge[[2]]={WL, SU[2], left,        g2,True};
Gauge[[3]]={WR, SU[2], right,       g2,True};
Gauge[[4]]={G,  SU[3], color,       g3,False};

Fields[[1]] = {{uL,  dL},                           3, qL,       1/6, 2, 1, 3};
Fields[[2]] = {{conj[dR], - conj[uR]},              3, qR,      -1/6, 1, 2,-3}; 
Fields[[3]] = {{vL,  eL},                           3, lL,      -1/2, 2, 1, 1};
Fields[[4]] = {{conj[eR],  - conj[vR]},             3, lR,       1/2, 1, 2, 1};
Fields[[5]] = {{{Hd0, Hup},{Hdm, Hu0}},             2, Phi,        0, 2,-2, 1};
Fields[[6]] = {{{deltaLp/Sqrt[2], deltaLpp},
                {deltaL0, - deltaLp/Sqrt[2]}},      1, deltaL,     1, 3, 1, 1};
Fields[[7]] = {{{deltaLbarm/Sqrt[2], deltaLbar0},
              {deltaLbarmm, - deltaLbarm/Sqrt[2]}}, 1, deltaLbar, -1, 3, 1, 1};
Fields[[8]] = {{{deltaRm/Sqrt[2], deltaR0},
              {deltaRmm, - deltaRm/Sqrt[2]}},       1, deltaR,    -1, 1, 3, 1};
Fields[[9]] = {{{deltaRbarp/Sqrt[2], deltaRbarpp},
              {deltaRbar0, - deltaRbarp/Sqrt[2]}},  1, deltaRbar,  1, 1, 3, 1};
Fields[[10]] = {{{omegaL0/Sqrt[2], omegaLp},
               {omegaLm, - omegaL0/Sqrt[2]}},       1, omegaL,     0, 3, 1, 1};
Fields[[11]] = {{{omegaR0/Sqrt[2], omegaRp},
               {omegaRm, - omegaR0/Sqrt[2]}},       1, omegaR,     0, 1, 3, 1};
\end{lstlisting}
The superpotential reads
\begin{lstlisting}
SuperPotential = { {{1, YQ},           {qL,qR,Phi}},
		   {{1, YL},           {lL,lR,Phi}}, 
		   {{1,  f},           {lL,deltaL,lL}},
		   {{1,conj[f]},       {lR,deltaR,lR}}, 
		   {{1,Mdelta},        {deltaL,deltaLbar}},
		   {{1,conj[Mdelta]},  {deltaR,deltaRbar}},
		   {{-1/2,Mu3},        {Phi,Phi}},
                   {{1,Momega},        {omegaL,omegaL}},
                   {{1,conj[Momega]},  {omegaR,omegaR}},
                   {{1,a},             {deltaL,omegaL,deltaLbar}},
                   {{1,conj[a]},       {deltaR,omegaR,deltaRbar}},
                   {{1,AlphaOm},       {omegaL,Phi,Phi}},
                   {{1,conj[AlphaOm]}, {omegaR,Phi,Phi}}   };
\end{lstlisting}
The gauge bosons and gauginos of the right sector decompose into
\begin{lstlisting}
DEFINITION[RSB][GaugeSector]= 
{{VWR,{1,{VWRm,  1/Sqrt[2]},  {conj[VWRm],1/Sqrt[2]}},
      {2,{VWRm, -I/Sqrt[2]},  {conj[VWRm],I/Sqrt[2]}},
      {3,{VWR0,  1}}},
 {fWR,{1,{fWRm,  1/Sqrt[2]},  {fWRp,1/Sqrt[2]}},
      {2,{fWRm, -I/Sqrt[2]},  {fWRp,I/Sqrt[2]}},
      {3,{fWR0,          1}}}};
\end{lstlisting}
after the $\Omega$ and $\Delta$ fields have received their VEV
\begin{lstlisting}
DEFINITION[RBLSB][VEVs]= 
{ {SomegaR0,    {vR,1/Sqrt[2]},  {sigmaOmR,I/Sqrt[2]}, {phiOmR,1/Sqrt[2]}},
  {SdeltaR0,    {vBL,1/Sqrt[2]}, {sigmaR,I/Sqrt[2]},   {phiR,1/Sqrt[2]}},
  {SdeltaRbar0, {vBL,1/Sqrt[2]}, {sigmaRbar,I/Sqrt[2]},{phiRbar,1/Sqrt[2]}}};
\end{lstlisting}
Finally, we need the rotations in the matter sector to the new mass eigenstates
\begin{lstlisting}
DEFINITION[RSB][MatterSector]= 
{    {{SdeltaRm, conj[SdeltaRbarp]},             {Hpm1R1,ZC1}},
     {{SomegaRm, conj[SomegaRp]},                {Hpm2R1,ZC2}},
     {{fB,fWR0,FdeltaR0, FdeltaRbar0, FomegaR0}, {L0, ZN}},
     {{{fWRm, FomegaRm}, {fWRp, FomegaRp}},      {{Lm,UM}, {Lp,UP}}},
     {{phiR, phiRbar, phiOmR},                   {hhR2, ZH}},
     {{sigmaR, sigmaRbar, sigmaOmR},             {AhR2, ZP}},
     {{FvL, conj[FvR]},                          {N0, Znu}},
     {{SHd0,conj[SHu0]},                         {SH0r1,UH0}},
     {{SHdm,conj[SHup]},                         {SHCr1,UHC}}, 
     {{SomegaLm,conj[SomegaLp]},                 {SO1r1,UO1}},
     {{SdeltaLp,conj[SdeltaLbarm]},              {SDLpR1,UDLp}},
     {{SdeltaLpp,conj[SdeltaLbarmm]},            {SDLppR1,UDLpp}},
     {{SdeltaL0,conj[SdeltaLbar0]},              {SDL0r1,UDL0}},
     {{SdeltaRmm,conj[SdeltaRbarpp]},            {SDRmmR1,UDRmm}},
     {{SdeltaR0,conj[SdeltaRbar0]},              {SDR0r1,UDR0}}
      };
\end{lstlisting}

\subsubsection*{Below $SU(2)_R \times U(1)_{B-L}$ breaking scale} 
Below the breaking scale, only the particles and gauge groups of the MSSM survive
as dynamical degrees of freedom. Therefore, the
model file is almost identical that for the MSSM. We only point out the
differences:
\begin{itemize}
 \item The superpotential contains the Weinberg operator
 \begin{lstlisting}
SuperPotential = {...,{{1,WOp},{l,Hu,l,Hu}}  }; 
\end{lstlisting}
 \item The neutrinos are massive and mix among each other
\begin{lstlisting}                                     
DEFINITION[EWSB][MatterSector]= { ...,{{FvL}, {FV, UV}}}; 
\end{lstlisting}
 \item These states form Majorana spinors
\begin{lstlisting}
dirac[[4]] = {Fv,  FV, conj[FV]};
\end{lstlisting}
\end{itemize}
\subsubsection{Model files for \SPheno output}
Two {\tt SPheno.m} files, one for each scale range, are neccessary in order
for \SARAH to create the Fortran source code. While the first one is rather
short, the second one includes all necessary boundary conditions.
\subsubsection*{Above $SU(2)_R \times U(1)_{B-L}$ breaking scale} 
First, the regime must be flagged as an intermediate scale, and \SARAH must be
told its position in the scale hierarchy (counted from GUT to low scale).
\begin{lstlisting}
RegimeNr = 1;
IntermediateScale = True; 
\end{lstlisting}
Afterwards, we give a list with all particles which are integrated out at the
threshold scale after gauge symmetry breaking.
\begin{lstlisting}
HeavyFields = {Hpm1R1, ChiR1, Cha1r1, hhR1, AhR1,
               FvR1, SVRr1, SH0r1[3], SHCr1[3],
               SO1r1, SDLpR1, SDLppR1,
               SDL0r1, SDRmmR1, DR3r1,
               DL1r1, DL2r1, DL3r1, H0r1, HCr1};
\end{lstlisting}
The numbers in square brackets indicate  that only
the third generation and above is integrated out.

In order to calculate the finite shifts of the gauge couplings and gaugino masses, it is
necessary to define the gauge sector of the next scale {\tt NextGauge} as well
as the quantum number of the fields which are integrated out with respect to
those gauge groups.
\begin{lstlisting}
NextGauge=           {U[1], SU[2], SU[3]};
NextQN = { {Hpm1R1,     -1,     1,    1},
           {ChiR1,       0,     1,    1},
	   {Cha1r1,     -1,     1,    1},
           {hhR1,        0,     1,    1},
           {AhR1,        0,     1,    1},
           {FvR1,        0,     1,    1},
           {SVRr1,       0,     1,    1},
	   {SH0r1,    -1/2,     1,    1},
	   {SHCr1,     1/2,     2,    1},
	   {SO1r1,       0,     1,    1},
	   {SDLpR1,      1,     1,    1},
	   {SDLppR1,     2,     1,    1},
	   {SDL0r1,      1,     3,    1},
	   {SDRmmR1,    -2,     1,    1},
	   {DR3r1,      -2,     1,    1},
	   {DL1r1,       1,     1,    1},
	   {DL2r1,       1,     1,    1},
	   {DL3r1,       1,     3,    1},
	   {H0r1,     -1/2,     1,    1},
	   {HCr1,      1/2,     2,    1}
};
\end{lstlisting}

Finally, \SARAH needs information on the vacuum conditions.
There are two different VEVs, and therefore, we need to choose two parameters
which are fixed by the tadpole equations. As there is no closed analytical
solution for them, we give an approximation obtained by neglecting the
soft-breaking terms.
\begin{lstlisting}
ParametersToSolveTadpoles = {Mdelta, Momega};
UseGivenTadpoleSolution = True;

SubSolutionsTadpolesTree={
  Mdelta -> - SignumMdelta ac1 vR/Sqrt[2],
  Momega -> - SignumMomega ac1 vBL^2/(2 Sqrt[2] vR)  
};
SubSolutionsTadpolesLoop={}; 
\end{lstlisting}

\subsubsection*{Below $SU(2)_R \times U(1)_{B-L}$ breaking scale} 
The second scale is not an intermediate scale, and hence
\begin{lstlisting}
RegimeNr = 2;
IntermediateScale = False; 
\end{lstlisting}
We make the following choice of free parameters of that model: to
the set of standard mSugra parameters (\(m_0,M_{1/2},A_0,\tan\beta,\text{sign}\mu\)),
we add \(B_0\), the superpotential parameter \(a\), the signs
of \(M_\Omega\) and \(M_\Delta\), the two VEVs \(v_R\) and \(v_{BL}\), and
the threshold scale. These are defined in the blocks {\tt MINPAR} and {\tt
EXTPAR}.
\begin{lstlisting}
MINPAR= {
         {1,  m0},
         {2,  m12},
         {3,  TanBeta},
         {4,  SignumMu},
         {5,  Azero},
         {6,  Bzero},
         {7,  SignumMomega},
         {8,  SignumMdelta},
         {9,  aInput}};

EXTPAR = {
          {100, vRinput},
	  {101, vBLinput},
	  {200, TScale}};
\end{lstlisting}
As in the, MSSM we fix the numerical values of \(\mu\) and \(B_\mu\) by the
solving the two tadpole equations,
\begin{lstlisting}
ParametersToSolveTadpoles = {\[Mu],B[\[Mu]]}; 
\end{lstlisting}
Furthermore, we use also the common definitions for the SUSY scale, see also
sec.~\ref{sec:Example_SPheno_MSSM}.
\begin{lstlisting}
RenormalizationScaleFirstGuess = m0^2 + 4 m12^2;
RenormalizationScale = MSu[1]*MSu[6];
\end{lstlisting}

We use as condition for the GUT scale
\begin{lstlisting}
ConditionGUTscale = {gBL==g2, g1 == g2}; 
\end{lstlisting}
The second entry is necessary because it might be that
the masses of the fields breaking the left-right symmetry
are above the unification scale which has to be handled
seperatly.
 At the GUT scale we use boundary
conditions motivated by minimal supergravity:
all scalar soft-breaking masses are proportional to \(m_0\), the gaugino masses
are proportional to \(M_{1/2}\), the trilinear soft-breaking couplings are
given by the corresponding superpotential parameter times \(A_0\), and the
bilinear soft-breaking couplings are set to \(B_0\) times the superpotential
parameter. The values for the coupling matrices \(f\),
\(\alpha\)  as well as for \(\mu\) are read from the LesHouches input file.
\begin{lstlisting}
BoundaryHighScale={
   {T[YQ],       Azero*YQ},
   {T[YL],       Azero*YL},
   {f,           LHInput[f]},
   {T[f],        Azero*LHInput[fm]},
   {AlphaOm,     LHInput[AlphaOm]},
   {T[AlphaOm],  Azero*LHInput[AlphaOm]},
   {T[a],        Azero*aInput},
   {B[Mdelta],   Bzero*Mdelta},
   {B[Momega],   Bzero*Momega},
   {B[Mu3],      Bzero*LHInput[Mu3]},
   {mqL2,        DIAGONAL m0^2},
   {mqR2,        DIAGONAL m0^2},
   {mlL2,        DIAGONAL m0^2},
   {mlR2,        DIAGONAL m0^2},
   {mPhi2,       DIAGONAL m0^2},
   {mdeltaL2,    m0^2},
   {mdeltaLbar2, m0^2},
   {mdeltaR2,    m0^2},
   {mdeltaRbar2, m0^2},
   {momegaL2,    m0^2},
   {momegaR2,    m0^2},
   {MassB,       m12},
   {MassWL,      m12},
   {MassG,       m12}
}; 
\end{lstlisting}
To glue the both regimes, we need to define the appropriate boundary
conditions. First, we initialize the arrays
\begin{lstlisting}
ThresholdScales = {TSCALE};

BoundaryConditionsUp = Table[{},{Length[ThresholdScales]}];
BoundaryConditionsDown = Table[{},{Length[ThresholdScales]}]; 
\end{lstlisting}
and then encode the equations
eqs.~(\ref{eq:boundary_LR_1})-(\ref{eq:boundary_LR_2}). In order to keep
the code short, we define
\begin{lstlisting}
  ST1  = Sin[Theta1];
  CT1  = Cos[Theta1];
  ST2  = Sin[Theta2];
  CT2  = Cos[Theta2];
  ST21 = Sin[Theta2-Theta1];
  CT21 = Cos[Theta2-Theta1];
\end{lstlisting}
Using these abbreviations, the boundary conditions can be written as
\begin{lstlisting}
BoundaryConditionsUp[[1]] = { 
 {YQ[index1,index2,1],   (Yu[index1,index2] ST1 + Yd[index1,index2]ST2)/ST21 },
 {YQ[index1,index2,2],   (Yu[index1,index2] CT1 + Yd[index1,index2]CT2)/ST21 },
 {YL[index1,index2,1],   (Yv[index1,index2] ST1 + Ye[index1,index2]ST2)/ST21 },
 {YL[index1,index2,2],   (Yv[index1,index2] CT1 + Ye[index1,index2]CT2)/ST21 },
 {gBL,                    Sqrt[2] g1 g2 /Sqrt[5 g2^2 -3 g1^2]},
 {Yv,                      LHInput[Yv]}
};

BoundaryConditionsDown[[1]] = {
 {vR,           vRinput}, 
 {vBL,          vBLinput},
 {a,            aInput},
 {Theta1,       ArcTan[RealPart[((vR*AlphaOm[1,2])/2 + Mu3[1,2])/Mu3[2,2]]]},
 {Theta2,       ArcTan[RealPart[(-(vR*AlphaOm[1,2])/2 + Mu3[1,2])/Mu3[2,2]]]},
 {g1,           Sqrt[5] g2 gBL/Sqrt[2 g2^2 + 3 gBL^2]},
 {MassB,        (2 g2^2 MassB + 3 gBL^2 MassWL)/(2 g2^2 + 3 gBL^2)},
 {MassWB,       MassWL},
 {Yd[index1,index2], YQ[index1,index2,1] CT1 - YQ[index1,index2,2] ST1},
 {Yu[index1,index2], - YQ[index1,index2,1] CT2 + YQ[index1,index2,2] ST2},
 {Ye[index1,index2], YL[index1,index2,1] CT1 - YL[index1,index2,2] ST1},
 {Yv[index1,index2], - YL[index1,index2,1] CT2 + YL[index1,index2,2] ST2},
 {T[Yd][index1,index2], T[YQ][index1,index2,1] CT1 -T[YQ][index1,index2,2]ST1},
 {T[Yu][index1,index2], -T[YQ][index1,index2,1] CT2+T[YQ][index1,index2,2] ST2},
 {T[Ye][index1,index2], T[YL][index1,index2,1] CT1 -T[YL][index1,index2,2] ST1},
 {mu2,          mqR2},
 {md2,          mqR2},
 {mq2,          mqR2},
 {me2,          mlR2},
 {ml2,          mlR2},
 {mHd2, CT1^2 mPhi2[1,1] + ST1^2 mPhi2[2,2] - ST1 CT1(mPhi2[1,2] + mPhi2[2,1])},
 {mHu2, CT2^2 mPhi2[1,1] + ST2^2 mPhi2[2,2] - ST2 CT2(mPhi2[1,2] + mPhi2[2,1])},
 {WOp,          MatMul2[MatMul2[Yv,InverseMatrix[f],FortranFalse],Transpose[Yv],
                   FortranFalse]/vR}
 };
\end{lstlisting}
Note that \(Y_\nu\) does not appear in any of the model files and it is
therefore
necessary to fix the dimension of that matrix by hand
\begin{lstlisting}
AdditionalVariablesSPheno={Yv[3,3]};
\end{lstlisting}
Several parameters are restricted to be real. In addition, it is
helpful to choose initialization values for some parameters to stabilize
the numerics in the first iteration
\begin{lstlisting}
RealParameters = {TanBeta, vRinput,vBLinput,Theta1,Theta2,TScale};

InitializationValues = {
 {Mu3IN[1,1], (Mu3IN[1,2]^2 - AlphaOmIN[1,2]^2 vRInput^2/4)/Mu3IN[2,2]},
 {Theta1,   ArcTan[RealPart[-(Mu3IN[1,2]+AlphaOmIN[1,2]vRInput/2)/Mu3IN[2,2]]]},
 {Theta2,   ArcTan[RealPart[(Mu3IN[1,2]-AlphaOmIN[1,2]vRInput/2)/Mu3IN[2,2]]]},
 {Mdelta,   aInput*SignumMdelta*vRinput/2 },
 {Momega,   SignumMomega*(aInput^2*vBLinput^2)/(8 Mdelta)}
}; 
\end{lstlisting}

 \section{Example for using \SSP in the MSSM: $(m_0$,$M_{1/2})$ grid scan}
\label{sec:Examples_SSP}
In this section we give an example on how to generate scans and plots within in the MSSM.
The example shown here is a short version of the file {\tt Example1.m} included
in \SSP. The goal will be varying $m_0$ and $M_{1/2}$ between 0
and 1000~GeV and thus investigating how different masses depend on these two
parameters.

First, it is necessary to define the location of all tools\footnote{If the
environment is set up using our provided script (c.f. \ref{sec:UsingTheScripts}),
the necessary file is created automatically in the process.}
in a file called 
{\tt DefaultSettings.m}. The content of {\tt DefaultSettings.m} should look
like
\begin{lstlisting} 
DEFAULT[SPheno] = "[Directory]/SPheno3.1.4/bin/SPhenoMSSM";
DEFAULT[SPhenoInputFile] = "LesHouches.in.MSSM";
DEFAULT[SPhenoSpectrumFile] = "SPheno.spc.MSSM";

DEFAULT[MicroOmegas] ="[Directory]/micromegas_2.4.1/MSSM/CalcOmega";
DEFAULT[MicroOmegasInputFile] = "SPheno.spc.MSSM";
DEFAULT[MicroOmegasOutputFile] = "omg.out";

DEFAULT[WHIZARD] = "[Directory]/whizard/bin/whizard";
DEFAULT[WHIZARDparFile] = "WHIZARD.par.MSSM";

DEFAULT[HiggsBounds] = "[Directory]/HiggsBounds/HiggsBounds LandH effC 3 1";
\end{lstlisting}
{\tt [Directory]} stands for the installation directory of the framework
presented here. In principle, for this example it would have been sufficient to
set only the information for \SPheno.

The properties of the desired scan have to be defined in a second
input file located in {\tt SSP/Input} which we will call
{\tt Exampple\_MSSM.m}.
The first thing to into the file is a list with identifiers for the
different scans, providing the possibility to perform several scans in a
row using only one input file. Here, we will restrict ourself to
just one scan
\begin{lstlisting} 
RunScans = {ScanM0M12};
\end{lstlisting}
In order to actually perform the scans,
the necessary information for creating the LesHouches input has to be
provided, which comprises all appearing block names as well as numerical
values for the parameters.
\begin{lstlisting} 
DEFINITION[ScanM0M12][Blocks]={MODSEL,SMINPUTS,MINPAR,SPhenoInput};

DEFINITION[ScanM0M12][MODSEL]={
  {{1},{Value->1}},
  {{6},{Value->1}}
};

DEFINITION[ScanM0M12][SMINPUTS]={
  {{2},{Value->1.166390*10^-5}},
  {{3},{Value->0.1172}},
  {{4},{Value->91.18760}},
  {{5},{Value->4.2}},
  {{6},{Value->172.9}},
  {{7},{Value->1.777}}
};

DEFINITION[ScanM0M12][MINPAR]={
  {{1}, {Min->0,Max->1000, Steps->10,Distribution->LINEAR}},
  {{2}, {Min->0,Max->1000, Steps->10,Distribution->LINEAR}},
  {{3}, {Value->10}},
  {{4}, {Value->1}},
  {{5}, {Value->0}}
  };

DEFINITION[ScanM0M12][SPhenoInput]={
  {{1},{Value->-1}},
  {{2},{Value->1}},
  {{11},{Value->0}},
  {{12},{Value->12}},
  {{21},{Value->0}}
};
\end{lstlisting} 
Observe that all parameters are assigned constant value, except \(m_0\) and
\(M_{1/2}\) which are to be linearly varied.

Finally, we define the plot variables. In this example, we choose
 \begin{itemize}
   \item All three neutral Higgs masses for different $m_0$ as functions of $M_{1/2}$
   \item All six selectron masses for different $m_0$ as functions of $M_{1/2}$
   \item All four neutralino masses for different $M_{1/2}$ as functions of $m_0$
   \item All three neutral Higgs masses for different $M_{1/2}$ as functions of $m_0$
 \item Contour plot of the light Higgs mass in the $(m_0,M_{1/2})$-plane
 \item Contour plot of the heavy Higgs mass in the $(m_0,M_{1/2})$-plane
\end{itemize}
\begin{figure}[p]
\begin{minipage}{\textwidth}
\includegraphics[width=0.45\textwidth]{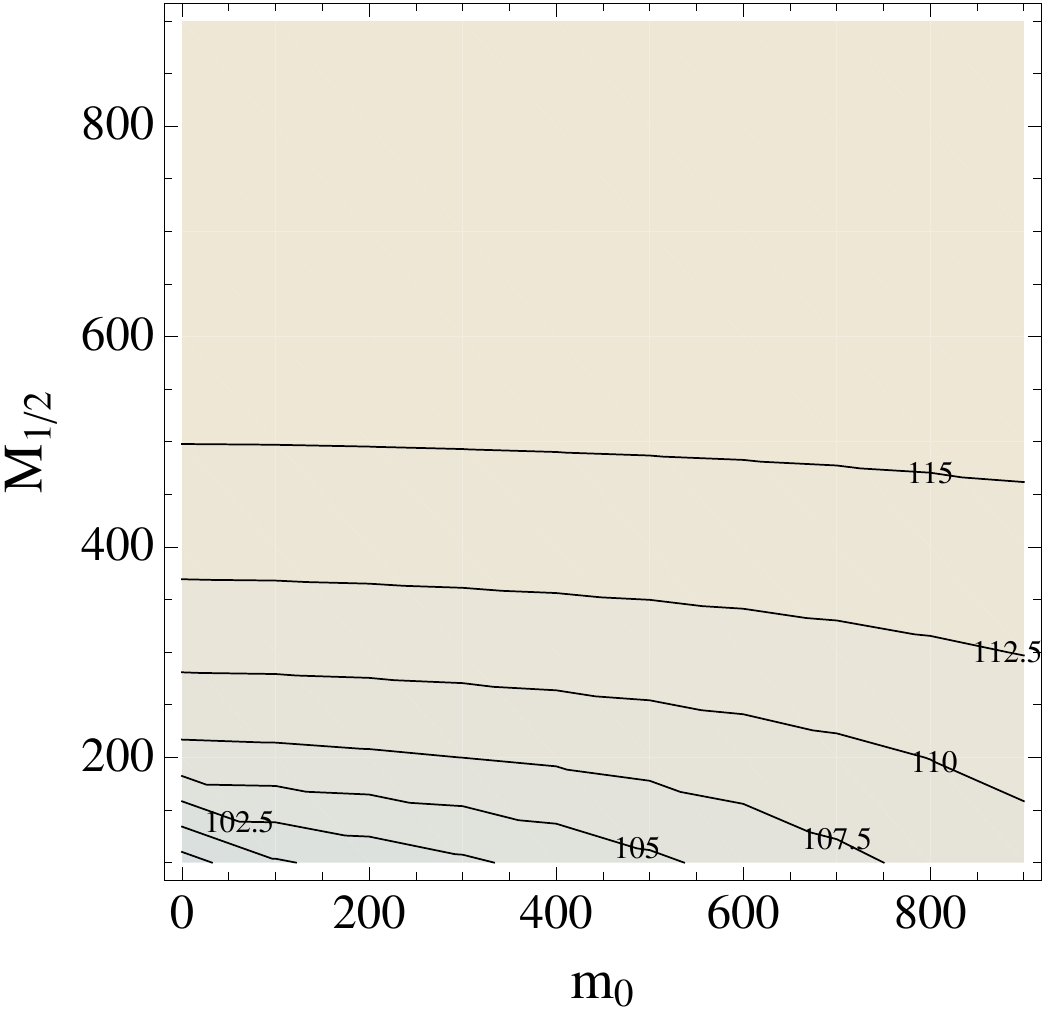} \hfill 
\includegraphics[width=0.45\textwidth]{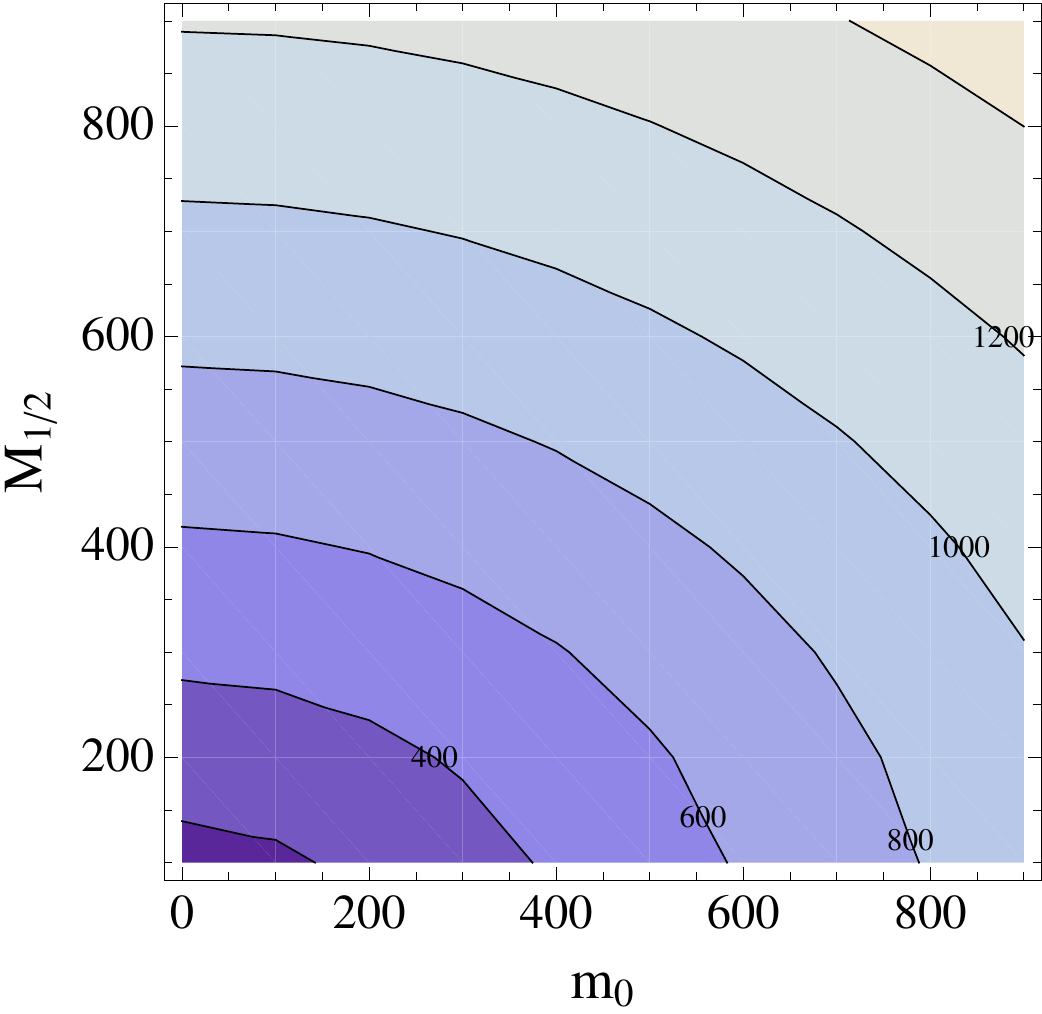} \\
\\
\includegraphics[width=0.45\textwidth]{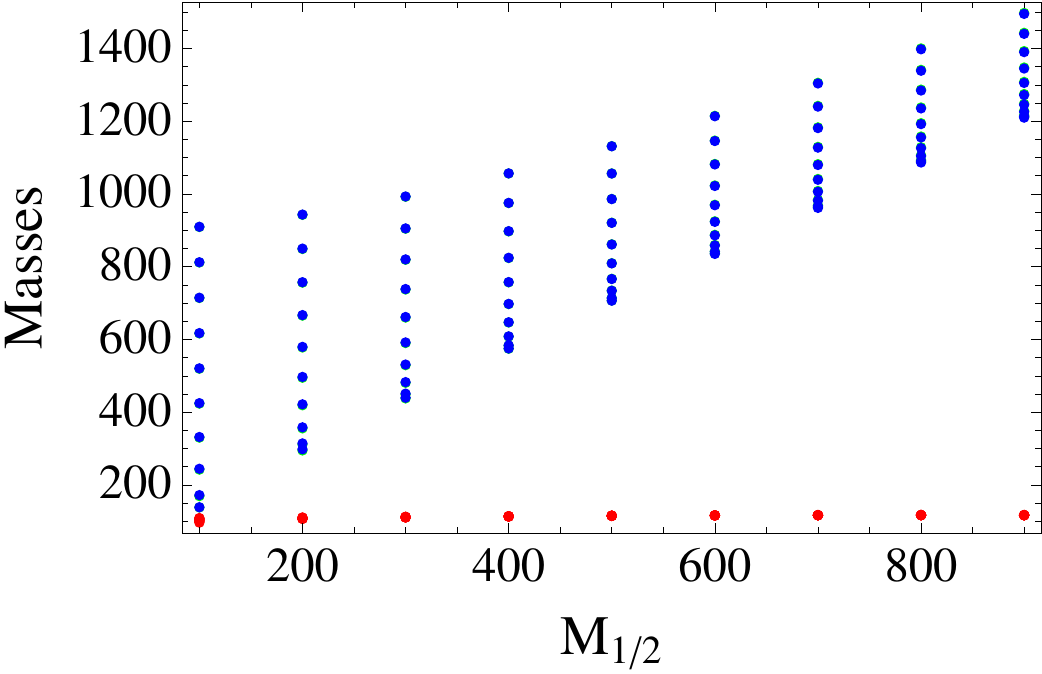} \hfill
\includegraphics[width=0.45\textwidth]{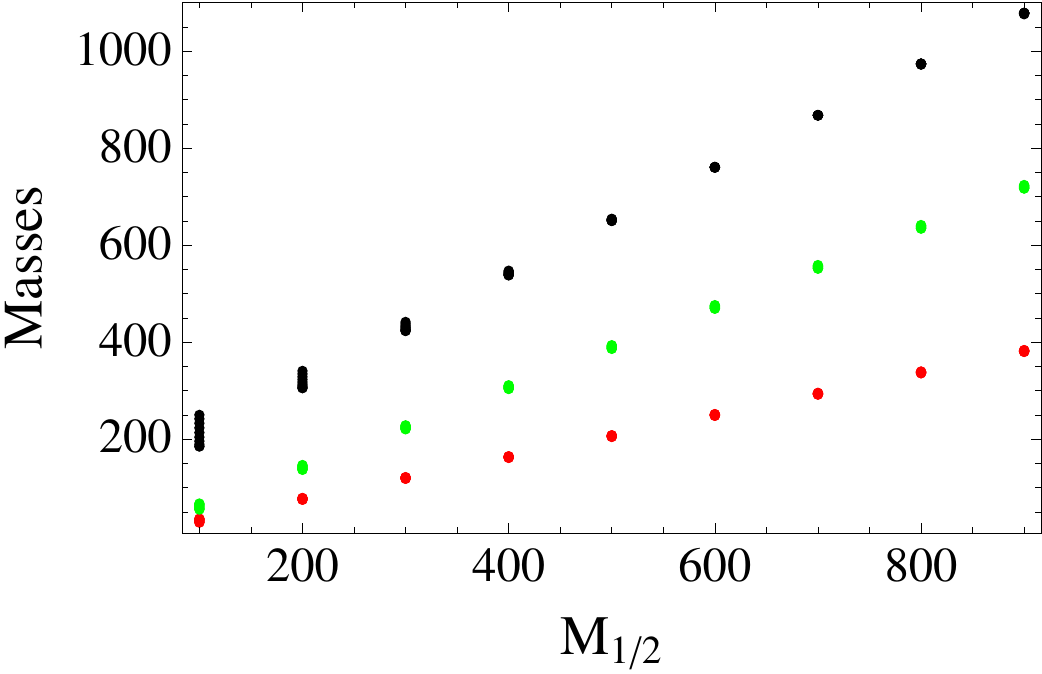} \\
\\
\includegraphics[width=0.45\textwidth]{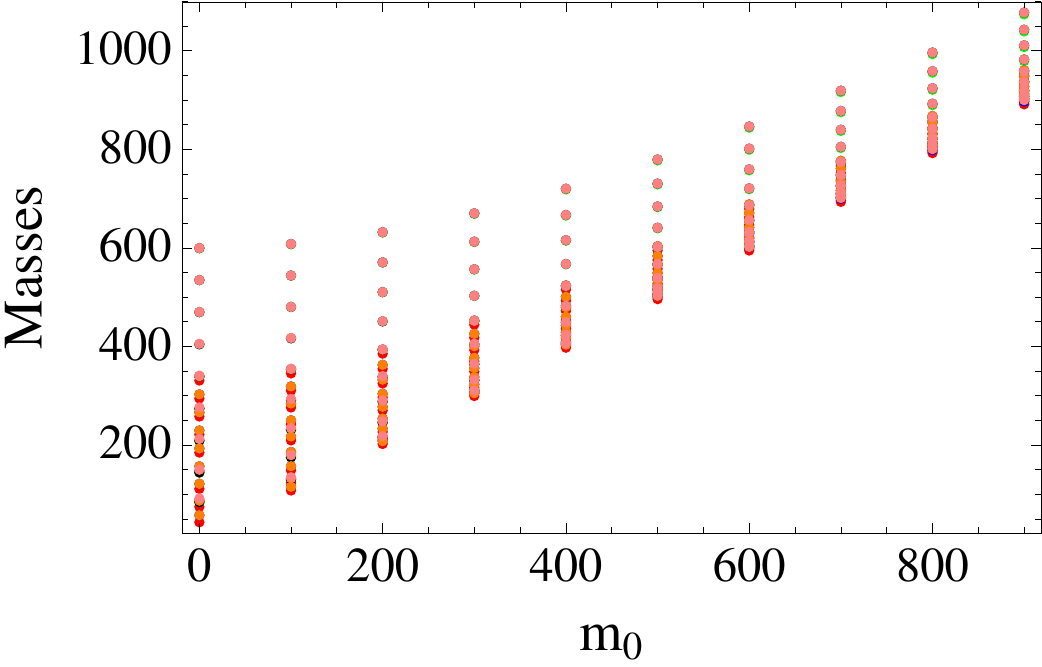} \hfill
\includegraphics[width=0.45\textwidth]{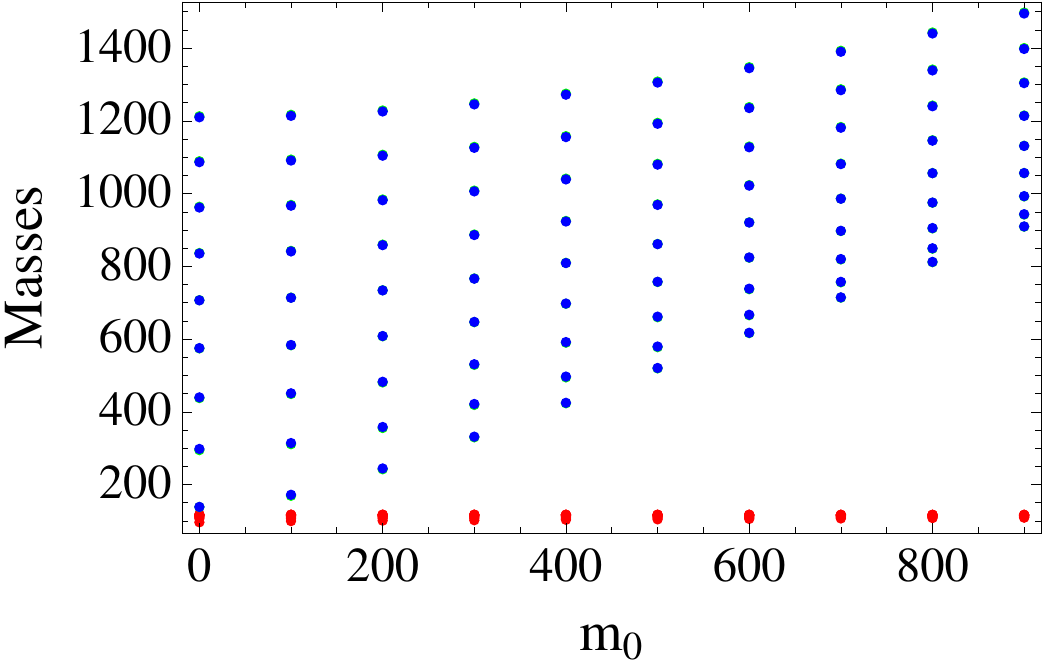} \\
\end{minipage}
\caption{From top to bottom, left to right: mass of the light Higgs in the
 $(m_0,M_{1/2})$-plane, mass of the heavy Higgs in the $(m_0,M_{1/2})$-plane,
masses of the scalar Higgs fields as functions of $M_{1/2}$ for different values of
$m_0$, masses of the neutralinos as functions of $M_{1/2}$ for different values of
$m_0$,  masses of the heavy Higgs in the $(m_0,M_{1/2})$ plane, masses of the scalar
Higgs fields as functions of $m_0$ for different values of $M_{1/2}$, masses of the
charged sleptons as functions of $m_0$ for different values of $M_{1/2}$.}
\label{fig:example_SSP}
\end{figure}
\begin{lstlisting} 
DEFINITION[ScanM0M12][Plots]={
  {P2D, {MINPAR[1],{MASS[25],MASS[35],MASS[36]}},
            Style1,"m0_Higgs_different_m12.eps"},
  {P2D, {MINPAR[1],{MASS[1000011],MASS[2000011],MASS[1000013],
            MASS[2000013],MASS[1000015],MASS[2000015]}},
            Style2, "m0_Selectrons_different_m12.eps"},
   {P2D, {MINPAR[2],{MASS[25],MASS[35],MASS[36]}},
            Style3,"m12_Higgs_different_m0.eps"},
  {P2D, {MINPAR[2],{MASS[1000022],MASS[1000023],
            MASS[1000034],MASS[1000035]}},
            Style4, "m12_Neutralinos_different_m0.eps"},
  {P3D, {MINPAR[1],MINPAR[2],MASS[25]},Style5,"m0_m12_Mass25.eps"},
  {P3D, {MINPAR[1],MINPAR[2],MASS[35]},Style5,"m0_m12_Mass35.eps"}
};
\end{lstlisting} 
Finally, we polish the plots by adjusting {\Mathematica}'s plot options
\begin{lstlisting} 
Style5= {Frame->True,Axes->False,
      FrameLabel->{Style[Subscript["m","0"],16],Style[Subscript["M","1/2"],16]},
      FrameTicksStyle -> Directive[Black, 14],ContourLabels->True};
\end{lstlisting}
For a detailed description of those options, we refer to the manual of {\tt
Mathematica}. The final plots are shown in Fig.~\ref{fig:example_SSP}.

\clearpage

 \section{Validation of \SPheno output}
\subsection{Mass spectrum of the MSSM}
\begin{figure}[p]
\begin{minipage}{\linewidth}
\includegraphics[width=0.45\linewidth]{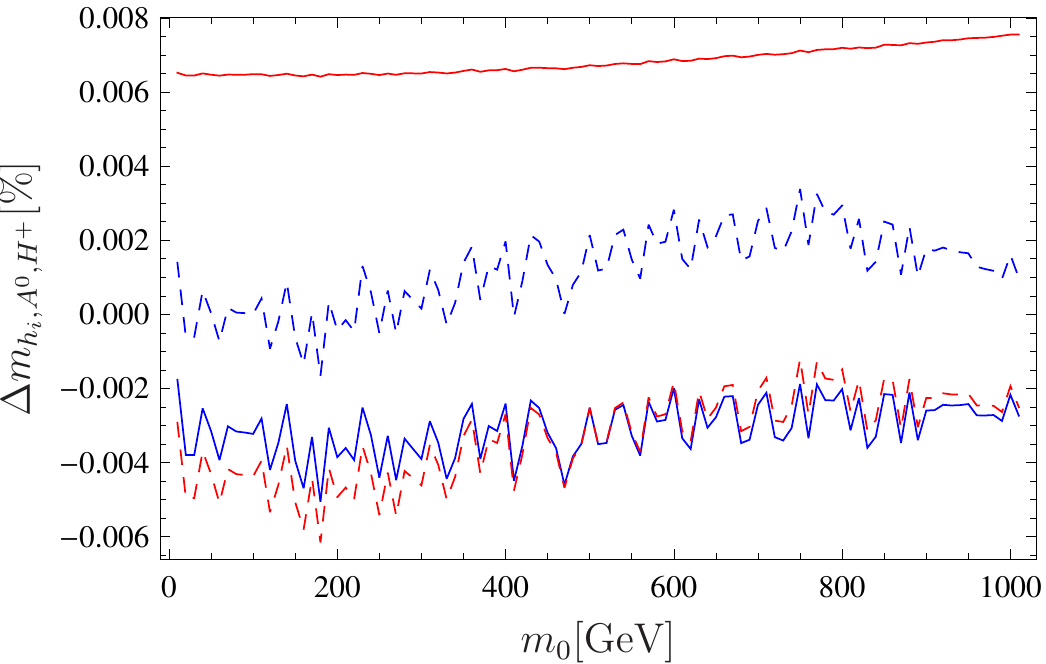} 
\hfill
\includegraphics[width=0.45\linewidth]{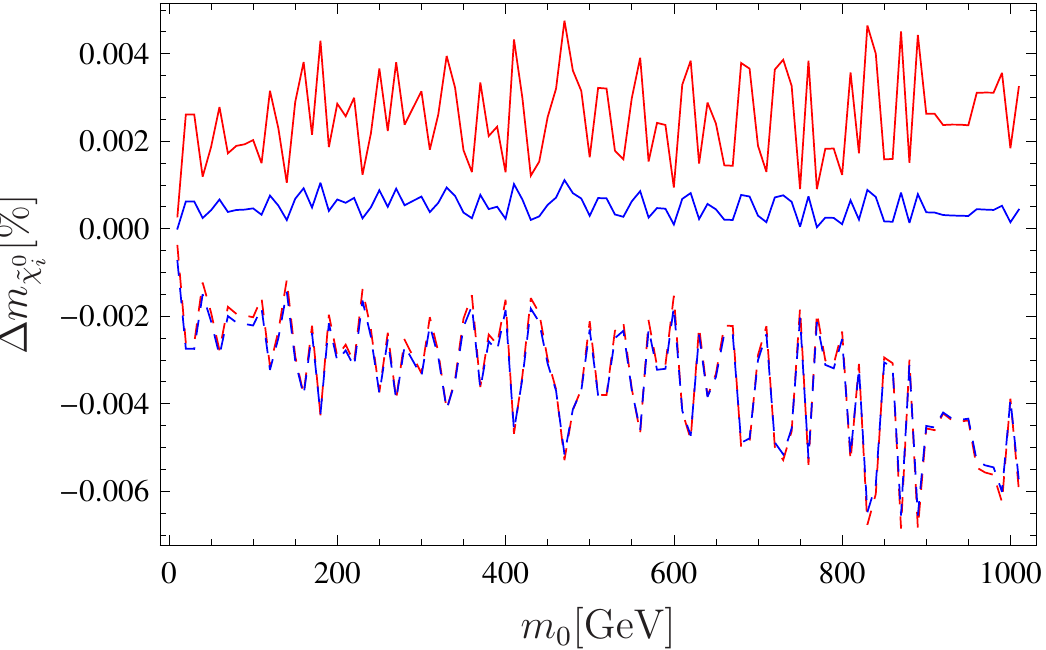} \\
\end{minipage}
\caption{Comparison of the masses of the Higgs fields and neutralinos
between \SPheno ``stock'' and \SPheno/\SARAH. The input values are chosen as $M_{1/2} =
500$~GeV, $A_0=0$~GeV, $\tan\beta=10$,
$\mbox{sign}\mu=1$. \SPheno ``stock'' is based on  \SPheno 3.1.4 with small
changes as explained in the text. 
}
\label{fig:SPhenoSARAH1}
\end{figure}

\begin{figure}[p]
\begin{minipage}{\linewidth}
\includegraphics[width=0.45\linewidth]{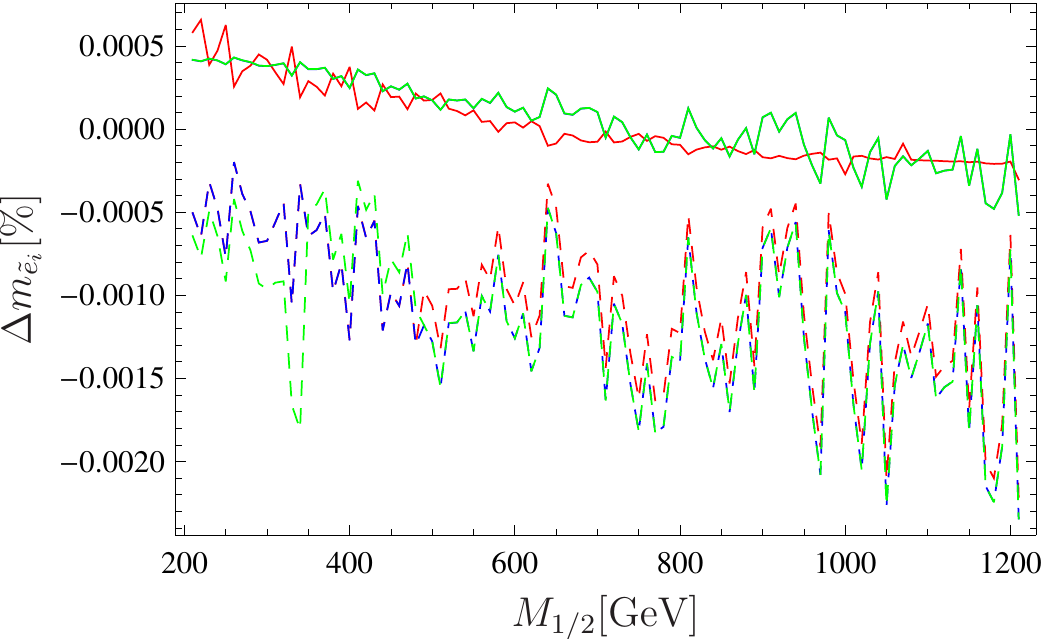} 
\hfill
\includegraphics[width=0.45\linewidth]{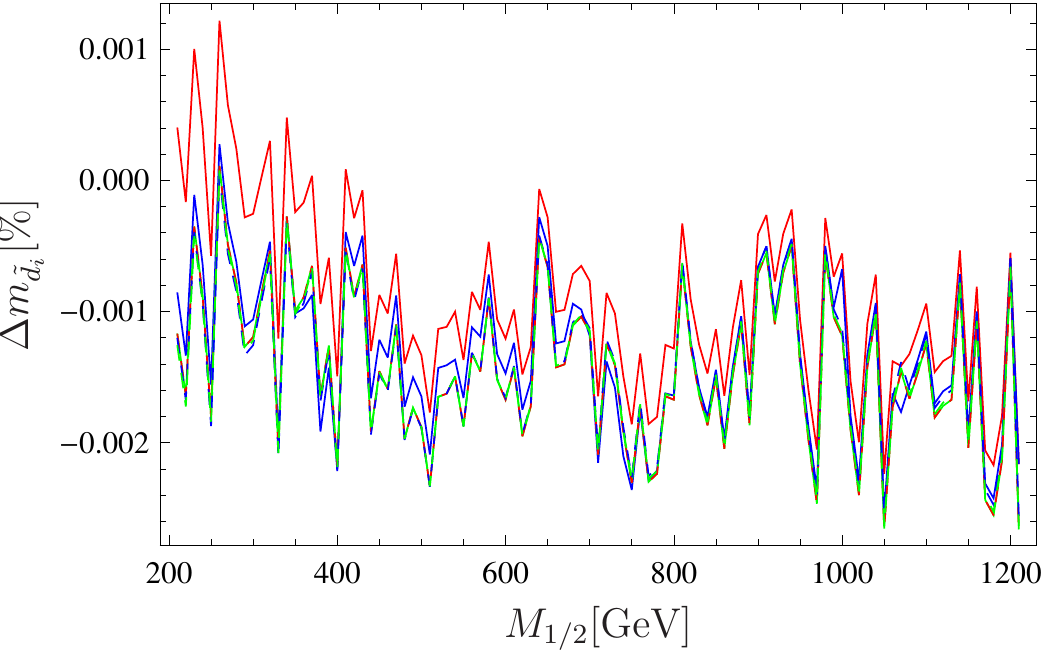} \\
\end{minipage}
\caption{Comparison of the masses of all six charged sleptons and down
squarks. Input values: $M_0 = 250$~GeV, $A_0=-300$~GeV,
$\tan\beta=20$, $\mbox{sign}\mu=1$.}
\label{fig:SPhenoSARAH2}
\end{figure}

In order to demonstrate the agreement between the ``stock'' \SPheno MSSM and the \SPheno
version generated by \SARAH, the relative mass difference 
\begin{equation}
\Delta m = \frac{m_{\rm \SPheno} - m_{\rm \SPheno-\SARAH}}{m_{\rm \SPheno}} 
\end{equation}
for two different points as functions of \(m_0\) resp. \(M_{1/2}\) is shown in
Fig.~\ref{fig:SPhenoSARAH1} and Fig.~\ref{fig:SPhenoSARAH2}

All comparisons were done using \SPheno\ 3.1.4\ .
For obtaining comparable results, the ``stock'' \SPheno was slightly modified
in order to compensate for the  systematic differences shown in
Tab.~\ref{tab:SPheno_SARAH}. Specifically, we restricted the calculation of the
mass spectrum
to two-loop RGEs and one-loop  self-energies, thus
neglecting the two-loop mass corrections in the Higgs sector normally
included by \SPheno. In addition we switched off the loop induced decays of
neutralinos and gluinos. We used a relative precision of \(O(10^{-4})\) to calculate
the spectrum with both \SPheno versions.

\subsection{Decay widths and branching ratios in the MSSM}
\begin{figure}[p]
\begin{minipage}{\linewidth}
\includegraphics[width=0.45\linewidth]{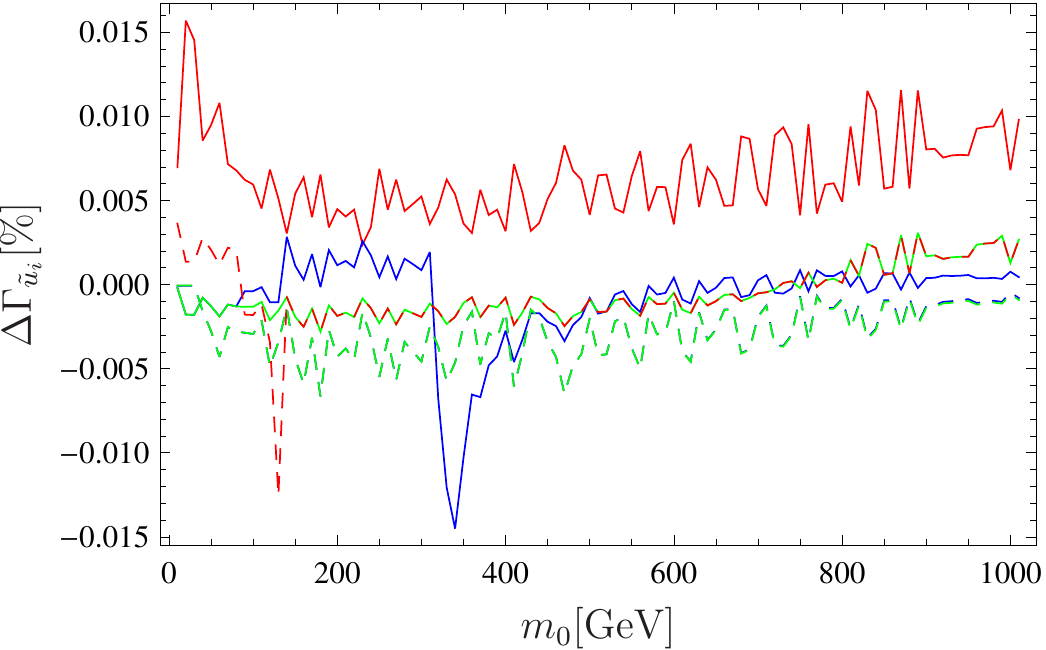} 
\hfill
\includegraphics[width=0.45\linewidth]{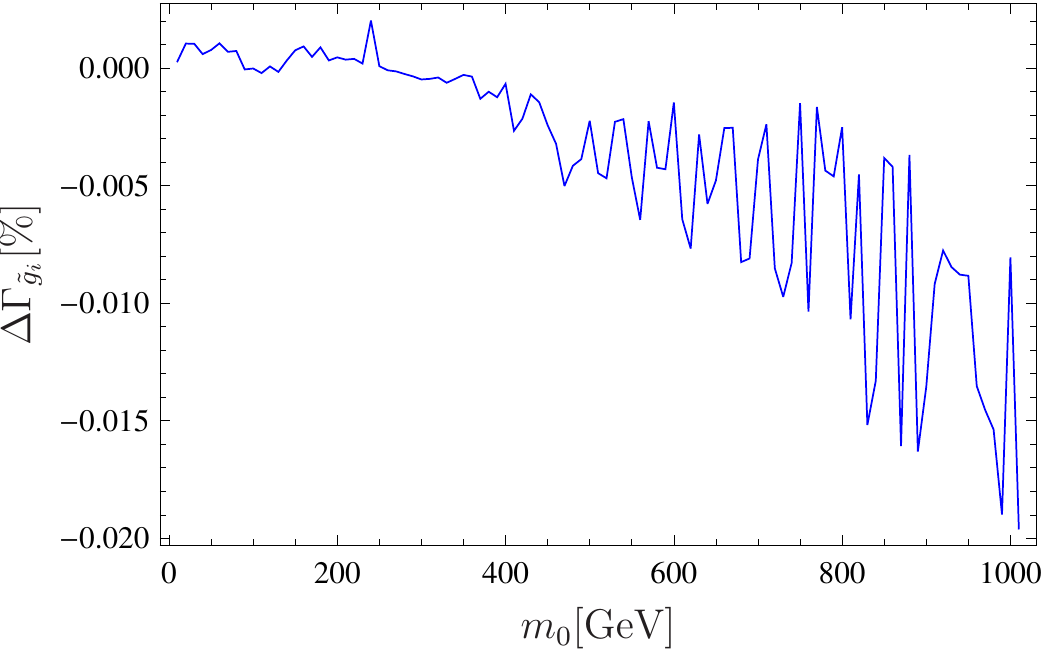} \\
\hfill
\end{minipage}
 \caption{The dependence of the total decay widths of the up-squarks
and the gluino on \(m_0\). Input values: $M_{1/2} = 500$~GeV, $A_0=0$~GeV, $\tan\beta=10$,
$\mbox{sign}\;\mu=1$.
As explained in the text, loop induced decays 
$\tilde{g} \rightarrow \Neu g$ were disabled in \SPheno ``stock''. }
 \label{fig:Decay_SPS4}
\end{figure}

As discussed in Sec.\ref{sec:SPheno}, \SARAH generates code
for \SPheno to handle all two-body decays as well as the fermionic
three-body decays. Shown in is
Tab.~\ref{tab:HiggsDecay} is a comparison of the Higgs decays in the MSSM between
\SPheno and \SPheno-\SARAH, including all branching ratios which are larger than
\(10^{-4}\). A comparison of the neutralino decays into two- and three-body final
states is given in Tab.~\ref{tab:NeutralinoDecay}. Furthermore, in
Fig.~\ref{fig:Decay_SPS4} we show the relative differences between the total
decay widths in \SPheno ``stock'' vs. \SPheno/\SARAH as functions of $m_0$
for a scenario based on SPS 4.

\begin{table}[p]
\centering
\begin{tabular}{|l | c c|c c|c c|}
\hline 
 & \multicolumn{2}{|c|}{\(h_1\)} & \multicolumn{2}{|c|}{\(h_2\)}  &
\multicolumn{2}{|c|}{\(A^0\)} \\ 
\hline
\(\Gamma\)~[GeV] & \(2.21\cdot 10^{-3}\)   & \(2.21\cdot 10^{-3}\)     &  
\(7.30 \cdot 10^{-1}\)    & \(7.30 \cdot 10^{-1}\)    & 1.16    & 1.16     \\
\hline
\multicolumn{7}{|c|}{Branching Ratios} \\
\hline
\(s \bar{s}\)                          & 0.0003 & 0.0003 & 0.0003         &
0.0003         & 0.0002         & 0.0002 \\
\(b \bar{b}\)                          & 0.7763 & 0.7760 & 0.6402         &
0.6403         & 0.4034         & 0.4032 \\
\(\mu \bar{\mu}\)                      & 0.0005 & 0.0005 & 0.0004         &
0.0004         & 0.0002         & 0.0002 \\
\(\tau \bar{\tau}\)                    & 0.1287 & 0.1288 & 0.1051         &
0.1051         & 0.0662         & 0.0662 \\
\(c \bar{c}\)                          & 0.0380 & 0.0379 & \(< 10^{-4}\)  & \(<
10^{-4}\)  & \(< 10^{-4}\)  &  \(< 10^{-4}\) \\
\(t \bar{t}\)                          & -      & -      & 0.0667         &
0.0668         & 0.1224         & 0.1223 \\
\(\Cha_1 \Chap_1\)                     & -      & -      & 0.0400         &
0.0400         & 0.2010         & 0.2010 \\
\(\Neu_1 \Neu_1\)                      & -      & -      & 0.0234         &
0.0234         & 0.0230         & 0.0230 \\
\(\Neu_1 \Neu_2\)                      & -      & -      & 0.0643         &
0.0643         & 0.0962         & 0.0962 \\
\(\Neu_2 \Neu_2\)                      & -      & -      & 0.0145         &
0.0145         & 0.0758         & 0.0758 \\
\(h_1 h_1\)                            & -      & -      & 0.0143         &
0.0143         & -              &  - \\
\(h_1 Z\)                              & -      & -      & -              & -   
          & 0.0021         & 0.0021 \\
\(\tilde{e}_1^* \tilde{e}_1\)          & -      & -      & 0.0062         &
0.0062         & 0.0049         & 0.0048 \\
\(\tilde{e}_1^* \tilde{e}_6\)          & -      & -      & 0.0052         &
0.0052         & \(< 10^{-4}\)  &  \(< 10^{-4}\)  \\
\(\tilde{e}_2^* \tilde{e}_2\)          & -      & -      & 0.0006         &
0.0006         & \(< 10^{-4}\)  &  \(< 10^{-4}\) \\
\(\tilde{e}_3^* \tilde{e}_3\)          & -      & -      & 0.0006         &
0.0006         & \(< 10^{-4}\)  &  \(< 10^{-4}\) \\
\(\tilde{e}_6^* \tilde{e}_1\)        & -      & -      & 0.0052         &
0.0052         & 0.0049         & 0.0048 \\
\(\tilde{\nu}_1^* \tilde{\nu}_1\)      & -      & -      & 0.0012         &
0.0012         & \(< 10^{-4}\)  &  \(< 10^{-4}\) \\
\(\tilde{\nu}_2^* \tilde{\nu}_2\)      & -      & -      & 0.0011         &
0.0011         & \(< 10^{-4}\)  &  \(< 10^{-4}\) \\
\(\tilde{\nu}_3^* \tilde{\nu}_3\)      & -      & -      & 0.0011         &
0.0011         & \(< 10^{-4}\)  &  \(< 10^{-4}\) \\
\(\gamma \gamma\)                      & 0.0025 & 0.0025 & \(< 10^{-4}\)  & \(<
10^{-4}\)  & \(< 10^{-4}\)  & \(< 10^{-4}\) \\
\(g g\)                                & 0.0310 & 0.0309 & 0.0007         &
0.0007         & \(< 10^{-4}\)  & \(< 10^{-4}\) \\
\(Z Z^*\)                              & 0.0011 & 0.0010 & -              & -   
          & -              & - \\
\(Z Z\)                                & -      & -      & 0.0029         &
0.0029         & -              & - \\
\(W^- (W^+)^*\)                        & 0.0110 & 0.0110 & -              & -   
          & -              & - \\
\((W^-)^* W^+\)                        & 0.0110 & 0.0110 & -              & -   
          & -              & - \\
\(W^- W^+\)                            & -      & -      & 0.0060         &
0.0060         & -              & - \\ 
\hline
\end{tabular}
\caption{Left column: \SPheno/\SARAH, right column: \SPheno ``stock''. Input
values:
$m_0=100$~GeV,$M_{1/2}=250$~GeV, $A_0=-100$~GeV, $\tan\beta$=10,
$\mbox{sign}\mu$=1. }
\label{tab:HiggsDecay}
\end{table}

\begin{table}[p]
\centering
\begin{tabular}{|l | c c|c c|c c|}
\hline 
 & \multicolumn{2}{|c|}{\(\Neu_2\)} & \multicolumn{2}{|c|}{\(\Neu_3\)}  &
\multicolumn{2}{|c|}{\(\Neu_4\)} \\ 
\hline
\(\Gamma\)~[GeV] & \(6.57\cdot 10^{-6}\)   & \(6.57\cdot 10^{-6}\)     &   1.39 
  & 1.39    & 1.60    & 1.60     \\
\hline
\multicolumn{7}{|c|}{Branching Ratios} \\
\hline
\(\Neu_1 d \bar{d}\)                   & 0.1742 & 0.1742      &\(< 10^{-4}\)&\(<
10^{-4}\)&\(< 10^{-4}\)&\(< 10^{-4}\) \\
\(\Neu_1 s \bar{s}\)                   & 0.1742 & 0.1742      &\(< 10^{-4}\)&\(<
10^{-4}\)&\(< 10^{-4}\)&\(< 10^{-4}\) \\
\(\Neu_1 b \bar{b}\)                   & 0.1762 & 0.1763      &\(< 10^{-4}\)&\(<
10^{-4}\)&\(< 10^{-4}\)&\(< 10^{-4}\) \\
\(\Neu_1 e \bar{e}\)                   & 0.0227 & 0.0227      &\(< 10^{-4}\)&\(<
10^{-4}\)&\(< 10^{-4}\)&\(< 10^{-4}\) \\
\(\Neu_1 \mu \bar{\mu}\)               & 0.0227 & 0.0227      &\(< 10^{-4}\)&\(<
10^{-4}\)&\(< 10^{-4}\)&\(< 10^{-4}\) \\
\(\Neu_1 \tau \bar{\tau}\)             & 0.0230 & 0.0231      &\(< 10^{-4}\)&\(<
10^{-4}\)&\(< 10^{-4}\)&\(< 10^{-4}\) \\
\(\Neu_1 u \bar{u}\)                   & 0.1293 & 0.1293      &\(< 10^{-4}\)&\(<
10^{-4}\)&\(< 10^{-4}\)&\(< 10^{-4}\) \\
\(\Neu_1 c \bar{c}\)                   & 0.1289 & 0.1287      &\(< 10^{-4}\)&\(<
10^{-4}\)&\(< 10^{-4}\)&\(< 10^{-4}\) \\
\(\Neu_1 \nu_i \bar{\nu}_i\)           & 0.1487 & 0.1486      &\(< 10^{-4}\)&\(<
10^{-4}\)&\(< 10^{-4}\)&\(< 10^{-4}\) \\
\(\Cha_1 W^+ \)                        & -      & -           & 0.3137      &
0.3137      & 0.3519      & 0.3519 \\
\(\Chap_1 W^-\)                        & -      & -           & 0.3137      &
0.3137      & 0.3519      & 0.3519 \\
\(\Neu_1 h_1\)                         & -      & -           & 0.0269      &
0.0260      & 0.0904      & 0.0904 \\
\(\Neu_2 h_1\)                         & -      & -           & 0.1228      &
0.1228      & 0.1467      & 0.1455 \\
\(\Neu_1 Z\)                           & -      & -           & 0.1359      &
0.1359      & 0.0327      & 0.0327  \\
\(\Neu_2 Z\)                           & -      & -           & 0.1985      &
0.1985      & 0.0264      & 0.0264 \\
\hline
\end{tabular}
\caption{Left column: \SPheno/\SARAH, right column: \SPheno ``stock''. Input
values:
$m_0=550$~GeV,$M_{1/2}=200$~GeV, $A_0=-100$~GeV, $\tan\beta$=10,
$\mbox{sign}\mu$=1.}
\label{tab:NeutralinoDecay}
\end{table}

\subsection{Masses in Seesaw type I -- III}
In order to demonstrate that the automatic inclusion of thresholds in \SARAH
works and that the 1-loop boundary conditions are calculated correctly,
we show a comparison for the seesaw type I--III between \SPheno ``stock''
and \SPheno/\SARAH in Tab.\ref{tab:seesawSphenoVsSarah}.
The model and {\tt SPheno.m} files for this scenario are shown in
appendix~\ref{app:Seesaw}.

\begin{table}[p]
 \begin{tabular}{|c||c|c||c|c||c|c|}
\hline
 & \multicolumn{2}{|c||}{Seesaw I} & \multicolumn{2}{|c||}{Seesaw II} &
\multicolumn{2}{|c|}{Seesaw III} \\
\hline 
 particle & \SPheno & \SARAH & \SPheno & \SARAH & \SPheno & \SARAH \\  
 \hline
\(M_{\rm GUT}\)   & \(2.36 \cdot 10^{16}\)  &\(2.36 \cdot 10^{16}\) &  \(2.75
\cdot 10^{16}\) & \(2.75 \cdot 10^{16}\) & \(3.61 \cdot 10^{16}\) & \(3.64 \cdot
10^{16}\)  \\
\hline
\(\Neu_1\)        &  97.59 &  97.61&  76.07 &  76.17&  47.04  &  46.92  \\
\(\Neu_2\)        & 180.26 & 180.30& 139.44 & 139.51&  84.42  &  84.21  \\
\(\Neu_3\)        & 345.69 & 345.93& 294.96 & 294.47& 209.78  & 209.56  \\
\(\Neu_4\)        & 365.60 & 365.81& 315.62 & 315.22& 234.36  & 234.39  \\
\(\Cha_1\)        & 179.74 & 179.78& 138.59 & 138.65&  81.91  &  81.70  \\
\(\Cha_2\)        & 366.57 & 366.78& 317.00 & 316.60& 236.90  & 236.69  \\
\hline
\(\tilde{g}\)     & 621.94 & 622.17& 509.89 & 509.78& 366.28  & 365.81  \\
\hline
\(h_1\)           & 105.73 & 105.73& 104.11 & 104.11& 101.41  & 101.40  \\
\(h_2\)           & 443.91 & 444.07& 384.29 & 385.31& 320.91  & 320.74  \\
\(A^0\)           & 443.64 & 443.81& 384.03 & 385.31& 320.64  & 320.46  \\
\(H^+\)           & 451.26 & 451.41& 392.65 & 393.65& 330.69  & 330.56  \\
\hline 
\(\tilde{e}_1\)   & 263.89 & 263.85& 258.62 & 258.77& 253.34  & 253.38  \\
\(\tilde{e}_2\)   & 270.25 & 270.25& 264.94 & 265.09& 259.33  & 259.31 \\
\(\tilde{e}_3\)   & 270.25 & 270.25& 264.97 & 265.11& 259.33  & 259.31  \\
\(\tilde{e}_4\)   & 304.59 & 304.58& 288.40 & 288.46& 270.48  & 270.42  \\
\(\tilde{e}_5\)   & 304.60 & 304.59& 288.41 & 288.47& 270.48  & 270.43  \\
\(\tilde{e}_6\)   & 306.38 & 306.22& 290.41 & 290.56& 272.41  & 272.36  \\
\hline
\(\tilde{d}_1\)   & 551.93 & 552.14& 466.24 & 466.39& 357.84  & 357.51  \\
\(\tilde{d}_2\)   & 592.29 & 592.50& 505.04 & 505.11& 396.81  & 396.48 \\
\(\tilde{d}_3\)   & 593.97 & 594.18& 506.27 & 506.35& 398.12  & 397.78 \\
\(\tilde{d}_4\)   & 593.98 & 594.18& 506.27 & 506.35& 398.12  & 307.79  \\
\(\tilde{d}_5\)   & 614.46 & 614.82& 522.97 & 523.06& 409.38  & 409.03  \\
\(\tilde{d}_6\)   & 614.46 & 614.82& 522.97 & 523.07& 409.38  & 409.04  \\
\hline
\(\tilde{u}_1\)   & 439.39 & 439.62& 367.16 & 367.10& 277.73  & 277.47  \\
\(\tilde{u}_2\)   & 594.19 & 594.40& 505.87 & 505.93& 396.56  & 396.22  \\
\(\tilde{u}_3\)   & 594.20 & 594.41& 505.88 & 505.94& 396.56  & 396.22  \\
\(\tilde{u}_4\)   & 609.64 & 609.84& 517.06 & 517.16& 401.71  & 401.35  \\
\(\tilde{u}_5\)   & 609.64 & 609.84& 517.07 & 517.16& 401.72  & 401.36  \\
\(\tilde{u}_6\)   & 612.32 & 612.56& 530.93 & 530.92& 426.07  & 425.71  \\
\hline
\(\tilde{\nu}_1\) & 292.48 & 292.29& 275.68 & 275.74& 256.86  & 256.81  \\
\(\tilde{\nu}_2\) & 293.79 & 293.78& 276.94 & 277.00& 258.14  & 258.09\\
\(\tilde{\nu}_3\) & 293.79 & 293.78& 276.94 & 277.00& 258.15  & 258.09  \\
\hline
 \end{tabular}
\caption{Comparison for Seesaw type~I--III. Input values: $m_0=250$~GeV,$M_{1/2}=250$~GeV,
$A_0=-0$~GeV, $\tan\beta$=10, $\mbox{sign}\mu$=1 and a (degenerated) Seesaw
scale of $5\cdot 10^{14}$~GeV was used.}
\label{tab:seesawSphenoVsSarah}
\end{table}

\subsection{Masses in the NMSSM}
We have compared the results of a \SPheno version for the NMSSM created by
\SARAH with {\tt NMSSM-Tools} and found good agreement. For
detailed discussion including numbers, see \cite{Staub:2010ty}.

\clearpage

 \section{Validation of the \WHIZARD model files}
We compared the \WHIZARD model generated by \SARAH from {\tt MSSM-NoFV}
to the MSSM implementation included in \WHIZARD. The result is shown in 
Tab.\ref{tab:WHIZARD_check} for a selection of $2\rightarrow2$ processes
and shows complete agreement between the two models.

For the comparison, the input parameters were chosen as defined by the
following \SINDARIN code: 
\begin{lstlisting}
mh     = 1.07507347E+02       # light Higgs mass   
wh     = 3.228E-3             # light Higgs width     
mHH    = 4.22642020E+02       # heavy Higgs mass   
mHA    = 4.22511683E+02       # axial Higgs mass
mHpm   = 4.30427526E+02       # charged Higgs mass            

al_h   = -1.10771279E-01      # Higgs mixing angle alpha
mu_h   = 3.94354160E+02       # Higgs mu parameter
tanb_h = 10                   # Higgs mixing angle tan(beta)

msu1   = 5.64102822E+02       # u-squark mass
msd1   = 5.69542907E+02	      # d-squark mass
msc1   = 5.64114921E+02	      # c-squark mass
mss1   = 5.69542932E+02	      # s-squark mass
mstop1 = 3.65489513E+02       # t-squark mass
msb1   = 5.05397426E+02	      # b-squark mass
msu2   = 5.46608780E+02	      # u-squark mass
msd2   = 5.46345296E+02	      # d-squark mass
msc2   = 5.46592401E+02	      # c-squark mass
mss2   = 5.46340940E+02	      # s-squark mass
mstop2 = 5.84545812E+02       # t-squark mass
msb2   = 5.45062834E+02	      # b-squark mass
mse1   = 1.89600608E+02	      # selectron1 mass
msne   = 1.72254490E+02	      # electron-sneutrino mass
msmu1  = 1.89622816E+02	      # smuon1 mass
msnmu  = 1.72247165E+02	      # muon-sneutrino mass
mstau1 = 1.08017023E+02	      # stau1 mass    
msntau = 1.70180182E+02	      # tau-sneutrino mass
mse2   = 1.25413287E+02       # selectron2 mass
msmu2  = 1.25349593E+02	      # smuon2 mass
mstau2 = 1.94644276E+02       # stau2 mass   
mgg    = 6.14828189E+02	      # gluino mass
mch1   = 1.83712010E+02	      # chargino1 mass (signed)
mch2   = 4.13907598E+02	      # chargino2 mass (signed)
mneu1  = 9.79820056E+01	      # neutralino1 mass (signed)
mneu2  = 1.83937651E+02	      # neutralino2 mass (signed)
mneu3  = -3.98749572E+02      # neutralino3 mass (signed)
mneu4  = 4.12379874E+02	      # neutralino4 mass (signed)

mt_11  =  5.64067874E-01      # Re[R_st(1,1)]
mt_12  =  8.25728426E-01      # Re[R_st(1,2)]
mt_21  = -8.25728426E-01      # Re[R_st(2,1)]
mt_22  =  5.64067874E-01      # Re[R_st(2,2)]

mb_11  =  9.39288883E-01      # Re[R_sb(1,1)]
mb_12  =  3.43127372E-01      # Re[R_sb(1,2)]
mb_21  = -3.43127372E-01      # Re[R_sb(2,1)]
mb_22  =  9.39288883E-01      # Re[R_sb(2,2)]

ml_11 =  3.14734480E-01       # Re[R_sta(1,1)]
ml_12 =  9.49179755E-01       # Re[R_sta(1,2)]
ml_21 = -9.49179755E-01       # Re[R_sta(2,1)]
ml_22 =  3.14734480E-01       # Re[R_sta(2,2)]

mn_11 =  -9.86903446E-01      # Re[N(1,1)]
mn_12 =   5.57184333E-02      # Re[N(1,2)]
mn_13 =  -1.42488513E-01      # Re[N(1,3)]
mn_14 =   5.11279651E-02      # Re[N(1,4)]
mn_21 =  -9.81519379E-02      # Re[N(2,1)]
mn_22 =  -9.50857577E-01      # Re[N(2,2)]
mn_23 =   2.57046449E-01      # Re[N(2,3)]
mn_24 =  -1.41997141E-01      # Re[N(2,4)]
mn_31 =  -5.87383665E-02      # Re[N(3,1)]
mn_32 =   8.82056923E-02      # Re[N(3,2)]
mn_33 =   6.96112966E-01      # Re[N(3,3)]
mn_34 =   7.10067813E-01      # Re[N(3,4)]
mn_41 =  -1.13743529E-01      # Re[N(4,1)]
mn_42 =   2.91522006E-01      # Re[N(4,2)]
mn_43 =   6.55019760E-01      # Re[N(4,3)]
mn_44 =  -6.87769179E-01      # Re[N(4,4)]

mu_11  = -9.34453143E-01      # Re[U(1,1)]
mu_12  =  3.56086119E-01      # Re[U(1,2)]
mu_21  =  3.56086119E-01      # Re[U(2,1)]
mu_22  =  9.34453143E-01      # Re[U(2,2)]

mv_11  = -9.81098181E-01      # Re[V(1,1)]
mv_12  =  1.93510619E-01      # Re[V(1,2)]
mv_21  =  1.93510619E-01      # Re[V(2,1)]
mv_22  =  9.81098181E-01      # Re[V(2,2)]
\end{lstlisting}%
The intergration was performed for a CMS energy of $2\;\text{GeV}$, with cuts
and integration parameters chosen as exemplified by the following sample input
(with the obvious replacement of \texttt{I1}/\texttt{2} and
\texttt{F1}/\texttt{F2} by the actual particle identifiers):
\begin{lstlisting}
model = MSSM
include("par_MSSM.txt")
process scatter = I1,I2 => F1,F2
 
compile 
sqrts = 2000 GeV 
cuts = all E >= 10 GeV [F1] and 
all E >= 10 GeV [F2] and 
all 0.1<Theta<2 [F1] and 
all 0.1<Theta<2 [F2] 

integrate(scatter) {iterations  = 10:5000,10:25000,10:25000:""}  
\end{lstlisting}
\newpage
{
\setlength{\LTcapwidth}{0.98\textwidth}
\begin{longtable}{|l|cc|cc|c|} 
\hline 
Process & \(\sigma_S\) [fb] & \(\delta_S\) [fb] & \(\sigma_W\) [fb] &
\(\delta_W\) [fb] & \(\Delta\) [fb] \\\hline\endfirsthead
\multicolumn{6}{|c|}{Tab.\ref{tab:WHIZARD_check}: continued}\\\hline
Process & \(\sigma_S\) [fb] & \(\delta_S\) [fb] & \(\sigma_W\) [fb] &
\(\delta_W\) [fb] & \(\Delta\) [fb] \\\hline\endhead
\hline\multicolumn{6}{|c|}{(continued on next page)}\endfoot
\endlastfoot
\(\tilde{\chi}^0_1 \tilde{\chi}^0_1 \rightarrow Z Z\) & \(1.227\times
10^{\text{-1}}\) & \(9.87\times 10^{\text{-6}}\) & \(1.227\times
10^{\text{-1}}\) & \(1.35\times 10^{\text{-5}}\) & \(3.28\times 10^{\text{-5}}\)
\\ 
\(\tilde{\chi}^0_1 \tilde{\chi}^0_1 \rightarrow \tilde{\chi}^0_1
\tilde{\chi}^0_1\) & \(2.538\times 10^{\text{-2}}\) & \(1.34\times
10^{\text{-6}}\) & \(2.538\times 10^{\text{-2}}\) & \(3.65\times
10^{\text{-6}}\) & \(-2.99\times 10^{\text{-6}}\) \\ 
\(\tilde{\chi}^0_1 \tilde{\chi}^0_1 \rightarrow \tilde{\chi}^0_2
\tilde{\chi}^0_2\) & \(2.219\times 10^{\text{-1}}\) & \(1.14\times
10^{\text{-5}}\) & \(2.219\times 10^{\text{-1}}\) & \(1.03\times
10^{\text{-5}}\) & \(3.4\times 10^{\text{-6}}\) \\ 
\(\tilde{\chi}^0_1 \tilde{\chi}^0_1 \rightarrow \tilde{\chi}^0_3
\tilde{\chi}^0_3\) & \(1.762\) & \(1.24\times 10^{\text{-4}}\) & \(1.762\) &
\(1.24\times 10^{\text{-4}}\) & \(-2.7\times 10^{\text{-4}}\) \\ 
\(\tilde{\chi}^0_1 \tilde{\chi}^0_1 \rightarrow \tilde{\chi}^0_4
\tilde{\chi}^0_4\) & \(1.105\) & \(8.23\times 10^{\text{-5}}\) & \(1.104\) &
\(6.36\times 10^{\text{-5}}\) & \(1.37\times 10^{\text{-4}}\) \\ 
\(\tilde{\chi}^0_1 \tilde{\chi}^0_1 \rightarrow h A^0\) & \(3.913\times
10^{\text{-1}}\) & \(3.46\times 10^{\text{-5}}\) & \(3.913\times
10^{\text{-1}}\) & \(2.36\times 10^{\text{-5}}\) & \(1.43\times 10^{\text{-5}}\)
\\ 
\(\tilde{\chi}^0_1 \tilde{\chi}^0_1 \rightarrow H A^0\) & \(1.843\) &
\(2.10\times 10^{\text{-4}}\) & \(1.841\) & \(1.68\times 10^{\text{-4}}\) &
\(1.38\times 10^{\text{-3}}\) \\ 
\(\tilde{\chi}^0_1 \tilde{\chi}^0_1 \rightarrow \tilde{\chi}^+_1
\tilde{\chi}^-_1\) & \(1.069\times 10^{\text{-1}}\) & \(6.69\times
10^{\text{-6}}\) & \(1.068\times 10^{\text{-1}}\) & \(1.10\times
10^{\text{-5}}\) & \(4.44\times 10^{\text{-5}}\) \\ 
\(\tilde{\chi}^0_1 \tilde{\chi}^0_1 \rightarrow \tilde{\chi}^+_2
\tilde{\chi}^-_2\) & \(1.542\) & \(1.69\times 10^{\text{-4}}\) & \(1.541\) &
\(1.34\times 10^{\text{-4}}\) & \(1.19\times 10^{\text{-3}}\) \\ 
\(\tilde{\chi}^0_1 \tilde{\chi}^0_1 \rightarrow \tilde{\chi}^0_1
\tilde{\chi}^0_2\) & \(1.504\times 10^{\text{-1}}\) & \(1.32\times
10^{\text{-5}}\) & \(1.504\times 10^{\text{-1}}\) & \(2.04\times
10^{\text{-5}}\) & \(-6.14\times 10^{\text{-5}}\) \\ 
\(\tilde{\chi}^0_1 \tilde{\chi}^0_1 \rightarrow W^- W^+\) & \(1.57\) &
\(1.73\times 10^{\text{-4}}\) & \(1.57\) & \(1.55\times 10^{\text{-4}}\) &
\(-5.73\times 10^{\text{-5}}\) \\ 
\(\tilde{\chi}^0_1 \tilde{\chi}^0_1 \rightarrow d \bar{d}\) & \(6.836\times
10^{\text{-1}}\) & \(8.44\times 10^{\text{-5}}\) & \(6.838\times
10^{\text{-1}}\) & \(5.21\times 10^{\text{-5}}\) & \(-2.12\times
10^{\text{-4}}\) \\ 
\(\tilde{\chi}^0_1 \tilde{\chi}^0_1 \rightarrow t \bar{t}\) & \(1.234\times
10^1\) & \(1.18\times 10^{\text{-3}}\) & \(1.234\times 10^1\) & \(8.07\times
10^{\text{-4}}\) & \(6.67\times 10^{\text{-4}}\) \\ 
\(\tilde{\chi}^0_1 \tilde{\chi}^0_1 \rightarrow b \bar{b}\) & \(7.878\times
10^{\text{-1}}\) & \(9.42\times 10^{\text{-5}}\) & \(7.879\times
10^{\text{-1}}\) & \(8.46\times 10^{\text{-5}}\) & \(-3.97\times
10^{\text{-5}}\) \\ 
\(\tilde{\chi}^0_1 \tilde{\chi}^0_1 \rightarrow \tilde{t}_1 \tilde{t}_1^*\) &
\(9.561\) & \(1.02\times 10^{\text{-3}}\) & \(9.563\) & \(8.33\times
10^{\text{-4}}\) & \(-2.5\times 10^{\text{-3}}\) \\ 
\(\tilde{\chi}^0_1 \tilde{\chi}^0_1 \rightarrow \tilde{t}_2 \tilde{t}_2^*\) &
\(5.421\times 10^{\text{-1}}\) & \(3.56\times 10^{\text{-5}}\) & \(5.421\times
10^{\text{-1}}\) & \(4.61\times 10^{\text{-5}}\) & \(6.33\times 10^{\text{-5}}\)
\\ 
\(\tilde{\chi}^0_1 \tilde{\chi}^0_1 \rightarrow \tilde{t}_1 \tilde{t}_2^*\) &
\(2.096\) & \(1.88\times 10^{\text{-4}}\) & \(2.098\) & \(2.22\times
10^{\text{-4}}\) & \(-1.92\times 10^{\text{-3}}\) \\ 
\(\tilde{\chi}^0_1 \tilde{\chi}^0_1 \rightarrow \tilde{t}_2 \tilde{t}_1^*\) &
\(2.097\) & \(1.70\times 10^{\text{-4}}\) & \(2.097\) & \(1.63\times
10^{\text{-4}}\) & \(-3.18\times 10^{\text{-5}}\) \\ 
\(\tilde{\chi}^0_1 \tilde{\chi}^0_1 \rightarrow \tilde{d}_1 \tilde{d}_1^*\) &
\(1.844\times 10^{\text{-1}}\) & \(1.20\times 10^{\text{-5}}\) & \(1.843\times
10^{\text{-1}}\) & \(1.58\times 10^{\text{-5}}\) & \(6.5\times 10^{\text{-5}}\)
\\ 
\(\tilde{\chi}^0_1 \tilde{\chi}^0_1 \rightarrow \tilde{d}_2 \tilde{d}_2^*\) &
\(7.987\times 10^{\text{-1}}\) & \(7.22\times 10^{\text{-5}}\) & \(7.986\times
10^{\text{-1}}\) & \(7.23\times 10^{\text{-5}}\) & \(1.01\times 10^{\text{-4}}\)
\\ 
\(\tilde{\chi}^0_1 \tilde{\chi}^0_1 \rightarrow \tilde{b}_1 \tilde{b}_2^*\) &
\(2.059\times 10^{\text{-1}}\) & \(1.54\times 10^{\text{-5}}\) & \(2.058\times
10^{\text{-1}}\) & \(2.20\times 10^{\text{-5}}\) & \(3.32\times 10^{\text{-6}}\)
\\ 
\(\tilde{\chi}^0_1 \tilde{\chi}^0_1 \rightarrow \tilde{b}_2 \tilde{b}_1^*\) &
\(2.057\times 10^{\text{-1}}\) & \(2.08\times 10^{\text{-5}}\) & \(2.06\times
10^{\text{-1}}\) & \(2.82\times 10^{\text{-5}}\) & \(-2.11\times
10^{\text{-4}}\) \\ 
\(\tilde{\chi}^0_1 \tilde{\chi}^0_1 \rightarrow \nu_e \bar{\nu}_e\) & \(2.209\)
& \(2.56\times 10^{\text{-4}}\) & \(2.21\) & \(1.65\times 10^{\text{-4}}\) &
\(-1.26\times 10^{\text{-3}}\) \\ 
\(\tilde{\chi}^0_1 \tilde{\chi}^0_1 \rightarrow \nu_\mu \bar{\nu}_\mu\) &
\(2.211\) & \(1.59\times 10^{\text{-4}}\) & \(2.21\) & \(1.36\times
10^{\text{-4}}\) & \(6.83\times 10^{\text{-4}}\) \\ 
\(\tilde{\chi}^0_1 \tilde{\chi}^0_1 \rightarrow \tilde{e}_1 \tilde{e}_1^*\) &
\(1.074\) & \(7.01\times 10^{\text{-5}}\) & \(1.073\) & \(7.70\times
10^{\text{-5}}\) & \(2.87\times 10^{\text{-4}}\) \\ 
\(\tilde{\chi}^0_1 \tilde{\chi}^0_1 \rightarrow \tilde{e}_2 \tilde{e}_2^*\) &
\(2.605\times 10^1\) & \(2.79\times 10^{\text{-3}}\) & \(2.604\times 10^1\) &
\(1.59\times 10^{\text{-3}}\) & \(9.78\times 10^{\text{-3}}\) \\ 
\(\tilde{\chi}^0_1 \tilde{\chi}^0_1 \rightarrow \tilde{\tau}_1
\tilde{\tau}_1^*\) & \(2.148\times 10^1\) & \(1.47\times 10^{\text{-3}}\) &
\(2.148\times 10^1\) & \(2.17\times 10^{\text{-3}}\) & \(5.8\times
10^{\text{-3}}\) \\ 
\(\tilde{\chi}^0_1 \tilde{\chi}^0_1 \rightarrow \tilde{\tau}_2 \tilde{\tau}_2\)
& \(4.171\times 10^{\text{-1}}\) & \(2.00\times 10^{\text{-4}}\) & \(4.169\times
10^{\text{-1}}\) & \(3.14\times 10^{\text{-5}}\) & \(1.9\times 10^{\text{-4}}\)
\\ 
\(\tilde{\chi}^0_1 \tilde{\chi}^0_1 \rightarrow \tilde{\tau}_1 \tilde{\tau}_2\)
& \(2.64\) & \(2.34\times 10^{\text{-4}}\) & \(2.64\) & \(2.11\times
10^{\text{-4}}\) & \(5.81\times 10^{\text{-4}}\) \\ 
\(\tilde{\chi}^0_1 \tilde{\chi}^0_1 \rightarrow \tilde{\tau}_2
\tilde{\tau}_1^*\) & \(2.639\) & \(1.95\times 10^{\text{-4}}\) & \(2.64\) &
\(2.05\times 10^{\text{-4}}\) & \(-7.78\times 10^{\text{-4}}\) \\ 
\(\tilde{\chi}^0_1 \tilde{\chi}^0_1 \rightarrow H^- H^+\) & \(1.14\) &
\(8.12\times 10^{\text{-5}}\) & \(1.14\) & \(1.00\times 10^{\text{-4}}\) &
\(-4.47\times 10^{\text{-4}}\) \\ 
\(\tilde{\chi}^0_1 \tilde{\chi}^0_1 \rightarrow h h\) & \(7.951\times
10^{\text{-2}}\) & \(8.62\times 10^{\text{-6}}\) & \(7.948\times
10^{\text{-2}}\) & \(7.51\times 10^{\text{-6}}\) & \(2.98\times 10^{\text{-5}}\)
\\ 
\(\tilde{\chi}^0_1 \tilde{\chi}^0_1 \rightarrow H H\) & \(5.380\times
10^{\text{-2}}\) & \(9.95\times 10^{\text{-6}}\) & \(5.380\times
10^{\text{-2}}\) & \(5.76\times 10^{\text{-6}}\) & \(3.25\times 10^{\text{-6}}\)
\\ 
\(\tilde{\chi}^0_1 \tilde{\chi}^0_1 \rightarrow h H\) & \(2.643\times
10^{\text{-1}}\) & \(2.19\times 10^{\text{-5}}\) & \(2.643\times
10^{\text{-1}}\) & \(1.79\times 10^{\text{-5}}\) & \(8.54\times 10^{\text{-5}}\)
\\ 
\(\tilde{\chi}^0_1 \tilde{\chi}^0_1 \rightarrow A^0 A^0\) & \(5.487\times
10^{\text{-2}}\) & \(6.71\times 10^{\text{-6}}\) & \(5.492\times
10^{\text{-2}}\) & \(1.10\times 10^{\text{-5}}\) & \(-5.63\times
10^{\text{-5}}\) \\ 
\(e \bar{e} \rightarrow Z Z\) & \(2.325\) & \(2.42\times 10^{\text{-4}}\) &
\(2.325\) & \(1.82\times 10^{\text{-4}}\) & \(-1.77\times 10^{\text{-4}}\) \\ 
\(e \bar{e} \rightarrow \gamma \gamma\) & \(1.636\times 10^1\) & \(1.52\times
10^{\text{-3}}\) & \(1.637\times 10^1\) & \(2.81\times 10^{\text{-3}}\) &
\(-1.09\times 10^{\text{-2}}\) \\ 
\(e \bar{e} \rightarrow \tilde{e}_1 \tilde{e}_1^*\) & \(5.823\) & \(4.33\times
10^{\text{-4}}\) & \(5.825\) & \(4.47\times 10^{\text{-4}}\) & \(-1.97\times
10^{\text{-3}}\) \\ 
\(e \bar{e} \rightarrow \tilde{e}_2 \tilde{e}_2^*\) & \(4.955\) & \(6.10\times
10^{\text{-4}}\) & \(4.953\) & \(5.79\times 10^{\text{-4}}\) & \(1.74\times
10^{\text{-3}}\) \\ 
\(e \bar{e} \rightarrow \tilde{\chi}^0_1 \tilde{\chi}^0_1\) & \(1.143\times
10^1\) & \(7.15\times 10^{\text{-4}}\) & \(1.143\times 10^1\) & \(1.01\times
10^{\text{-3}}\) & \(-9.57\times 10^{\text{-4}}\) \\ 
\(e \bar{e} \rightarrow \tilde{\chi}^0_2 \tilde{\chi}^0_2\) & \(8.503\) &
\(6.86\times 10^{\text{-4}}\) & \(8.503\) & \(6.09\times 10^{\text{-4}}\) &
\(-9.71\times 10^{\text{-5}}\) \\ 
\(e \bar{e} \rightarrow \tilde{\chi}^0_3 \tilde{\chi}^0_3\) & \(1.488\times
10^{\text{-3}}\) & \(1.37\times 10^{\text{-7}}\) & \(1.488\times
10^{\text{-3}}\) & \(1.01\times 10^{\text{-7}}\) & \(1.28\times 10^{\text{-7}}\)
\\ 
\(e \bar{e} \rightarrow \tilde{\chi}^0_4 \tilde{\chi}^0_4\) & \(4.235\times
10^{\text{-2}}\) & \(2.38\times 10^{\text{-6}}\) & \(4.234\times
10^{\text{-2}}\) & \(2.45\times 10^{\text{-6}}\) & \(8.23\times 10^{\text{-7}}\)
\\ 
\(e \bar{e} \rightarrow h A^0\) & \(1.932\times 10^{\text{-4}}\) & \(1.41\times
10^{\text{-8}}\) & \(1.932\times 10^{\text{-4}}\) & \(3.38\times
10^{\text{-9}}\) & \(1.21\times 10^{\text{-8}}\) \\ 
\(e \bar{e} \rightarrow H A^0\) & \(1.351\) & \(1.06\times 10^{\text{-4}}\) &
\(1.351\) & \(2.44\times 10^{\text{-5}}\) & \(-2.41\times 10^{\text{-4}}\) \\ 
\(e \bar{e} \rightarrow \tilde{\chi}^+_1 \tilde{\chi}^-_1\) & \(1.658\times
10^1\) & \(1.25\times 10^{\text{-3}}\) & \(1.658\times 10^1\) & \(1.12\times
10^{\text{-3}}\) & \(6.14\times 10^{\text{-3}}\) \\ 
\(e \bar{e} \rightarrow \tilde{\chi}^+_2 \tilde{\chi}^-_2\) & \(9.619\) &
\(7.71\times 10^{\text{-4}}\) & \(9.617\) & \(5.87\times 10^{\text{-4}}\) &
\(2.26\times 10^{\text{-3}}\) \\ 
\(e \bar{e} \rightarrow \tilde{\chi}^0_1 \tilde{\chi}^0_2\) & \(4.361\) &
\(5.20\times 10^{\text{-4}}\) & \(4.36\) & \(2.45\times 10^{\text{-4}}\) &
\(1.14\times 10^{\text{-3}}\) \\ 
\(e \bar{e} \rightarrow W^- W^+\) & \(2.656\times 10^1\) & \(5.08\times
10^{\text{-3}}\) & \(2.657\times 10^1\) & \(1.50\times 10^{\text{-3}}\) &
\(-1.21\times 10^{\text{-2}}\) \\ 
\(e \bar{e} \rightarrow d \bar{d}\) & \(7.563\) & \(8.01\times 10^{\text{-4}}\)
& \(7.558\) & \(4.25\times 10^{\text{-4}}\) & \(5.42\times 10^{\text{-3}}\) \\ 
\(e \bar{e} \rightarrow t \bar{t}\) & \(1.428\times 10^1\) & \(1.58\times
10^{\text{-4}}\) & \(1.428\times 10^1\) & \(2.03\times 10^{\text{-3}}\) &
\(-1.05\times 10^{\text{-3}}\) \\ 
\(e \bar{e} \rightarrow b \bar{b}\) & \(7.561\) & \(7.10\times 10^{\text{-4}}\)
& \(7.559\) & \(4.15\times 10^{\text{-4}}\) & \(2.13\times 10^{\text{-3}}\) \\ 
\(e \bar{e} \rightarrow \tilde{t}_1 \tilde{t}_1^*\) & \(3.695\) & \(1.15\times
10^{\text{-3}}\) & \(3.693\) & \(1.48\times 10^{\text{-4}}\) & \(1.92\times
10^{\text{-3}}\) \\ 
\(e \bar{e} \rightarrow \tilde{t}_2 \tilde{t}_2^*\) & \(3.091\) & \(5.84\times
10^{\text{-5}}\) & \(3.087\) & \(3.37\times 10^{\text{-4}}\) & \(3.88\times
10^{\text{-3}}\) \\ 
\(e \bar{e} \rightarrow \tilde{t}_1 \tilde{t}_2^*\) & \(7.906\times
10^{\text{-1}}\) & \(4.15\times 10^{\text{-6}}\) & \(7.906\times
10^{\text{-1}}\) & \(2.13\times 10^{\text{-6}}\) & \(-1.3\times 10^{\text{-5}}\)
\\ 
\(e \bar{e} \rightarrow \tilde{t}_2 \tilde{t}_1^*\) & \(7.907\times
10^{\text{-1}}\) & \(2.29\times 10^{\text{-5}}\) & \(7.907\times
10^{\text{-1}}\) & \(6.09\times 10^{\text{-6}}\) & \(6.73\times 10^{\text{-5}}\)
\\ 
\(e \bar{e} \rightarrow \tilde{d}_1 \tilde{d}_1^*\) & \(3.087\) & \(1.42\times
10^{\text{-5}}\) & \(3.086\) & \(1.09\times 10^{\text{-4}}\) & \(5.13\times
10^{\text{-4}}\) \\ 
\(e \bar{e} \rightarrow \tilde{d}_2 \tilde{d}_2^*\) & \(6.913\times
10^{\text{-1}}\) & \(3.26\times 10^{\text{-6}}\) & \(6.913\times
10^{\text{-1}}\) & \(2.64\times 10^{\text{-6}}\) & \(-6.95\times
10^{\text{-7}}\) \\ 
\(e \bar{e} \rightarrow \tilde{b}_1 \tilde{b}_2^*\) & \(3.482\times
10^{\text{-1}}\) & \(7.06\times 10^{\text{-5}}\) & \(3.484\times
10^{\text{-1}}\) & \(1.21\times 10^{\text{-6}}\) & \(-2.33\times
10^{\text{-4}}\) \\ 
\(e \bar{e} \rightarrow \tilde{b}_2 \tilde{b}_1^*\) & \(3.482\times
10^{\text{-1}}\) & \(5.59\times 10^{\text{-5}}\) & \(3.484\times
10^{\text{-1}}\) & \(9.92\times 10^{\text{-6}}\) & \(-2.07\times
10^{\text{-4}}\) \\ 
\(e \bar{e} \rightarrow \nu_\mu \bar{\nu}_\mu\) & \(2.038\) & \(2.50\times
10^{\text{-5}}\) & \(2.038\) & \(1.49\times 10^{\text{-5}}\) & \(-2.14\times
10^{\text{-5}}\) \\ 
\(e \bar{e} \rightarrow \tilde{\tau}_1 \tilde{\tau}_1^*\) & \(3.382\) &
\(1.47\times 10^{\text{-5}}\) & \(3.383\) & \(1.07\times 10^{\text{-3}}\) &
\(-1.55\times 10^{\text{-3}}\) \\ 
\(e \bar{e} \rightarrow \tilde{\tau}_2 \tilde{\tau}_2\) & \(3.818\) &
\(2.15\times 10^{\text{-4}}\) & \(3.816\) & \(1.73\times 10^{\text{-4}}\) &
\(1.62\times 10^{\text{-3}}\) \\ 
\(e \bar{e} \rightarrow \tilde{\tau}_1 \tilde{\tau}_2\) & \(1.56\times
10^{\text{-1}}\) & \(2.21\times 10^{\text{-6}}\) & \(1.56\times 10^{\text{-1}}\)
& \(3.58\times 10^{\text{-6}}\) & \(3.54\times 10^{\text{-7}}\) \\ 
\(e \bar{e} \rightarrow \tilde{\tau}_2 \tilde{\tau}_1^*\) & \(1.56\times
10^{\text{-1}}\) & \(2.12\times 10^{\text{-6}}\) & \(1.56\times 10^{\text{-1}}\)
& \(1.38\times 10^{\text{-6}}\) & \(-5.22\times 10^{\text{-6}}\) \\ 
\(e \bar{e} \rightarrow \gamma Z\) & \(1.205\times 10^1\) & \(1.72\times
10^{\text{-3}}\) & \(1.204\times 10^1\) & \(8.95\times 10^{\text{-3}}\) &
\(1.28\times 10^{\text{-2}}\) \\ 
\(e \bar{e} \rightarrow H^- H^+\) & \(3.148\) & \(1.85\times 10^{\text{-4}}\) &
\(3.147\) & \(2.94\times 10^{\text{-5}}\) & \(1.63\times 10^{\text{-3}}\) \\ 
\(\tau \bar{\tau} \rightarrow Z Z\) & \(2.325\) & \(2.39\times 10^{\text{-4}}\)
& \(2.325\) & \(1.67\times 10^{\text{-4}}\) & \(-4.3\times 10^{\text{-4}}\) \\ 
\(\tau \bar{\tau} \rightarrow \gamma \gamma\) & \(1.637\times 10^1\) &
\(1.27\times 10^{\text{-3}}\) & \(1.637\times 10^1\) & \(1.26\times
10^{\text{-3}}\) & \(5.26\times 10^{\text{-3}}\) \\ 
\(\tau \bar{\tau} \rightarrow \tilde{\chi}^0_1 \tilde{\chi}^0_1\) &
\(1.149\times 10^1\) & \(1.43\times 10^{\text{-3}}\) & \(1.148\times 10^1\) &
\(7.34\times 10^{\text{-4}}\) & \(4.91\times 10^{\text{-3}}\) \\ 
\(\tau \bar{\tau} \rightarrow \tilde{\chi}^0_2 \tilde{\chi}^0_2\) & \(8.577\) &
\(4.95\times 10^{\text{-4}}\) & \(8.576\) & \(6.90\times 10^{\text{-4}}\) &
\(1.19\times 10^{\text{-3}}\) \\ 
\(\tau \bar{\tau} \rightarrow \tilde{\chi}^0_3 \tilde{\chi}^0_3\) &
\(1.805\times 10^{\text{-2}}\) & \(1.42\times 10^{\text{-6}}\) & \(1.805\times
10^{\text{-2}}\) & \(1.32\times 10^{\text{-6}}\) & \(-1.36\times
10^{\text{-6}}\) \\ 
\(\tau \bar{\tau} \rightarrow \tilde{\chi}^0_4 \tilde{\chi}^0_4\) &
\(8.165\times 10^{\text{-2}}\) & \(6.90\times 10^{\text{-6}}\) & \(8.161\times
10^{\text{-2}}\) & \(7.69\times 10^{\text{-6}}\) & \(3.49\times 10^{\text{-5}}\)
\\ 
\(\tau \bar{\tau} \rightarrow h A^0\) & \(6.969\times 10^{\text{-4}}\) &
\(7.26\times 10^{\text{-8}}\) & \(6.967\times 10^{\text{-4}}\) & \(1.02\times
10^{\text{-7}}\) & \(1.57\times 10^{\text{-7}}\) \\ 
\(\tau \bar{\tau} \rightarrow H A^0\) & \(9.225\times 10^{\text{-1}}\) &
\(5.97\times 10^{\text{-5}}\) & \(9.223\times 10^{\text{-1}}\) & \(9.79\times
10^{\text{-5}}\) & \(1.57\times 10^{\text{-4}}\) \\ 
\(\tau \bar{\tau} \rightarrow \tilde{\chi}^+_1 \tilde{\chi}^-_1\) &
\(1.659\times 10^1\) & \(1.03\times 10^{\text{-3}}\) & \(1.659\times 10^1\) &
\(1.23\times 10^{\text{-3}}\) & \(-2.02\times 10^{\text{-3}}\) \\ 
\(\tau \bar{\tau} \rightarrow \tilde{\chi}^+_2 \tilde{\chi}^-_2\) & \(9.41\) &
\(5.50\times 10^{\text{-4}}\) & \(9.412\) & \(1.01\times 10^{\text{-3}}\) &
\(-1.6\times 10^{\text{-3}}\) \\ 
\(\tau \bar{\tau} \rightarrow \tilde{\chi}^0_1 \tilde{\chi}^0_2\) & \(4.329\) &
\(3.15\times 10^{\text{-4}}\) & \(4.326\) & \(3.45\times 10^{\text{-4}}\) &
\(2.57\times 10^{\text{-3}}\) \\ 
\(\tau \bar{\tau} \rightarrow W^- W^+\) & \(2.66\times 10^1\) & \(1.98\times
10^{\text{-3}}\) & \(2.66\times 10^1\) & \(1.93\times 10^{\text{-3}}\) &
\(-4.75\times 10^{\text{-3}}\) \\ 
\(\tau \bar{\tau} \rightarrow d \bar{d}\) & \(7.56\) & \(7.16\times
10^{\text{-4}}\) & \(7.56\) & \(4.47\times 10^{\text{-4}}\) & \(5.37\times
10^{\text{-4}}\) \\ 
\(\tau \bar{\tau} \rightarrow t \bar{t}\) & \(1.428\times 10^1\) & \(5.20\times
10^{\text{-4}}\) & \(1.428\times 10^1\) & \(1.45\times 10^{\text{-4}}\) &
\(-3.12\times 10^{\text{-3}}\) \\ 
\(\tau \bar{\tau} \rightarrow b \bar{b}\) & \(8.38\) & \(5.27\times
10^{\text{-4}}\) & \(8.377\) & \(4.90\times 10^{\text{-4}}\) & \(2.96\times
10^{\text{-3}}\) \\ 
\(\tau \bar{\tau} \rightarrow \tilde{t}_1 \tilde{t}_1^*\) & \(3.804\) &
\(1.07\times 10^{\text{-4}}\) & \(3.803\) & \(5.25\times 10^{\text{-4}}\) &
\(4.57\times 10^{\text{-4}}\) \\ 
\(\tau \bar{\tau} \rightarrow \tilde{t}_2 \tilde{t}_2^*\) & \(3.187\) &
\(1.26\times 10^{\text{-5}}\) & \(3.185\) & \(7.57\times 10^{\text{-4}}\) &
\(2.04\times 10^{\text{-3}}\) \\ 
\(\tau \bar{\tau} \rightarrow \tilde{t}_1 \tilde{t}_2^*\) & \(9.566\times
10^{\text{-1}}\) & \(1.22\times 10^{\text{-4}}\) & \(9.564\times
10^{\text{-1}}\) & \(8.73\times 10^{\text{-5}}\) & \(1.72\times 10^{\text{-4}}\)
\\ 
\(\tau \bar{\tau} \rightarrow \tilde{t}_2 \tilde{t}_1^*\) & \(9.572\times
10^{\text{-1}}\) & \(1.80\times 10^{\text{-6}}\) & \(9.569\times
10^{\text{-1}}\) & \(4.59\times 10^{\text{-5}}\) & \(2.8\times 10^{\text{-4}}\)
\\ 
\(\tau \bar{\tau} \rightarrow \tilde{d}_1 \tilde{d}_1^*\) & \(3.087\) &
\(7.22\times 10^{\text{-5}}\) & \(3.085\) & \(3.80\times 10^{\text{-5}}\) &
\(1.29\times 10^{\text{-3}}\) \\ 
\(\tau \bar{\tau} \rightarrow \tilde{d}_2 \tilde{d}_2^*\) & \(6.913\times
10^{\text{-1}}\) & \(2.90\times 10^{\text{-5}}\) & \(6.912\times
10^{\text{-1}}\) & \(4.63\times 10^{\text{-6}}\) & \(6.97\times 10^{\text{-5}}\)
\\ 
\(\tau \bar{\tau} \rightarrow \tilde{b}_1 \tilde{b}_2^*\) & \(3.484\times
10^{\text{-1}}\) & \(7.64\times 10^{\text{-7}}\) & \(3.485\times
10^{\text{-1}}\) & \(1.65\times 10^{\text{-5}}\) & \(-6.59\times
10^{\text{-5}}\) \\ 
\(\tau \bar{\tau} \rightarrow \tilde{b}_2 \tilde{b}_1^*\) & \(3.48\times
10^{\text{-1}}\) & \(1.70\times 10^{\text{-4}}\) & \(3.484\times
10^{\text{-1}}\) & \(1.28\times 10^{\text{-6}}\) & \(-4.05\times
10^{\text{-4}}\) \\ 
\(\tau \bar{\tau} \rightarrow \nu_e \bar{\nu}_e\) & \(2.036\) & \(4.42\times
10^{\text{-4}}\) & \(2.038\) & \(8.12\times 10^{\text{-5}}\) & \(-1.22\times
10^{\text{-3}}\) \\ 
\(\tau \bar{\tau} \rightarrow \nu_\mu \bar{\nu}_\mu\) & \(2.038\) & \(7.67\times
10^{\text{-5}}\) & \(2.038\) & \(3.27\times 10^{\text{-5}}\) & \(-2.33\times
10^{\text{-4}}\) \\ 
\(\tau \bar{\tau} \rightarrow \tilde{e}_1 \tilde{e}_1^*\) & \(4.052\) &
\(5.98\times 10^{\text{-5}}\) & \(4.052\) & \(2.06\times 10^{\text{-4}}\) &
\(-1.19\times 10^{\text{-4}}\) \\ 
\(\tau \bar{\tau} \rightarrow \tilde{e}_2 \tilde{e}_2^*\) & \(3.444\) &
\(3.51\times 10^{\text{-4}}\) & \(3.447\) & \(1.13\times 10^{\text{-5}}\) &
\(-2.49\times 10^{\text{-3}}\) \\ 
\(\tau \bar{\tau} \rightarrow \tilde{\tau}_1 \tilde{\tau}_1^*\) & \(3.454\) &
\(1.92\times 10^{\text{-4}}\) & \(3.453\) & \(1.62\times 10^{\text{-4}}\) &
\(1.39\times 10^{\text{-3}}\) \\ 
\(\tau \bar{\tau} \rightarrow \tilde{\tau}_2 \tilde{\tau}_2\) & \(4.131\) &
\(2.21\times 10^{\text{-4}}\) & \(4.131\) & \(1.98\times 10^{\text{-4}}\) &
\(1.63\times 10^{\text{-4}}\) \\ 
\(\tau \bar{\tau} \rightarrow \tilde{\tau}_1 \tilde{\tau}_2\) & \(1.426\) &
\(1.08\times 10^{\text{-4}}\) & \(1.426\) & \(1.04\times 10^{\text{-4}}\) &
\(-3.02\times 10^{\text{-5}}\) \\ 
\(\tau \bar{\tau} \rightarrow \tilde{\tau}_2 \tilde{\tau}_1^*\) & \(1.425\) &
\(9.58\times 10^{\text{-5}}\) & \(1.425\) & \(8.42\times 10^{\text{-5}}\) &
\(2.39\times 10^{\text{-4}}\) \\ 
\(\tau \bar{\tau} \rightarrow \gamma Z\) & \(1.205\times 10^1\) & \(3.18\times
10^{\text{-3}}\) & \(1.206\times 10^1\) & \(2.10\times 10^{\text{-3}}\) &
\(-4.33\times 10^{\text{-3}}\) \\ 
\(\tau \bar{\tau} \rightarrow H^- H^+\) & \(2.964\) & \(2.41\times
10^{\text{-4}}\) & \(2.963\) & \(3.02\times 10^{\text{-4}}\) & \(3.63\times
10^{\text{-4}}\) \\ 
\(\tau \bar{\tau} \rightarrow h h\) & \(3.539\times 10^{\text{-5}}\) &
\(1.86\times 10^{\text{-9}}\) & \(3.540\times 10^{\text{-5}}\) & \(2.75\times
10^{\text{-9}}\) & \(-1.12\times 10^{\text{-8}}\) \\ 
\(\tau \bar{\tau} \rightarrow H H\) & \(1.317\times 10^{\text{-3}}\) &
\(1.32\times 10^{\text{-7}}\) & \(1.316\times 10^{\text{-3}}\) & \(1.56\times
10^{\text{-7}}\) & \(6.19\times 10^{\text{-7}}\) \\ 
\(\tau \bar{\tau} \rightarrow h H\) & \(5.755\times 10^{\text{-4}}\) &
\(5.17\times 10^{\text{-8}}\) & \(5.757\times 10^{\text{-4}}\) & \(6.97\times
10^{\text{-8}}\) & \(-1.80\times 10^{\text{-7}}\) \\ 
\(\tau \bar{\tau} \rightarrow A^0 A^0\) & \(1.256\times 10^{\text{-3}}\) &
\(1.65\times 10^{\text{-7}}\) & \(1.257\times 10^{\text{-3}}\) & \(3.07\times
10^{\text{-7}}\) & \(-5.60\times 10^{\text{-7}}\) \\ 
\(d \bar{d} \rightarrow Z Z\) & \(3.043\) & \(2.57\times 10^{\text{-4}}\) &
\(3.042\) & \(1.50\times 10^{\text{-4}}\) & \(6.63\times 10^{\text{-4}}\) \\ 
\(d \bar{d} \rightarrow \gamma \gamma\) & \(6.740\times 10^{\text{-2}}\) &
\(1.01\times 10^{\text{-5}}\) & \(6.733\times 10^{\text{-2}}\) & \(9.16\times
10^{\text{-6}}\) & \(6.95\times 10^{\text{-5}}\) \\ 
\(d \bar{d} \rightarrow \tilde{\chi}^0_1 \tilde{\chi}^0_1\) & \(3.762\times
10^{\text{-2}}\) & \(2.54\times 10^{\text{-6}}\) & \(3.763\times
10^{\text{-2}}\) & \(3.39\times 10^{\text{-6}}\) & \(-9.14\times
10^{\text{-6}}\) \\ 
\(d \bar{d} \rightarrow \tilde{\chi}^0_2 \tilde{\chi}^0_2\) & \(1.547\) &
\(1.24\times 10^{\text{-4}}\) & \(1.547\) & \(1.91\times 10^{\text{-4}}\) &
\(5.74\times 10^{\text{-4}}\) \\ 
\(d \bar{d} \rightarrow \tilde{\chi}^0_3 \tilde{\chi}^0_3\) & \(9.423\times
10^{\text{-4}}\) & \(7.59\times 10^{\text{-8}}\) & \(9.424\times
10^{\text{-4}}\) & \(6.25\times 10^{\text{-8}}\) & \(-4.87\times
10^{\text{-8}}\) \\ 
\(d \bar{d} \rightarrow \tilde{\chi}^0_4 \tilde{\chi}^0_4\) & \(2.998\times
10^{\text{-2}}\) & \(2.27\times 10^{\text{-6}}\) & \(3.000\times
10^{\text{-2}}\) & \(3.31\times 10^{\text{-6}}\) & \(-1.77\times
10^{\text{-5}}\) \\ 
\(d \bar{d} \rightarrow h A^0\) & \(9.509\times 10^{\text{-5}}\) & \(7.01\times
10^{\text{-9}}\) & \(9.514\times 10^{\text{-5}}\) & \(1.83\times
10^{\text{-8}}\) & \(-4.92\times 10^{\text{-8}}\) \\ 
\(d \bar{d} \rightarrow H A^0\) & \(6.653\times 10^{\text{-1}}\) & \(3.82\times
10^{\text{-5}}\) & \(6.653\times 10^{\text{-1}}\) & \(1.87\times
10^{\text{-6}}\) & \(-3.64\times 10^{\text{-6}}\) \\ 
\(d \bar{d} \rightarrow \tilde{\chi}^+_1 \tilde{\chi}^-_1\) & \(4.149\) &
\(2.53\times 10^{\text{-4}}\) & \(4.148\) & \(2.74\times 10^{\text{-4}}\) &
\(8.91\times 10^{\text{-4}}\) \\ 
\(d \bar{d} \rightarrow \tilde{\chi}^+_2 \tilde{\chi}^-_2\) & \(1.532\) &
\(9.89\times 10^{\text{-5}}\) & \(1.532\) & \(1.02\times 10^{\text{-4}}\) &
\(2.56\times 10^{\text{-4}}\) \\ 
\(d \bar{d} \rightarrow \tilde{\chi}^0_1 \tilde{\chi}^0_2\) & \(1.732\times
10^{\text{-1}}\) & \(1.20\times 10^{\text{-5}}\) & \(1.731\times
10^{\text{-1}}\) & \(1.33\times 10^{\text{-5}}\) & \(3.95\times 10^{\text{-5}}\)
\\ 
\(d \bar{d} \rightarrow W^- W^+\) & \(8.039\) & \(8.13\times 10^{\text{-4}}\) &
\(8.037\) & \(8.19\times 10^{\text{-4}}\) & \(1.94\times 10^{\text{-3}}\) \\ 
\(d \bar{d} \rightarrow d \bar{d}\) & \(5.63\times 10^3\) & \(5.52\times
10^{\text{-1}}\) & \(5.63\times 10^3\) & \(3.47\times 10^{\text{-1}}\) &
\(-1.87\times 10^{\text{-2}}\) \\ 
\(d \bar{d} \rightarrow t \bar{t}\) & \(4.218\times 10^2\) & \(5.85\times
10^{\text{-2}}\) & \(4.222\times 10^2\) & \(5.25\times 10^{\text{-3}}\) &
\(-4.27\times 10^{\text{-1}}\) \\ 
\(d \bar{d} \rightarrow b \bar{b}\) & \(4.179\times 10^2\) & \(3.76\times
10^{\text{-2}}\) & \(4.176\times 10^2\) & \(8.24\times 10^{\text{-2}}\) &
\(3.12\times 10^{\text{-1}}\) \\ 
\(d \bar{d} \rightarrow \tilde{t}_1 \tilde{t}_1^*\) & \(1.493\times 10^2\) &
\(2.27\times 10^{\text{-3}}\) & \(1.493\times 10^2\) & \(6.30\times
10^{\text{-3}}\) & \(2.93\times 10^{\text{-2}}\) \\ 
\(d \bar{d} \rightarrow \tilde{t}_2 \tilde{t}_2^*\) & \(9.922\times 10^1\) &
\(1.34\times 10^{\text{-3}}\) & \(9.922\times 10^1\) & \(4.09\times
10^{\text{-4}}\) & \(-2.94\times 10^{\text{-3}}\) \\ 
\(d \bar{d} \rightarrow \tilde{t}_1 \tilde{t}_2^*\) & \(3.893\times
10^{\text{-1}}\) & \(4.45\times 10^{\text{-6}}\) & \(3.893\times
10^{\text{-1}}\) & \(1.49\times 10^{\text{-4}}\) & \(-3.14\times
10^{\text{-6}}\) \\ 
\(d \bar{d} \rightarrow \tilde{t}_2 \tilde{t}_1^*\) & \(3.893\times
10^{\text{-1}}\) & \(1.91\times 10^{\text{-5}}\) & \(3.892\times
10^{\text{-1}}\) & \(2.07\times 10^{\text{-6}}\) & \(2.72\times 10^{\text{-5}}\)
\\ 
\(d \bar{d} \rightarrow \tilde{d}_1 \tilde{d}_1^*\) & \(3.286\times 10^2\) &
\(1.63\times 10^{\text{-2}}\) & \(3.286\times 10^2\) & \(2.04\times
10^{\text{-2}}\) & \(4.91\times 10^{\text{-2}}\) \\ 
\(d \bar{d} \rightarrow \tilde{d}_2 \tilde{d}_2^*\) & \(3.973\times 10^2\) &
\(2.08\times 10^{\text{-2}}\) & \(3.973\times 10^2\) & \(2.62\times
10^{\text{-2}}\) & \(2.6\times 10^{\text{-2}}\) \\ 
\(d \bar{d} \rightarrow \tilde{b}_1 \tilde{b}_2^*\) & \(1.713\times
10^{\text{-1}}\) & \(1.58\times 10^{\text{-6}}\) & \(1.715\times
10^{\text{-1}}\) & \(1.97\times 10^{\text{-6}}\) & \(-2.02\times
10^{\text{-4}}\) \\ 
\(d \bar{d} \rightarrow \tilde{b}_2 \tilde{b}_1^*\) & \(1.715\times
10^{\text{-1}}\) & \(2.92\times 10^{\text{-6}}\) & \(1.715\times
10^{\text{-1}}\) & \(6.87\times 10^{\text{-6}}\) & \(-1.12\times
10^{\text{-5}}\) \\ 
\(d \bar{d} \rightarrow \nu_e \bar{\nu}_e\) & \(1.003\) & \(4.84\times
10^{\text{-5}}\) & \(1.002\) & \(6.54\times 10^{\text{-4}}\) & \(1.22\times
10^{\text{-3}}\) \\ 
\(d \bar{d} \rightarrow \nu_\mu \bar{\nu}_\mu\) & \(1.003\) & \(1.03\times
10^{\text{-4}}\) & \(1.003\) & \(1.63\times 10^{\text{-4}}\) & \(5.07\times
10^{\text{-4}}\) \\ 
\(d \bar{d} \rightarrow \tilde{e}_1 \tilde{e}_1^*\) & \(5.843\times
10^{\text{-1}}\) & \(4.91\times 10^{\text{-6}}\) & \(5.838\times
10^{\text{-1}}\) & \(1.38\times 10^{\text{-5}}\) & \(4.89\times 10^{\text{-4}}\)
\\ 
\(d \bar{d} \rightarrow \tilde{e}_2 \tilde{e}_2^*\) & \(1.28\times
10^{\text{-1}}\) & \(4.74\times 10^{\text{-6}}\) & \(1.28\times 10^{\text{-1}}\)
& \(3.03\times 10^{\text{-6}}\) & \(1.72\times 10^{\text{-5}}\) \\ 
\(d \bar{d} \rightarrow \tilde{\tau}_1 \tilde{\tau}_1^*\) & \(9.762\times
10^{\text{-2}}\) & \(3.05\times 10^{\text{-5}}\) & \(9.772\times
10^{\text{-2}}\) & \(3.57\times 10^{\text{-7}}\) & \(-1.01\times
10^{\text{-4}}\) \\ 
\(d \bar{d} \rightarrow \tilde{\tau}_2 \tilde{\tau}_2\) & \(4.616\times
10^{\text{-1}}\) & \(4.36\times 10^{\text{-5}}\) & \(4.617\times
10^{\text{-1}}\) & \(7.55\times 10^{\text{-5}}\) & \(-1.35\times
10^{\text{-4}}\) \\ 
\(d \bar{d} \rightarrow \tilde{\tau}_1 \tilde{\tau}_2\) & \(7.682\times
10^{\text{-2}}\) & \(9.73\times 10^{\text{-7}}\) & \(7.682\times
10^{\text{-2}}\) & \(4.04\times 10^{\text{-7}}\) & \(-3.85\times
10^{\text{-6}}\) \\ 
\(d \bar{d} \rightarrow \tilde{\tau}_2 \tilde{\tau}_1^*\) & \(7.679\times
10^{\text{-2}}\) & \(7.09\times 10^{\text{-6}}\) & \(7.683\times
10^{\text{-2}}\) & \(2.60\times 10^{\text{-6}}\) & \(-4.29\times
10^{\text{-5}}\) \\ 
\(d \bar{d} \rightarrow \gamma Z\) & \(6.594\times 10^{\text{-1}}\) &
\(5.08\times 10^{\text{-5}}\) & \(6.592\times 10^{\text{-1}}\) & \(4.78\times
10^{\text{-5}}\) & \(1.83\times 10^{\text{-4}}\) \\ 
\(d \bar{d} \rightarrow H^- H^+\) & \(4.539\times 10^{\text{-1}}\) &
\(2.27\times 10^{\text{-5}}\) & \(4.539\times 10^{\text{-1}}\) & \(7.42\times
10^{\text{-6}}\) & \(-6.51\times 10^{\text{-5}}\) \\ 
\(d \bar{d} \rightarrow \tilde{g} \tilde{g}\) & \(4.938\times 10^2\) &
\(2.83\times 10^{\text{-2}}\) & \(4.94\times 10^2\) & \(4.32\times
10^{\text{-2}}\) & \(-2.43\times 10^{\text{-1}}\) \\ 
\(d \bar{d} \rightarrow g g\) & \(1.118\times 10^3\) & \(1.03\times
10^{\text{-1}}\) & \(1.118\times 10^3\) & \(7.46\times 10^{\text{-2}}\) &
\(-1.83\times 10^{\text{-1}}\) \\ 
\(\gamma \gamma \rightarrow \tilde{\chi}^+_1 \tilde{\chi}^-_1\) & \(3.402\times
10^1\) & \(4.19\times 10^{\text{-3}}\) & \(3.401\times 10^1\) & \(2.50\times
10^{\text{-3}}\) & \(6.\times 10^{\text{-3}}\) \\ 
\(\gamma \gamma \rightarrow W^- W^+\) & \(3.22\times 10^2\) & \(2.21\times
10^{\text{-2}}\) & \(3.22\times 10^2\) & \(3.20\times 10^{\text{-2}}\) &
\(5.72\times 10^{\text{-2}}\) \\ 
\(\gamma \gamma \rightarrow \tilde{\chi}^+_2 \tilde{\chi}^-_2\) & \(3.696\times
10^1\) & \(2.23\times 10^{\text{-3}}\) & \(3.696\times 10^1\) & \(2.88\times
10^{\text{-3}}\) & \(4.95\times 10^{\text{-3}}\) \\ 
\(\gamma \gamma \rightarrow d \bar{d}\) & \(1.212\) & \(1.75\times
10^{\text{-4}}\) & \(1.213\) & \(2.79\times 10^{\text{-4}}\) & \(-6.44\times
10^{\text{-4}}\) \\ 
\(\gamma \gamma \rightarrow t \bar{t}\) & \(2.008\times 10^1\) & \(3.19\times
10^{\text{-3}}\) & \(2.009\times 10^1\) & \(2.72\times 10^{\text{-3}}\) &
\(-8.75\times 10^{\text{-3}}\) \\ 
\(\gamma \gamma \rightarrow b \bar{b}\) & \(1.213\) & \(1.69\times
10^{\text{-4}}\) & \(1.213\) & \(9.21\times 10^{\text{-5}}\) & \(1.13\times
10^{\text{-4}}\) \\ 
\(\gamma \gamma \rightarrow \tilde{t}_1 \tilde{t}_1^*\) & \(6.059\) &
\(6.91\times 10^{\text{-4}}\) & \(6.058\) & \(4.91\times 10^{\text{-4}}\) &
\(7.36\times 10^{\text{-4}}\) \\ 
\(\gamma \gamma \rightarrow \tilde{t}_2 \tilde{t}_2^*\) & \(3.777\) &
\(3.17\times 10^{\text{-4}}\) & \(3.778\) & \(3.37\times 10^{\text{-4}}\) &
\(-3.11\times 10^{\text{-4}}\) \\ 
\(\gamma \gamma \rightarrow \tilde{d}_1 \tilde{d}_1^*\) & \(2.44\times
10^{\text{-1}}\) & \(1.57\times 10^{\text{-5}}\) & \(2.44\times 10^{\text{-1}}\)
& \(1.55\times 10^{\text{-5}}\) & \(5.97\times 10^{\text{-6}}\) \\ 
\(\gamma \gamma \rightarrow \tilde{d}_2 \tilde{d}_2^*\) & \(2.57\times
10^{\text{-1}}\) & \(1.99\times 10^{\text{-5}}\) & \(2.569\times
10^{\text{-1}}\) & \(1.54\times 10^{\text{-5}}\) & \(7.42\times 10^{\text{-5}}\)
\\ 
\(\gamma \gamma \rightarrow \tilde{e}_1 \tilde{e}_1^*\) & \(1.319\times 10^1\) &
\(8.09\times 10^{\text{-4}}\) & \(1.319\times 10^1\) & \(9.91\times
10^{\text{-4}}\) & \(1.09\times 10^{\text{-3}}\) \\ 
\(\gamma \gamma \rightarrow \tilde{e}_2 \tilde{e}_2^*\) & \(1.391\times 10^1\) &
\(7.59\times 10^{\text{-4}}\) & \(1.391\times 10^1\) & \(1.32\times
10^{\text{-3}}\) & \(-2.98\times 10^{\text{-4}}\) \\ 
\(\gamma \gamma \rightarrow \tilde{\tau}_1 \tilde{\tau}_1^*\) & \(1.406\times
10^1\) & \(1.52\times 10^{\text{-3}}\) & \(1.406\times 10^1\) & \(1.13\times
10^{\text{-3}}\) & \(-6.26\times 10^{\text{-3}}\) \\ 
\(\gamma \gamma \rightarrow \tilde{\tau}_2 \tilde{\tau}_2\) & \(1.312\times
10^1\) & \(6.91\times 10^{\text{-4}}\) & \(1.312\times 10^1\) & \(1.17\times
10^{\text{-3}}\) & \(1.12\times 10^{\text{-3}}\) \\ 
\(\gamma \gamma \rightarrow H^- H^+\) & \(8.983\) & \(1.07\times
10^{\text{-3}}\) & \(8.978\) & \(8.81\times 10^{\text{-4}}\) & \(4.25\times
10^{\text{-3}}\)\\\hline
\caption{Comparision between the WHIZARD stock MSSM and the version generated
by SARAH for several for various $2\rightarrow 2$ processes. The calculated
values for the cross sections are shown together with the respective absolute
integration errors. See the text for details.}
\label{tab:WHIZARD_check}
\end{longtable}
}

 \end{appendix}

\bibliographystyle{ieeetr}

\end{document}